\documentclass{emulateapj}

\def\myputfigure#1#2#3#4#5%
{\vskip#5pt\makebox[0pt]{\hskip#2in
\includegraphics[width=#3\textwidth]{#1}}\vskip#4pt\hfill}

\newcommand\lsim{\mathrel{\rlap{\lower4pt\hbox{\hskip1pt$\sim$}}
         \raise1pt\hbox{$<$}}}
\newcommand\gsim{\mathrel{\rlap{\lower4pt\hbox{\hskip1pt$\sim$}}
         \raise1pt\hbox{$>$}}}

\newcommand{\Tvir}{T_{\rm vir}}

\newcommand{\nf}{x_{\rm HI}}

\newcommand{\Mmin}{M_{\rm min}}

\newcommand{\delc}{\delta_c(z)}
\newcommand{\delbias}{\delta_0}
\newcommand{\rhoO}{\langle \rho_0 \rangle}
\newcommand{\fcol}{F_{\rm col}(>\Mmin, z)}
\newcommand{\zinit}{z_{\rm init}}

\newcommand{\Msun}{M_\odot}

\newcommand{\fcd}{f_{\rm cd}}

\newcommand{\flash}{Flash}
\newcommand{\nothing}{NoUVB}
\newcommand{\heatinglow}{Heat0.08}
\newcommand{\heatinghigh}{Heat0.8}
\newcommand{\earlyheating}{EarlyHeat0.8}

\newcommand{\zuvbon}{z_{\rm UVB, on}}
\newcommand{\zuvboff}{z_{\rm UVB, off}}
\newcommand{\delN}{\delta_{N, {\rm cd}}(z)}
\newcommand{\delM}{\delta_{M, {\rm cd}}(z)}
\newcommand{\Nrun}{N_{\rm cd}^{{\rm run}i}}
\newcommand{\Mrun}{M_{\rm cd}^{{\rm run}i}}
\newcommand{\rhocut}{\rho_{\rm cutoff}}
\newcommand{\hMpc}{h^{-1} ~ {\rm Mpc}}
\newcommand{\ffit}{f_{\rm fit}}

\newcommand{\OMm}{\Omega_{\rm M}}
\newcommand{\OMb}{\Omega_b}
\newcommand{\tHtwo}{t_{\rm H_2}}
\newcommand{\tcomp}{t_{\rm C}}
\newcommand{\trec}{t_{\rm rec}}
\newcommand{\fdelay}{f_{\rm delay}}

\begin{document}
\title{UV Radiative Feedback on High--Redshift Proto--Galaxies}

\author{Andrei Mesinger, Greg L. Bryan, Zolt\'{a}n Haiman}
\affil{Department of Astronomy, Columbia University, 550 West 120th  
Street, New York, NY 10027}
\vspace{+0.4cm}

\submitted{Submitted to the ApJ}

\begin{abstract}
We use three-dimensional hydrodynamic simulations to investigate the
effects of a transient photoionizing ultraviolet (UV) flux on the
collapse and cooling of pregalactic clouds.  These clouds have masses
in the range $10^5$ -- $10^7~\Msun$, form at high redshifts
($z\gsim18$), and are assumed to lie within the short--lived
cosmological HII regions around the first generation of stars.  In
addition, we study the combined effects of this transient UV flux and
a persistent Lyman--Werner (LW) background (at photon energies below
13.6eV) from distant sources.  In the absence of a LW background, we
find that a critical specific intensity of $J_{\rm UV} \sim 0.1 \times
10^{-21}{\rm ergs~s^{-1}~cm^{-2}~Hz^{-1}~sr^{-1}}$ demarcates a
transition from net negative to positive feedback for the halo
population. A weaker UV flux stimulates subsequent star formation
inside the fossil HII regions, by enhancing the ${\rm H_2}$ molecule
abundance. A stronger UV flux significantly delays star--formation by
reducing the gas density, and increasing the cooling time, at the
centers of collapsing halos.  At a fixed $J_{\rm UV}$, the sign of the
feedback also depends strongly on the density of the gas at the time
of UV illumination.  Regardless of the whether the feedback is
positive or negative, we find that once the UV flux is turned off, its
impact stars to diminish after $\sim30\%$ of the Hubble time.  In the
more realistic case when a LW background is present, with $J_{\rm LW}
\gsim 0.01 \times 10^{-21}{\rm ergs~s^{-1}~cm^{-2}~Hz^{-1}~sr^{-1}}$,
strong suppression persists down to the lowest redshift ($z=18$) in
our simulations.  Finally, we find evidence that heating and
photoevaporation by the transient UV flux renders the $\sim 10^6~{\rm
M_\odot}$ halos inside fossil HII regions more vulnerable to
subsequent ${\rm H_2}$ photo--dissociation by a LW background.
\end{abstract}
\keywords{cosmology: theory -- early Universe -- galaxies:
high-redshift -- evolution}
\vspace{+0.5cm}

\section{Introduction}
\label{sec:intro}

Semi-analytical and numerical studies agree that the first generation
of stars are likely to have formed at very high redshifts, $z \gsim
20$, at locations corresponding to rare peaks of the fluctuating
primordial density field.  Numerical simulations suggest that the
first collapsed objects host high--mass ($\sim100\Msun$),
low--metalicity (so--called 'Population III', hereafter 'PopIII')
stars \citep{ABN02, BCL02}. In the absence of any feedback processes,
these stars, and their accreting remnant black holes (BHs), could
significantly reionize the intergalactic medium (IGM). Recent evidence
from the $z\sim6$ quasars discovered in the Sloan Digital Sky Survey
(SDSS), whose spectra exhibit dark and sharp Gunn-Peterson troughs
\citep{FCK06, MH04}, and from the cosmic microwave polarization
anisotropies measured by the {\it WMAP} satellite, which imply a
conservative electron scattering optical depth $\tau_e = 0.09\pm0.03$
\citep{Page06, Spergel06}, suggest that cosmic reionization was
delayed to a later stage of structure formation, $z\sim$ 6 -- 10.
Such a late reionization would require a substantial suppression of
very high redshift ionizing sources (e.g., \citealt{HB06}).

The impact that the first generation of stars would have on their
surroundings plays a crucial role in how reionization progresses at
high redshifts \citep{HH03, Cen03_postWMAP, WL03_postWMAP}, but is
poorly understood from first principles.  Several feedback mechanisms
are expected to be potentially important, including both direct
chemical and energetic input from the first supernovae, and radiative
processes due to UV and X--ray radiation from the first stars
themselves.  The enhanced metallicity due to the supernovae will
result in an increase in the cooling rate, leading to more efficient
star formation and lower stellar masses (e.g., \citealt{Omukai2000,
Bromm2003}).  The signature of the metals that are produced may be
seen in quasar absorption spectra, although semi-analytic work on the
subject suggests that chemical feedback from pregalactic objects did
not play a large role in setting the observed intergalactic
metallicity distribution at $z\lsim 6$ \citep{SFM02, SSF03}.  In this
paper we focus on radiative feedback, and postpone the study of PopIII
metal pollution and feedback to future work.

Radiative feedback can be either positive and negative, in that it can
enhance or suppress subsequent star--formation.  Positive feedback can
result when the enhanced free-electron fraction from ionizing photons
or from shocks catalyzes the formation of molecular hydrogen (H$_2$),
which can provide the dominant cooling channel at high-densities and
low temperatures.  The catalyst electrons can be produced by X-rays
emitted as a result of gas accretion onto early BHs
(e.g., \citealt{HRL96}), or by a previous epoch of photoionization
inside ``fossil'' HII regions \citep{RGS02b, OH03, OShea05}, or by
collisional ionization in protogalactic shocks \citep{SK87, Susa98,
OH02}. Indeed, cosmological simulations have noted net positive
feedback close to the edge of HII regions \citep{RGS02b, KM05}.
Negative feedback can result from chemical or thermodynamical effects.
UV photons in the Lyman-Werner (LW) band of ${\rm H_2}$ can dissociate
these molecules, thereby reducing their effectiveness in cooling the
gas (e.g., \citealt{HRL97, HAR00, CFA00, MBA01}).  Active radiative
heating can photo--evaporate gas in low---mass halos
\citep{Efstathiou92, BL99, SIR04}.  Additionally, a past episode of
photoionization heating in inactive ``fossil'' HII regions can leave
the gas with tenacious excess entropy, reducing the gas densities,  
hindering
${\rm H_2}$ formation, cooling and collapse \citep{OH03}.

The above feedback effects, their relative importance, and their net
outcome on star formation within the population of early halos, is not
well understood ab--initio, and is poorly constrained by observations
at high redshifts. Although invaluable in furthering our understanding
of the main physical concepts, semi-analytic studies (e.g.,
\citealt{HRL97, OH03, MST05}) do not fully take into account the
details of cosmological density structure and evolution, which can be
very important.  On the other hand, numerical studies are limited in
scope due to computational restrictions associated with full radiative
transfer on such large scales and large dynamical ranges (see, e.g.,
\citealt{Iliev06} for a recent review).  In particular, it is
difficult to couple radiative transfer and hydrodynamics --- most work
to date has focused on either one or the other issue (with some
exceptions, e.g., \citealt{GA01, RGS02a,SIR04}).  Techniques
approximating full radiative transfer can provide a crucial speed-up
of computing time, but such simulations still do not provide a large
statistical sample for a detailed study of radiative feedback, and/or
can be limited by a small dynamical range, thereby missing the
smallest halos which would be most susceptible to negative feedback
(e.g., \citealt{RGS02a}).

{\it The purpose of the present paper is to statistically investigate
UV radiative feedback associated with the first generation of stars}.
Prior numerical studies have either focused on radiation in a single
different band, such as LW photons \citep{MBA01} or X-rays
\citep{MBA03, KM05} or lacked quantitative statistics and have not
included photo--heating and photo--evaporation of low--mass halos
\citep{RGS02a, RGS02b, OShea05}.  The recent work by \citet{ABS05} has
studied the impact of photoionizing radiation in detail
within a single HII region, but without self-consistently modeling
the hydrodynamics.  In the present study, we quantify the
combined effects of UV photo-ionization and LW radiation from nearby
Pop III star formation. Rather than simulating the radiative transfer
within an individual HII region, we take a statistical approach.  In
particular, we examine how halos which are in the process of
collapsing are affected by spatially constant but potentially
short-lived radiation backgrounds at various
intensities.  This allows us to calibrate the sign and amplitude of
the resulting feedback, in order to include these effects in
future semi-analytic studies.  To this end, we carry out simulations
in which a large region is photo-ionized for a short period of time,  
and neglect radiative transfer effects (we discuss the impact of this  
approximation in more detail in \S~\ref{sec:results} below).

The rest of this paper is organized as follows.  In \S~\ref{sec:sims},
we describe the simulations. In \S~\ref{sec:results}, we present the
results of the simulations with photoionization heating, but without a
LW background. In \S~\ref{sec:lw}, we discuss simulation runs that
also include a LW background.  Finally, in \S~\ref{sec:conc}, we
summarize our conclusions and discuss the implications of this work.
For completeness and for reference, in the Appendix, we present the
dark matter halo mass functions found in our simulations.

Throughout this paper, we adopt the background cosmological parameters
($\Omega_\Lambda$, $\Omega_{\rm M}$, $\Omega_b$, n, $\sigma_8$, $H_0$)
= (0.7, 0.3, 0.047, 1, 0.92, 70 km s$^{-1}$ Mpc$^{-1}$), consistent
with the measurements of the power spectrum of CMB temperature
anisotropies by the first year of data from the {\it WMAP} satellite
\citep{Spergel03}.  The three--year data from {\it WMAP} favors
decreased small--scale power (i.e. lower values for $\sigma_8$ and
$n_s$; \citealt{Spergel05}), which would translate to a $\sim 15\%$
redshift delay, but would not change our conclusions.  Unless stated
otherwise, we quote all quantitities in comoving units.

\section{Simulations}
\label{sec:sims}

We use the Eulerian adaptive mesh refinement (AMR) code Enzo, which is
described in greater detail elsewhere \citep{Bryan99, NB99}.  Our
simulation volume is 1 $(\hMpc)^3$, initialized at $\zinit=99$ with
density perturbations drawn from the \citet{EH99} power spectrum.  We
first run a low resolution ($128^3$ particles), dark matter (DM) only
run down to $z=15$, to find the highest density peak in the box.  We
then re-center the box around the spatial location of that peak, and
rerun the simulations with the inclusion of gas and at a higher
resolution inside a 0.25 $\hMpc$ cube, centered in the 1 $\hMpc$ box.
This refined central region has an average physical overdensity of
$\delta (\zinit) \equiv \rho/\bar{\rho} - 1 =0.1637$, corresponding to
a 2.4 $ \sigma$ mass fluctuation of an equivalent spherical volume.
We use such a biased region for our analysis since it hosts a large
number of halos at high regshifts.  This not only provides good number
statistics, but also helps mimic a pristine, unpolluted region which
is likely to host the first generation of stars.  We postpone a lower
redshift analysis to a future work.

Our fiducial runs are shown in Table \ref{tbl:runs}.  Our root grid is
$128^3$.  We have two additional static levels of refinement inside
the central 0.25 $\hMpc$ cube.  Furthermore, grid cells inside the
central region are allowed to dynamically refine so that the Jeans
length is resolved by at least 4 grid zones and no grid cell contains
more than 4 times the initial gas mass element.  Each additional grid
level refines the mesh length of the parent grid cell by a factor of
2.  We allow for a maximum of 10 levels of refinement inside the
refined central region, granting us a spatial resolution of 7.63
$h^{-1}~{\rm pc}$.  This comoving resolution translates to 0.36
$h^{-1}~{\rm proper~pc}$ at $z=20$.  We find that this resolution is
sufficient to adequately resolve the gross physical processes in the
few $\times 10^5-10^7$ $\Msun$ halos of interest in this work by
comparing with higher mass and spatial resolution runs (not shown
here).  The dark matter particle mass is 747 $\Msun$.  We also include
the non-equilibrium reaction network of nine chemical species (H,
H$^+$, He, He$^+$, He$^{++}$, $e^-$, H$_2$, H$_2^+$, H$^-$) using the
algorithm of \citet{Anninos97}, and initialized with
post-recombination abundances from \citet{AN96}.  Our analysis below
is based on the central refined region; the low resolution dark matter
outside the refined region serves to provide the necessary tidal
forces to our refined region.  Readers interested in further details
concerning the simulation methodology are encouraged to consult, e.g.,
\citet{MBA01}.

As shown in Table \ref{tbl:runs}, we have performed five different
runs without a LW background, distinguished by the duration or
amplitude of the assumed UV background radiation (hereafter UVB), and
six additional runs that include an additional constant LW background.
For the UV radiation we assume an isotropic background flux with a
$T=2\times10^4$K blackbody spectral shape, normalized at the hydrogen
ionization frequency, $h \nu_H$ = 13.6 eV. The values of $J_{\rm UV}$
are shown in Table \ref{tbl:runs} in units of $10^{-21}{\rm
ergs~s^{-1}~cm^{-2}~Hz^{-1}~sr^{-1}}$.  The \nothing\ run contains no
UV radiation, and serves mainly as a reference run. The \heatinglow\
and \heatinghigh\ runs include a UVB with $J_{\rm UV} = 0.08, 0.8$,
respectively.  The value of $J_{\rm UV}=0.08$ was chosen to correspond
to the mean UV flux expected inside a typical HII region surrounding a
primordial star (e.g., \citealt{ABS05}).  As we do not include
dynamically expanding HII regions in our code, the \heatinghigh\ and
\flash\ runs can be viewed as extremes, corresponding to conditions
close to the center and close to the edge of the HII region,
respectively.  More generally, studying a range of values of $J_{\rm
UV}$ is useful, since the UV flux of a massive, Pop-III star is
uncertain.  In the latter two runs, the UVB is turned on at
$\zuvbon=25$ and turned off at $\zuvboff=24.64$. This redshift range
corresponds to a typical theoretical stellar lifetime, $\sim3$Myr, of
a $\sim 100 \Msun$ primordial (Pop-III) star \citep{Schaerer02}.  The
flash ionization run, \flash, instantaneously sets the gas temperature
to T=15000 K and the hydrogen neutral fraction to $\nf = 10^{-3}$
throughout the simulation volume, but involves no heating thereafter.
This allows us to compare our results to those of \citet{OShea05}, and
to identify the importance of including the dynamical effects of the
photo--heating.  We also include an early UVB run, \earlyheating, with
$J_{\rm UV} = 0.8$, $\zuvbon=33$, and $\zuvboff=33.23$, in order to
study how our results vary with redshift (or equivalently, with the
ionizing efficiency of the first sources).  Finally, the six runs in
the bottom half of the Table repeat pairs of the \nothing\ and
\heatinghigh\ runs with three different constant LW backgrounds
($J_{\rm LW}=$ 0.001, 0.01, and 0.1, normalized at 12.87eV in units of
$10^{-21}{\rm ergs~s^{-1}~cm^{-2}~Hz^{-1}~sr^{-1}}$, and assumed to be
frequency--independent within the narrow LW band).

We stress that we do not attempt to model reionization in this work.
Rather, we focus on the statistical analysis of the feedback
associated with the UV and LW backgrounds.  In particular, we want to
simulate the effect of a short-lived UV and a persistent LW background
on a range of halo masses, in order to calibrate the net impact of
fossil HII regions on subsequent star formation within these regions.
In this we are aided by the large number of halos (few hundred) in our
refined region, dozens of which manage to host cold, dense (CD) gas by
the end of our simulation runs at $z=18$.

\begin{table}[ht]
\vspace{0.2cm}
\caption{Summary of Simulation Runs}
\vspace{-0.2cm}
\label{tbl:runs}
\begin{center}
\begin{tabular}{ccccc}
\hline
Run Name & $J_{\rm UV}$ &  $\zuvbon$ & $\zuvboff$ &   $J_{\rm LW}$  \\
\hline
\hline
\multicolumn{5}{c}{Runs without a LW background}\\
\hline
\nothing      &   0   &    NA   &   NA    & 0 \\
\flash        &   --  &    25   &   25    & 0 \\
\heatinglow   &  0.08 &    25   &  24.62  & 0 \\
\heatinghigh  &  0.8  &    25   &  24.62  & 0 \\
\earlyheating &  0.8  &    33   &  32.23  & 0 \\
\hline
\multicolumn{5}{c}{Runs with a LW background}\\
\hline
\nothing      &   0   &    NA   &   NA    & 0.001\\
\heatinghigh  &  0.8  &    25   &  24.62  & 0.001\\
\nothing      &   0   &    NA   &   NA    & 0.01\\
\heatinghigh  &  0.8  &    25   &  24.62  & 0.01\\
\nothing      &   0   &    NA   &   NA    & 0.1\\
\heatinghigh  &  0.8  &    25   &  24.62  & 0.1\\
\hline
\end{tabular}\\
\end{center}
\end{table}

We use the HOP algorithm \citep{EH98} on the DM particles to identify
DM halos.  We then convert the resulting DM halo mass, $M_{DM}$, to a
total halo mass using the average conversion factor, $M_{\rm halo} =
M_{DM} ~ \OMm/(\OMm-\OMb)$.  We find that the halo masses defined in
this manner agree to within a factor of two with masses obtained by
integrating the densities over a sphere whose radius is the halo's
virial radius.

In the analysis below, it will be useful to define the fraction of
total gas within the virial radius which is cold and dense (CD),
$\fcd$.  By cold, we mean gas whose temperature is $<0.5 \Tvir$, where
$\Tvir$ is the halo's virial temperature (for how virial temperatures
are associated with halos in the simulation, see \citealt{MBA01}).  By
dense, we mean gas whose density is $> 10^{19}$ $\Msun$ Mpc$^{-3}$
$\approx$ 330 cm$^{-3}$, roughly corresponding to the density at which
the baryons become important to the gravitation potential at the core,
taken to be an immediate precursor to primordial star formation
\citep{ABN02}.  Henceforth, we treat $\fcd$ as a proxy for the
fraction of the halo's gas which is available for star formation.

Additionally, we discount halos which have been substantially
contaminated by the large (low-resolution) DM particles outside of our
refined region.  Specifically, we remove fromS our analysis halos with
an average DM particle mass greater than 115\% of the refined region's
DM mass resolution, $747\Msun$.  Another possible source of
contamination arises from closely separated halos.  If some CD gas
belonging to a halo is within another halo's virial radius (most
likely in the process of merging), the other halo could undeservedly
be flagged as containing CD gas as well.  To counteract this, we set
$\fcd=0$ for low--mass halos ($<2\times10^5\Msun$) whose centers are
less than $\sim$ 5 $h^{-1}$ kpc away from the center of a halo
containing CD gas.

\section{Results without a LW Background}
\label{sec:results}

\begin{figure*}
\vspace{+0\baselineskip}
{
\includegraphics[width=0.245\textwidth]{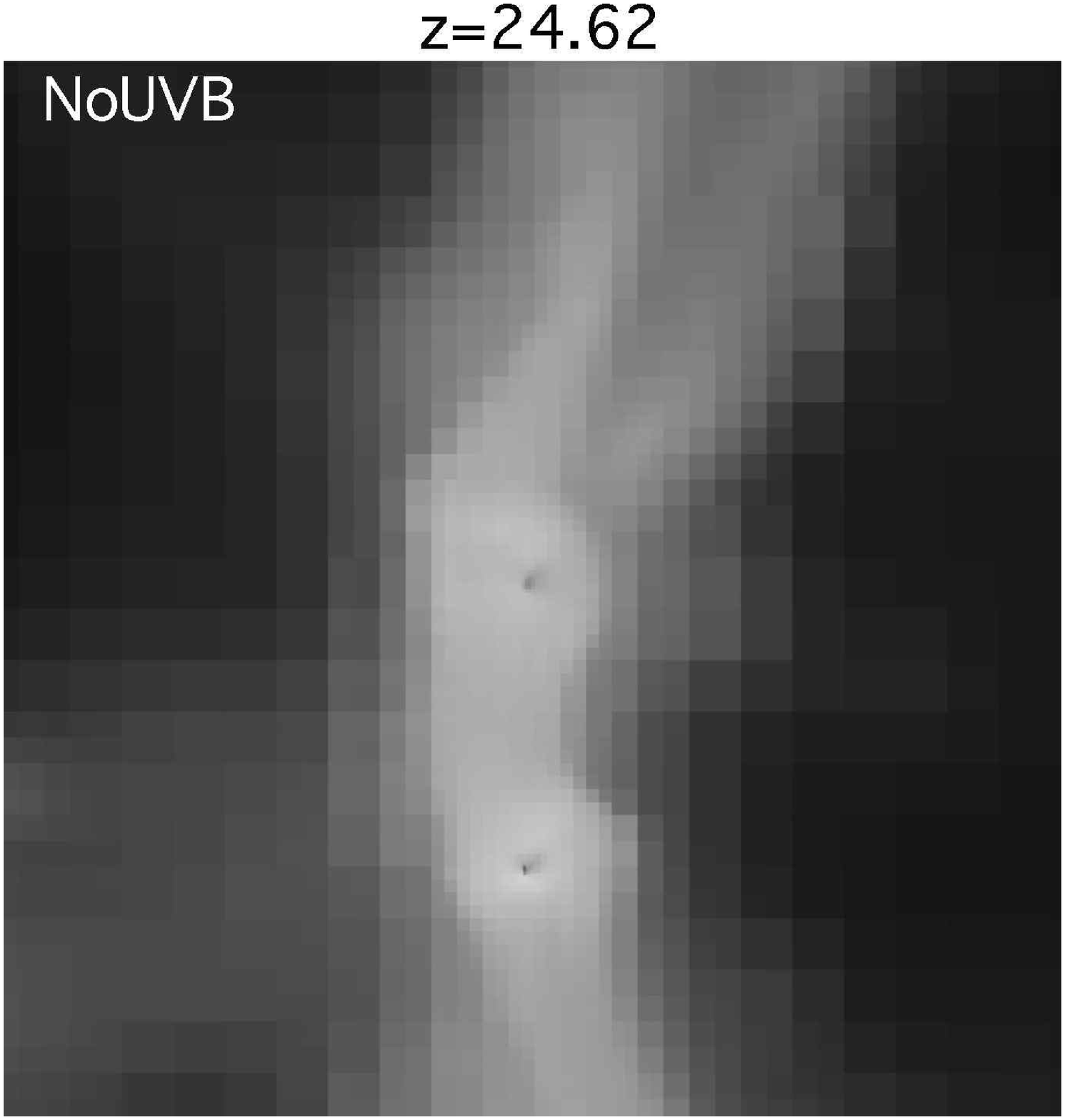}
\includegraphics[width=0.245\textwidth]{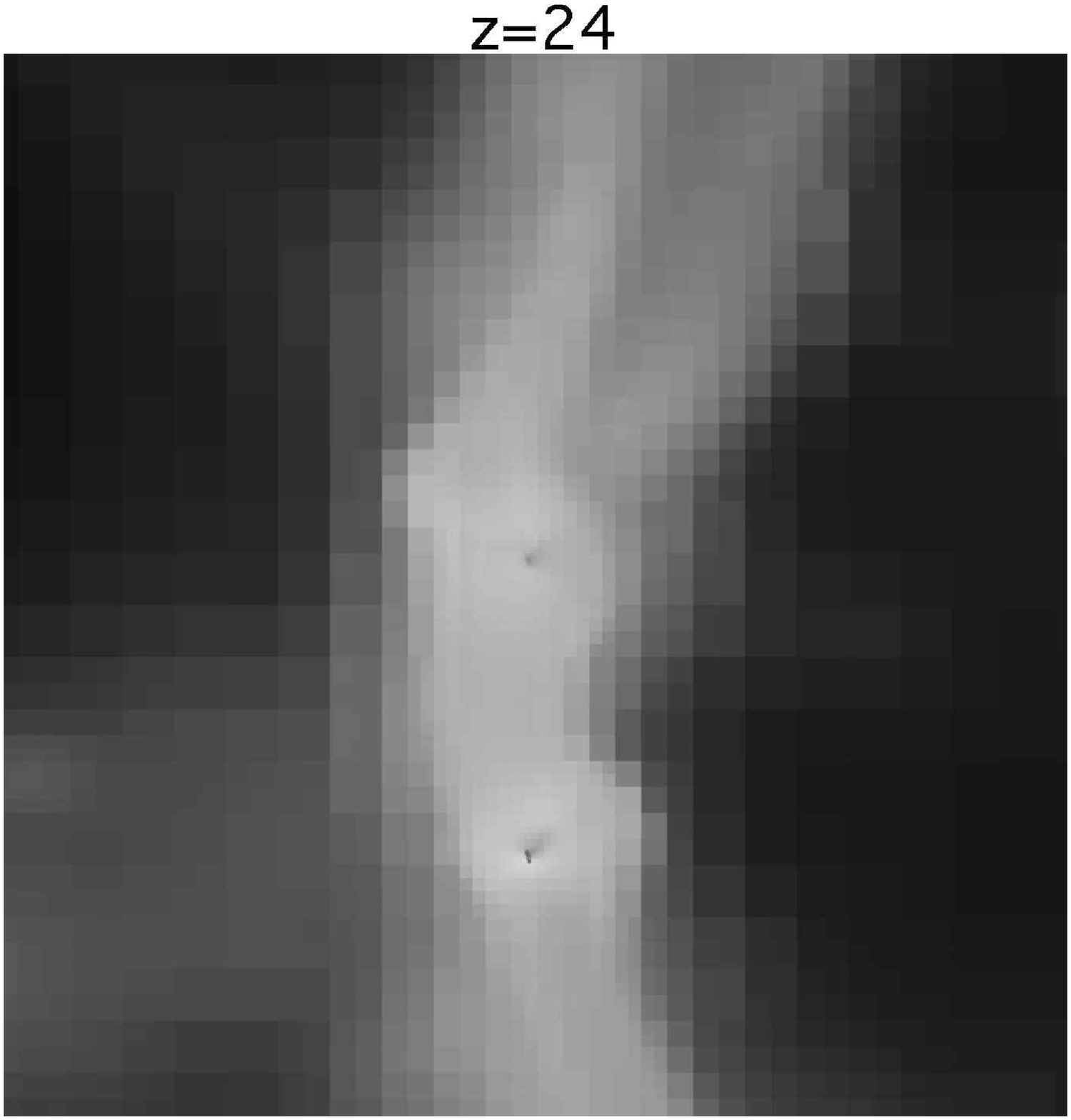}
\includegraphics[width=0.245\textwidth]{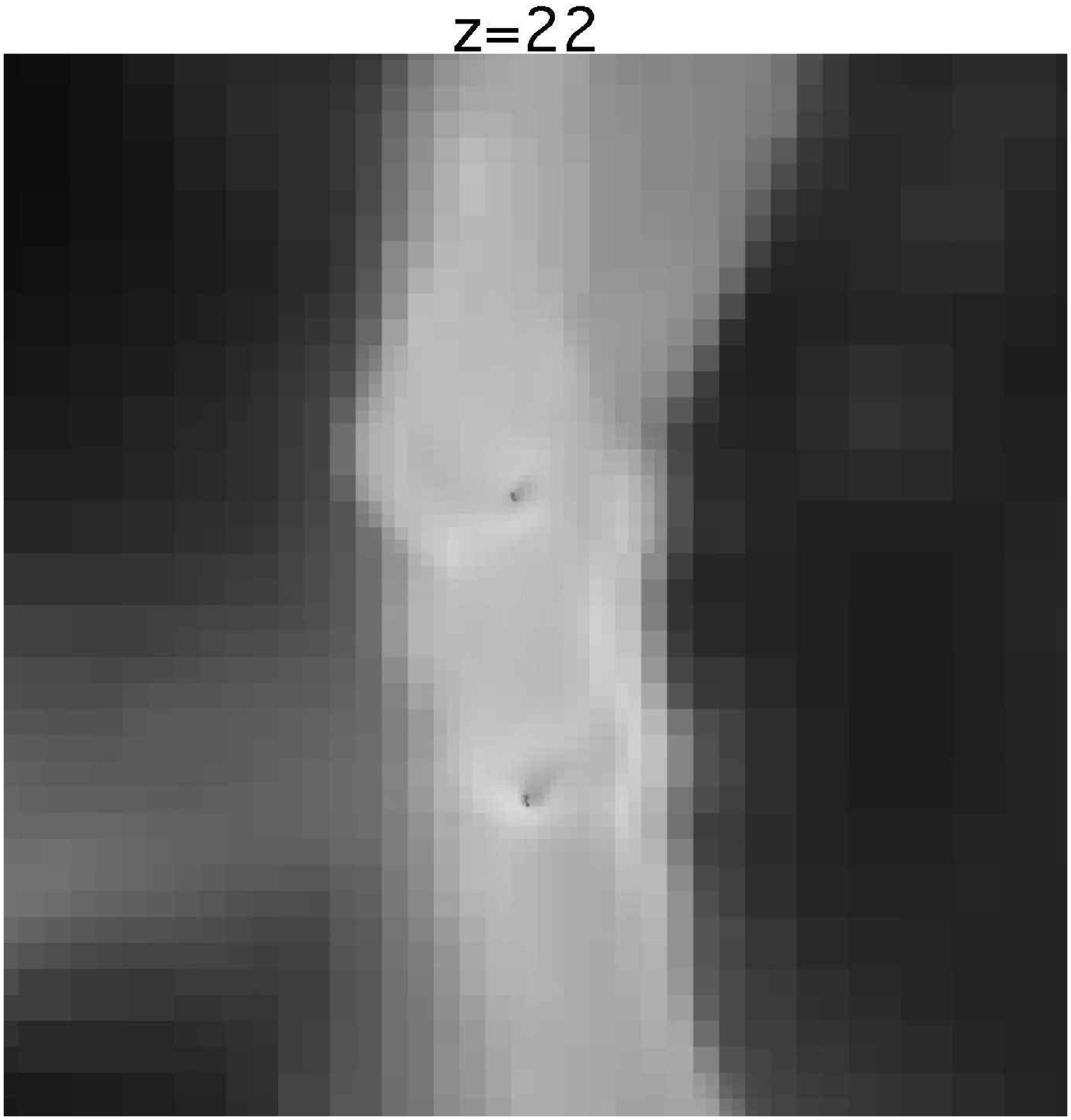}
\includegraphics[width=0.245\textwidth]{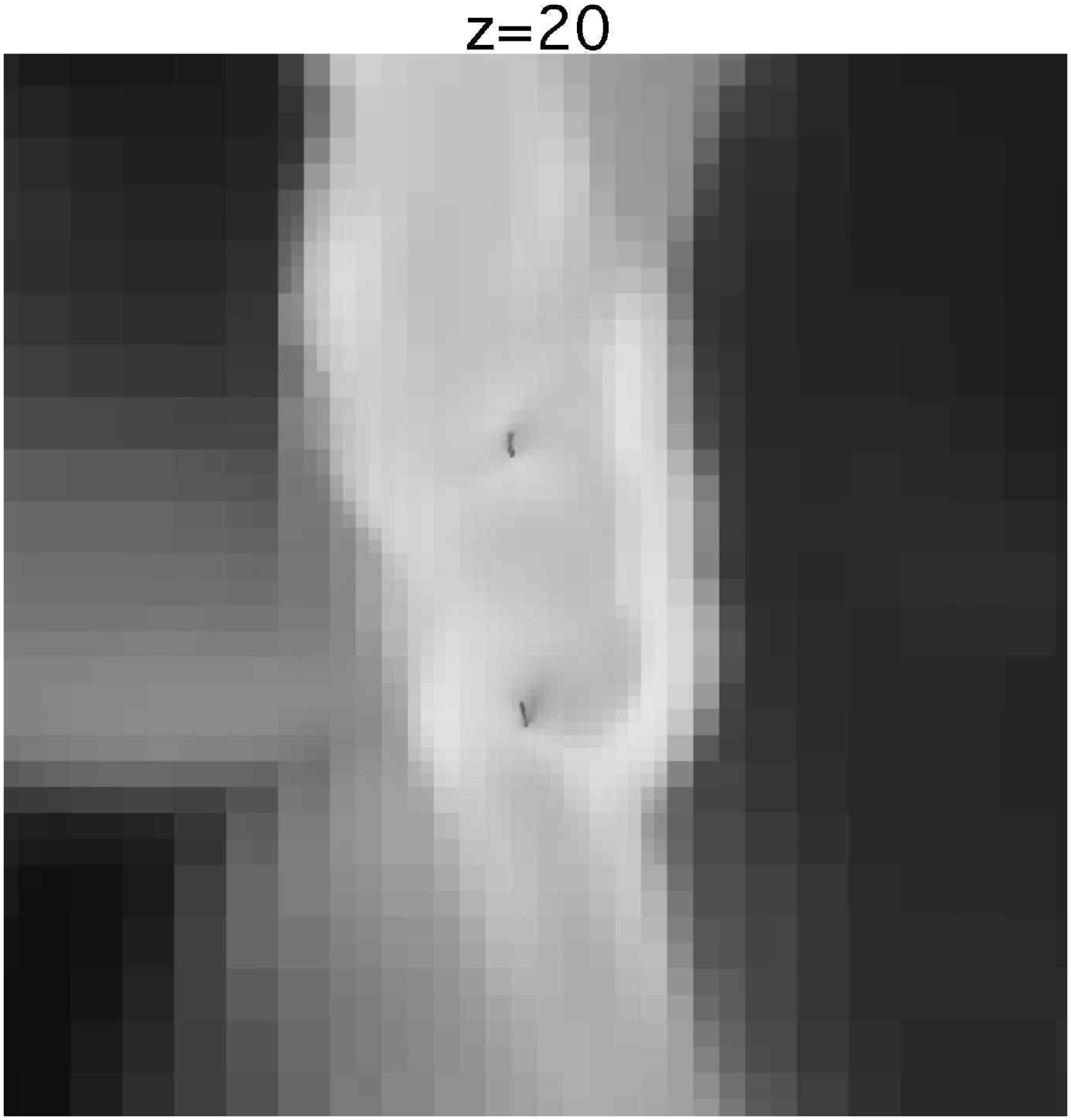}
}
{
\includegraphics[width=0.245\textwidth]{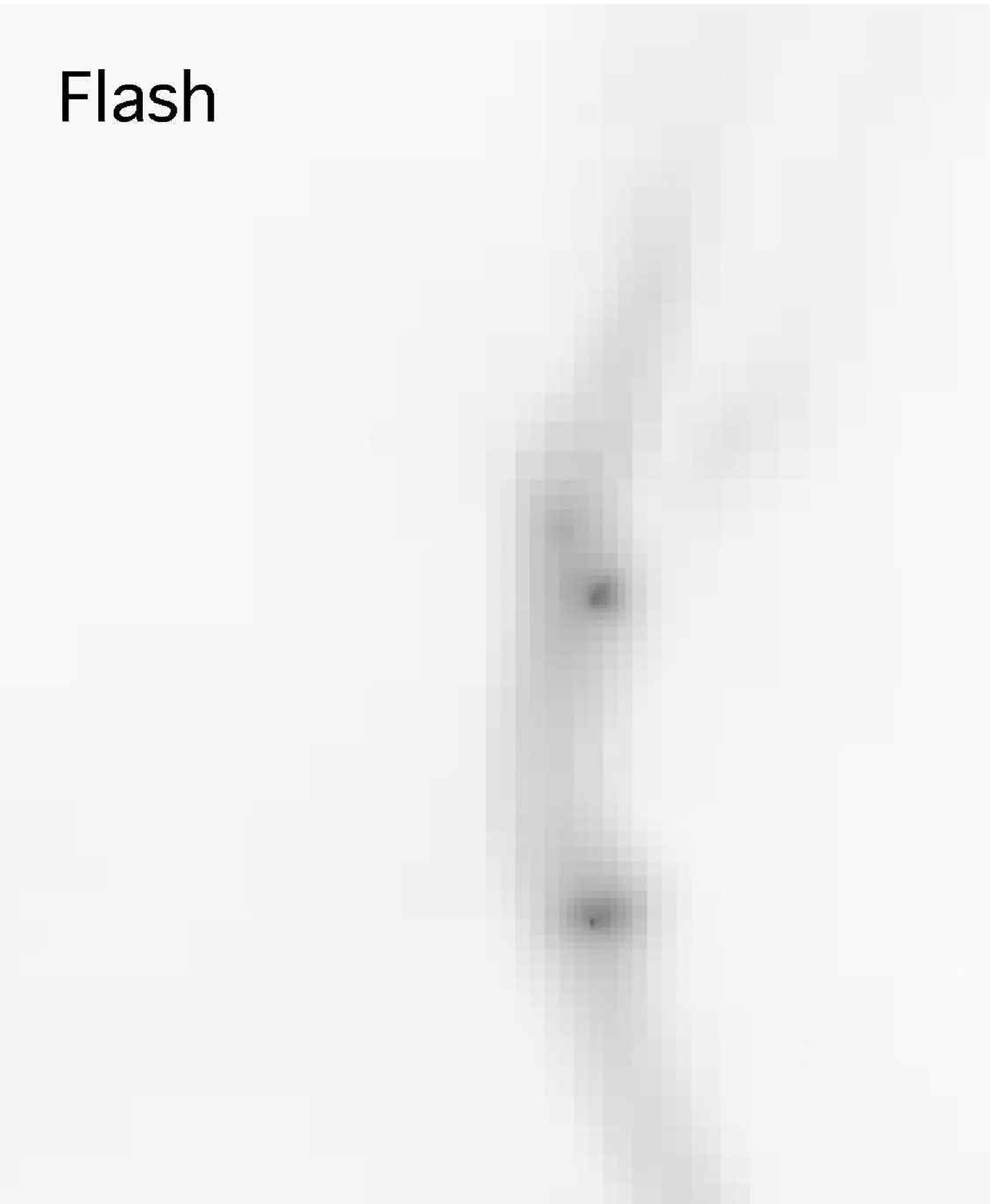}
\includegraphics[width=0.245\textwidth]{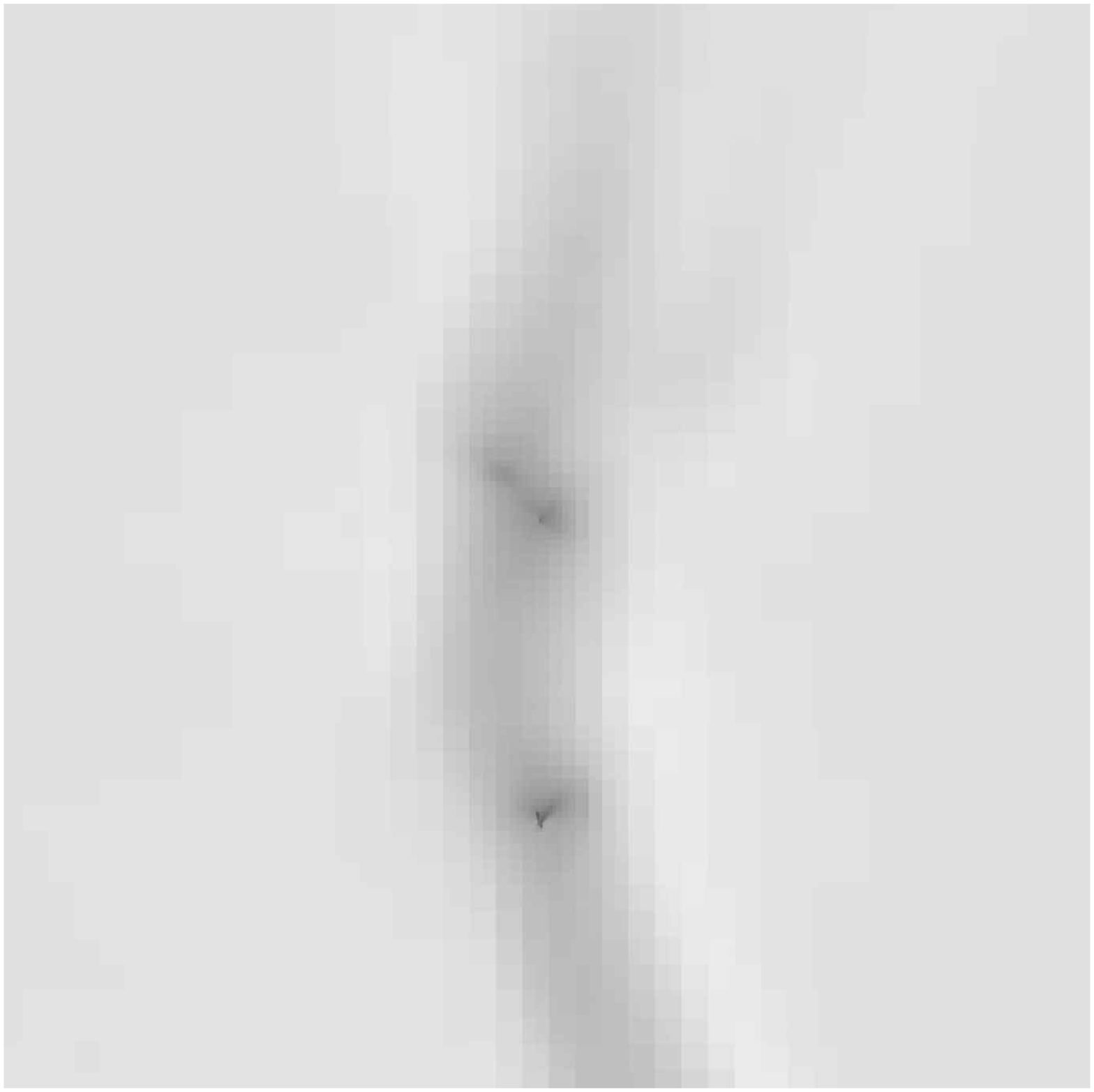}
\includegraphics[width=0.245\textwidth]{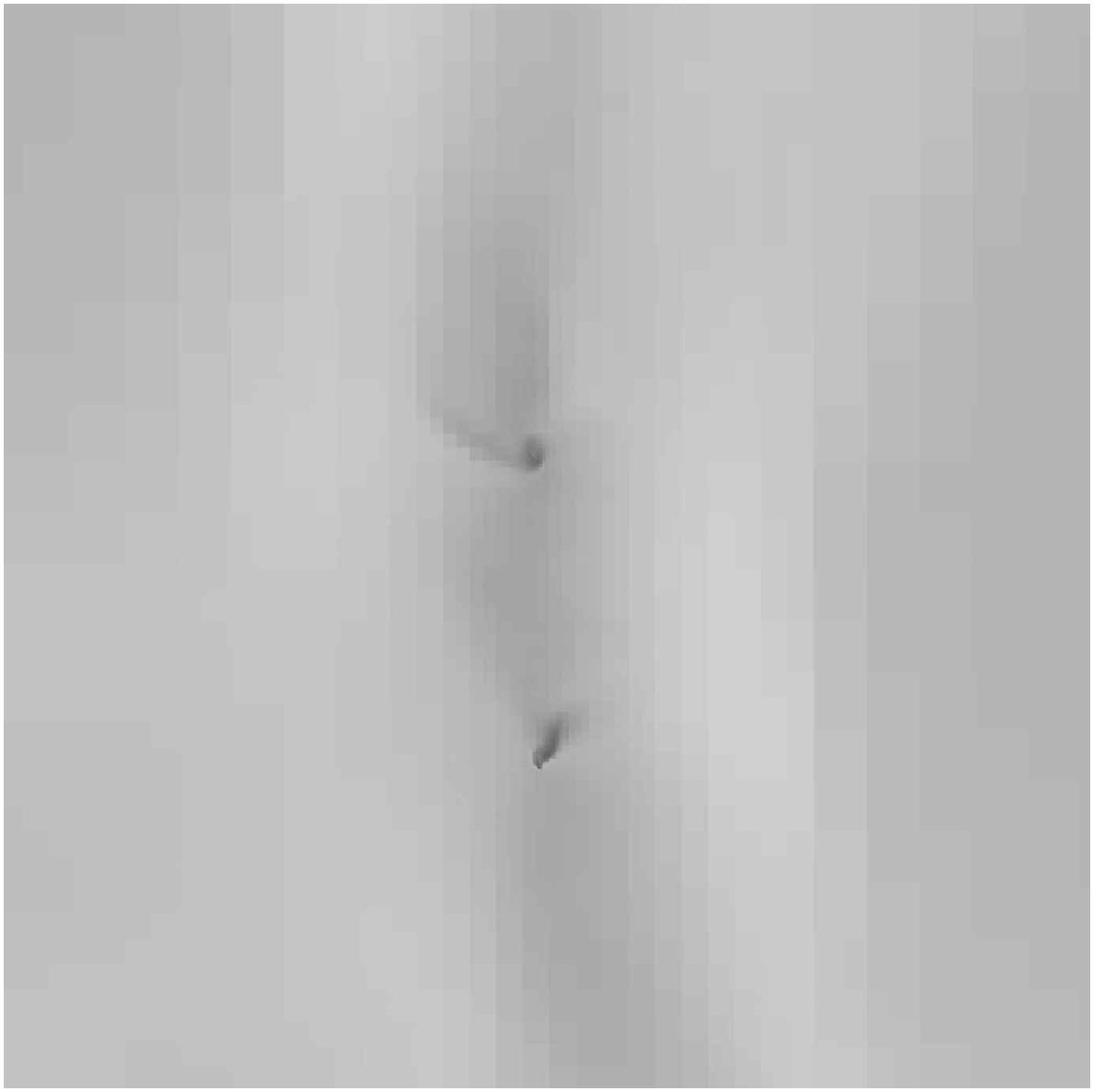}
\includegraphics[width=0.245\textwidth]{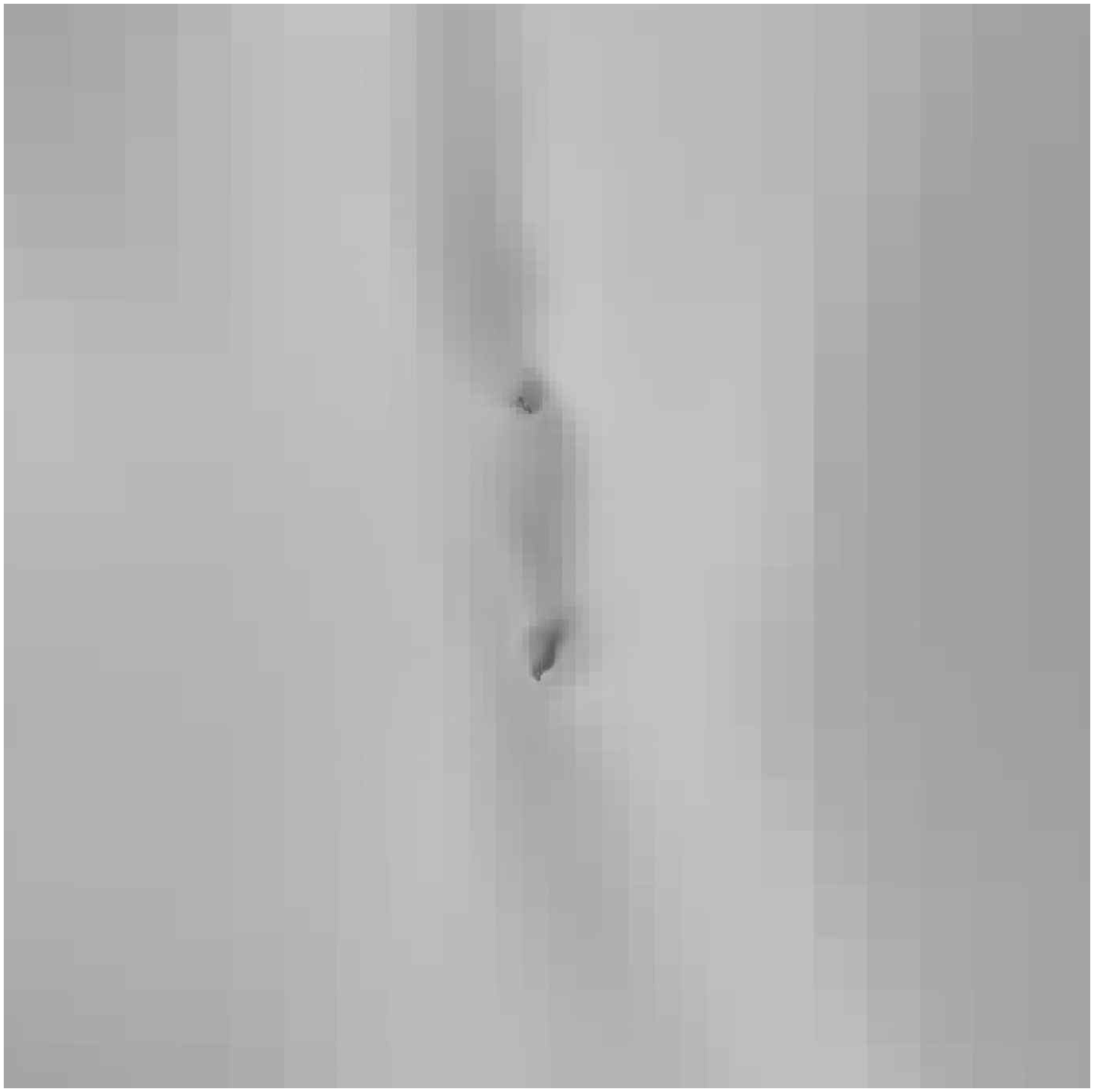}
}
{
\includegraphics[width=0.245\textwidth]{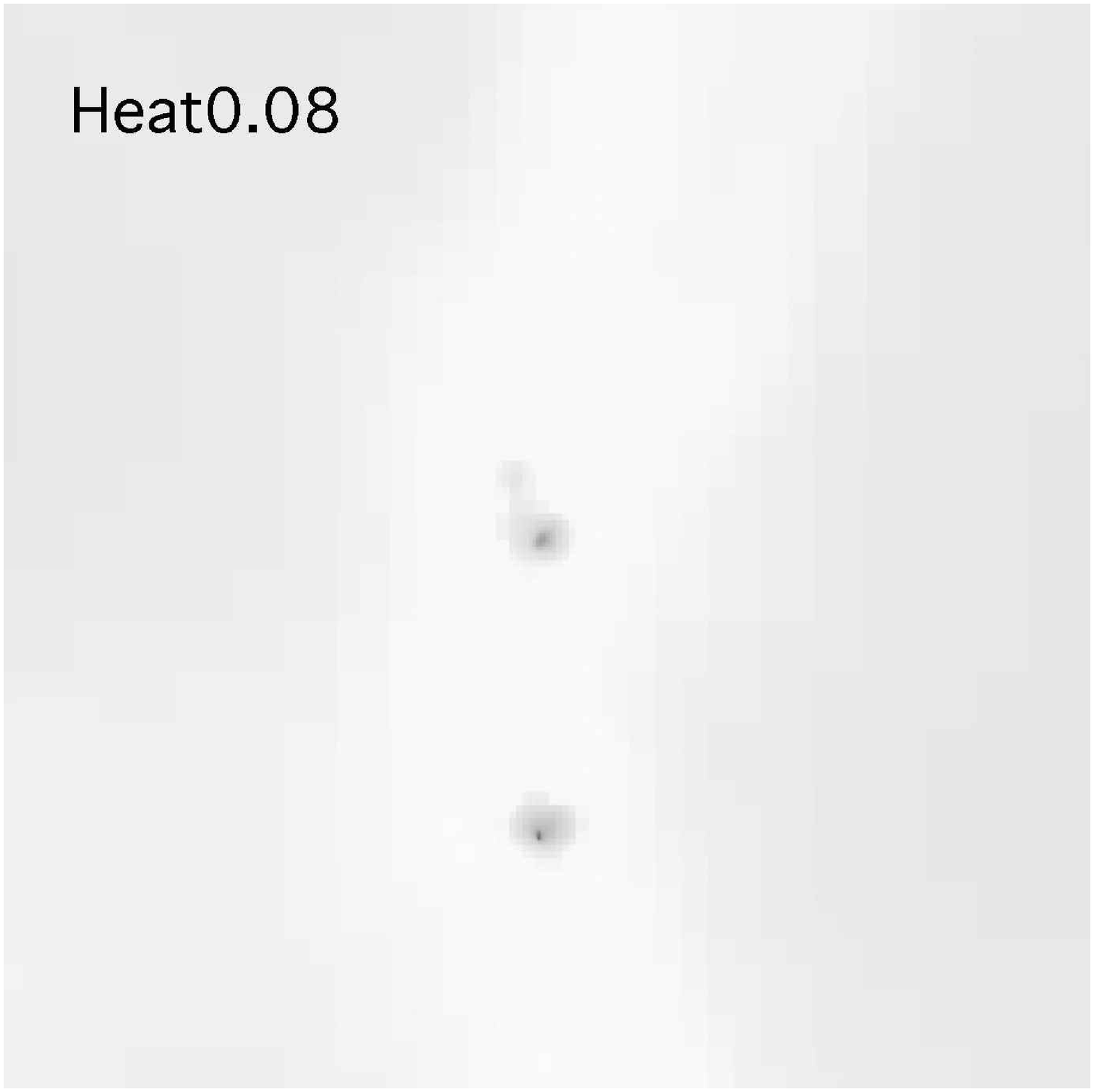}
\includegraphics[width=0.245\textwidth]{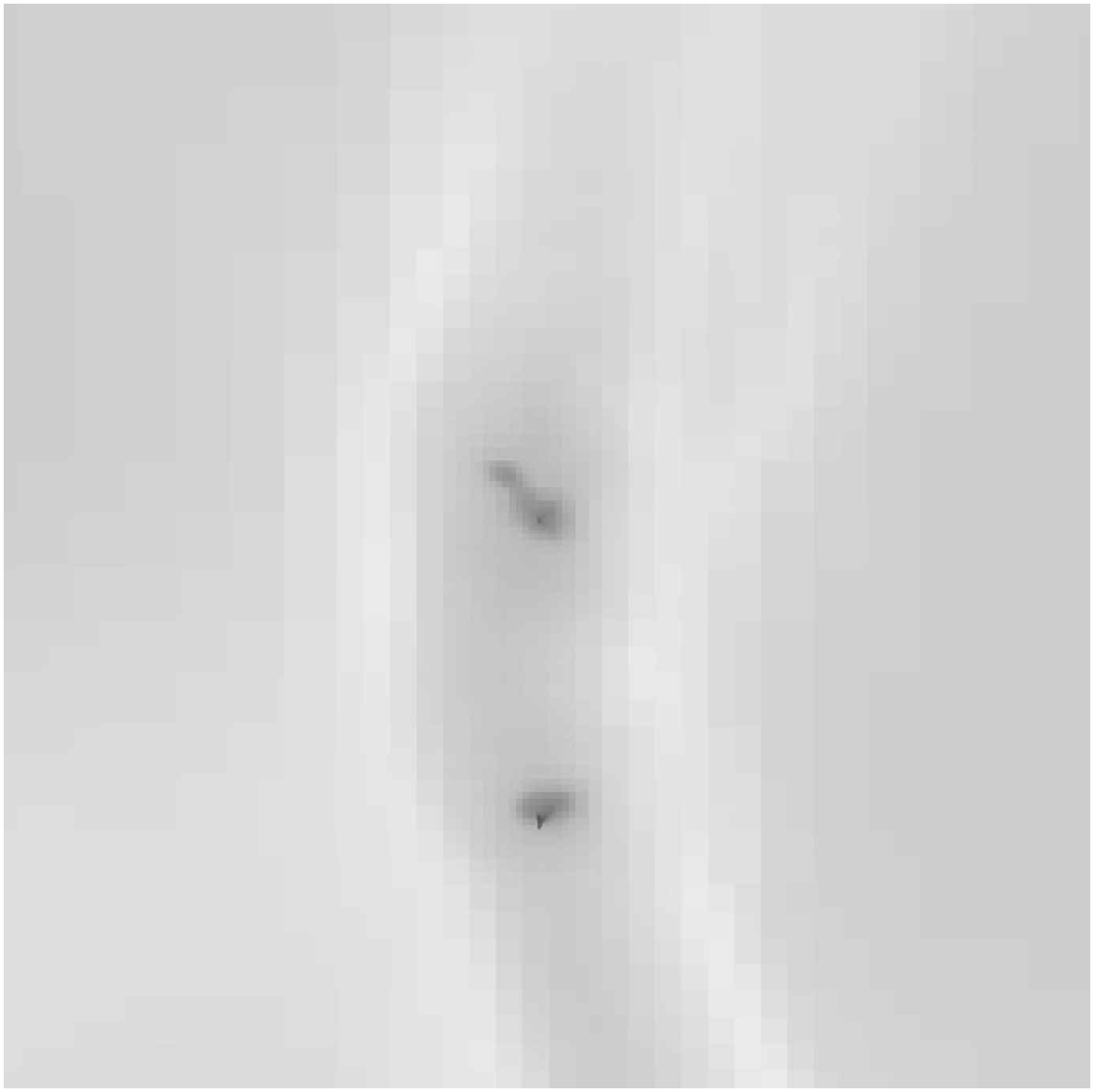}
\includegraphics[width=0.245\textwidth]{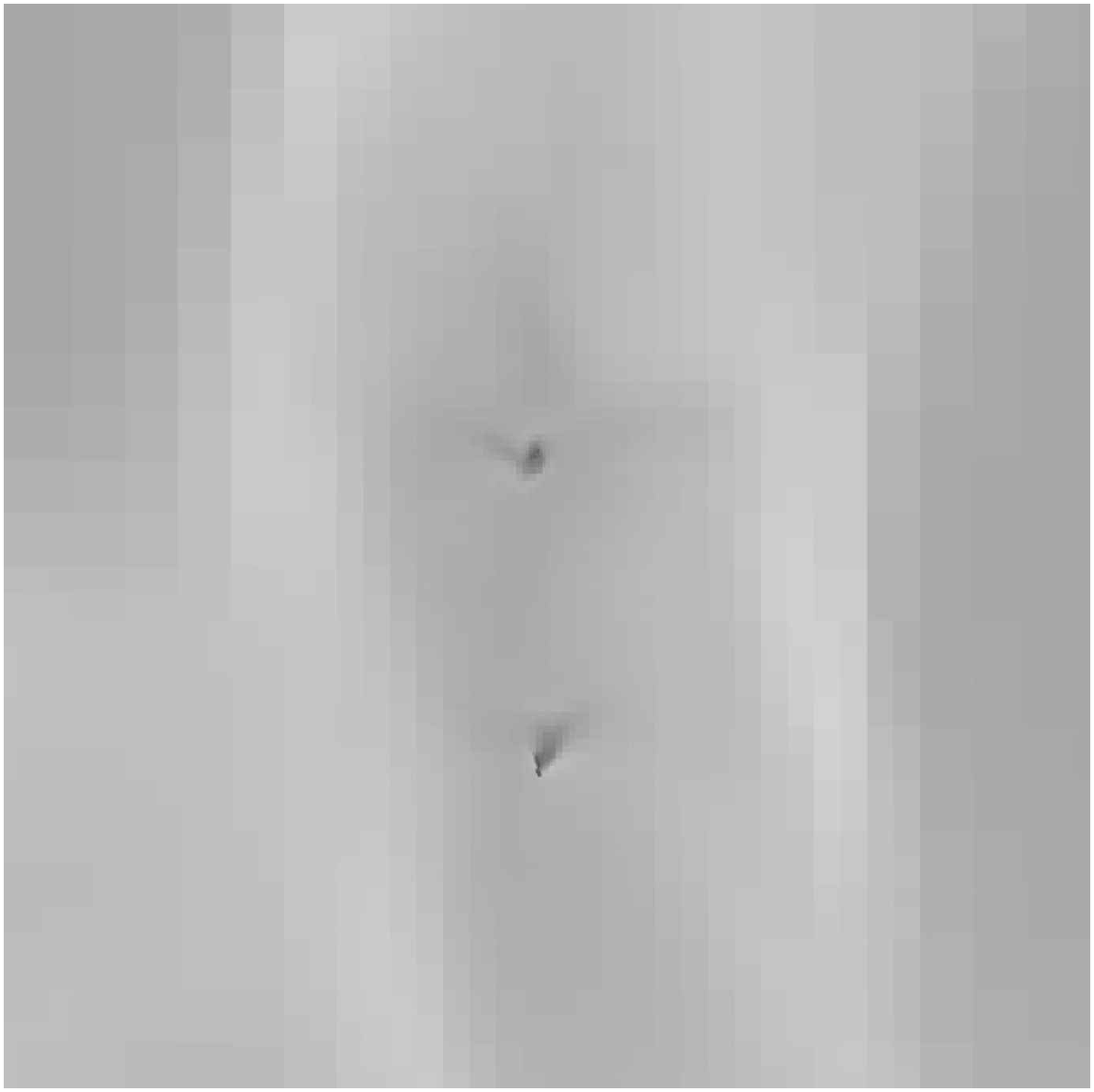}
\includegraphics[width=0.245\textwidth]{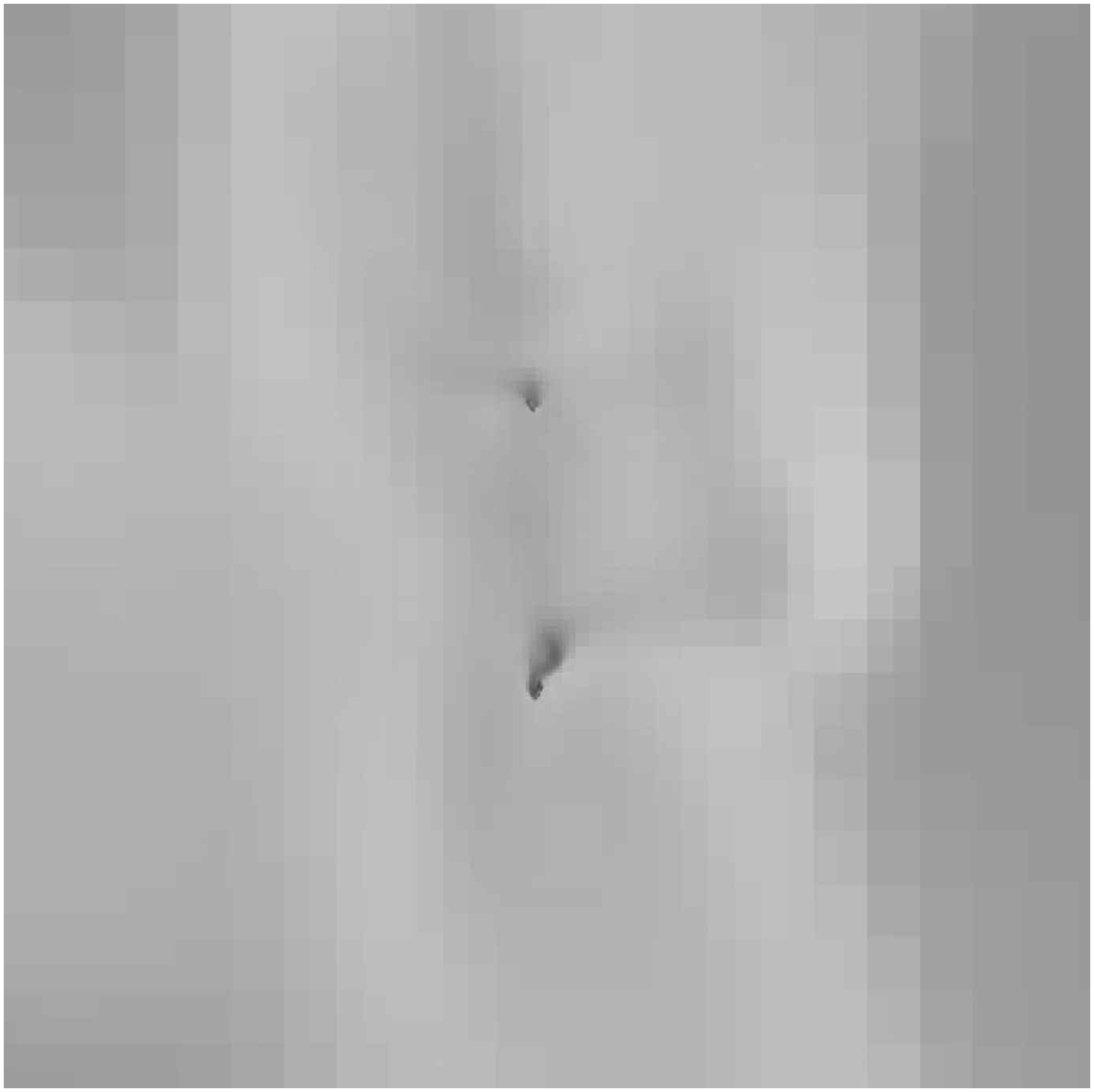}
}
{
\includegraphics[width=0.245\textwidth]{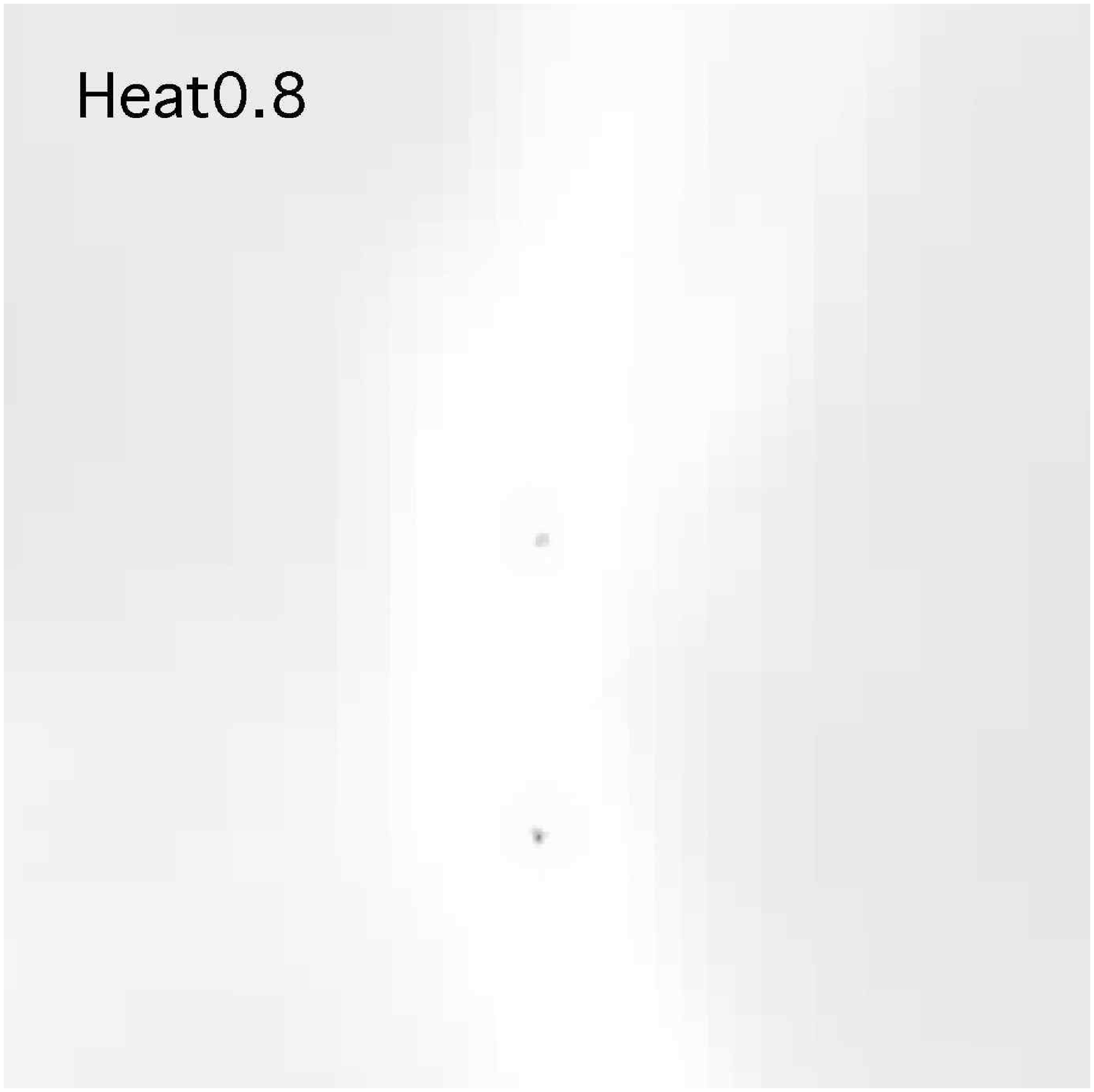}
\includegraphics[width=0.245\textwidth]{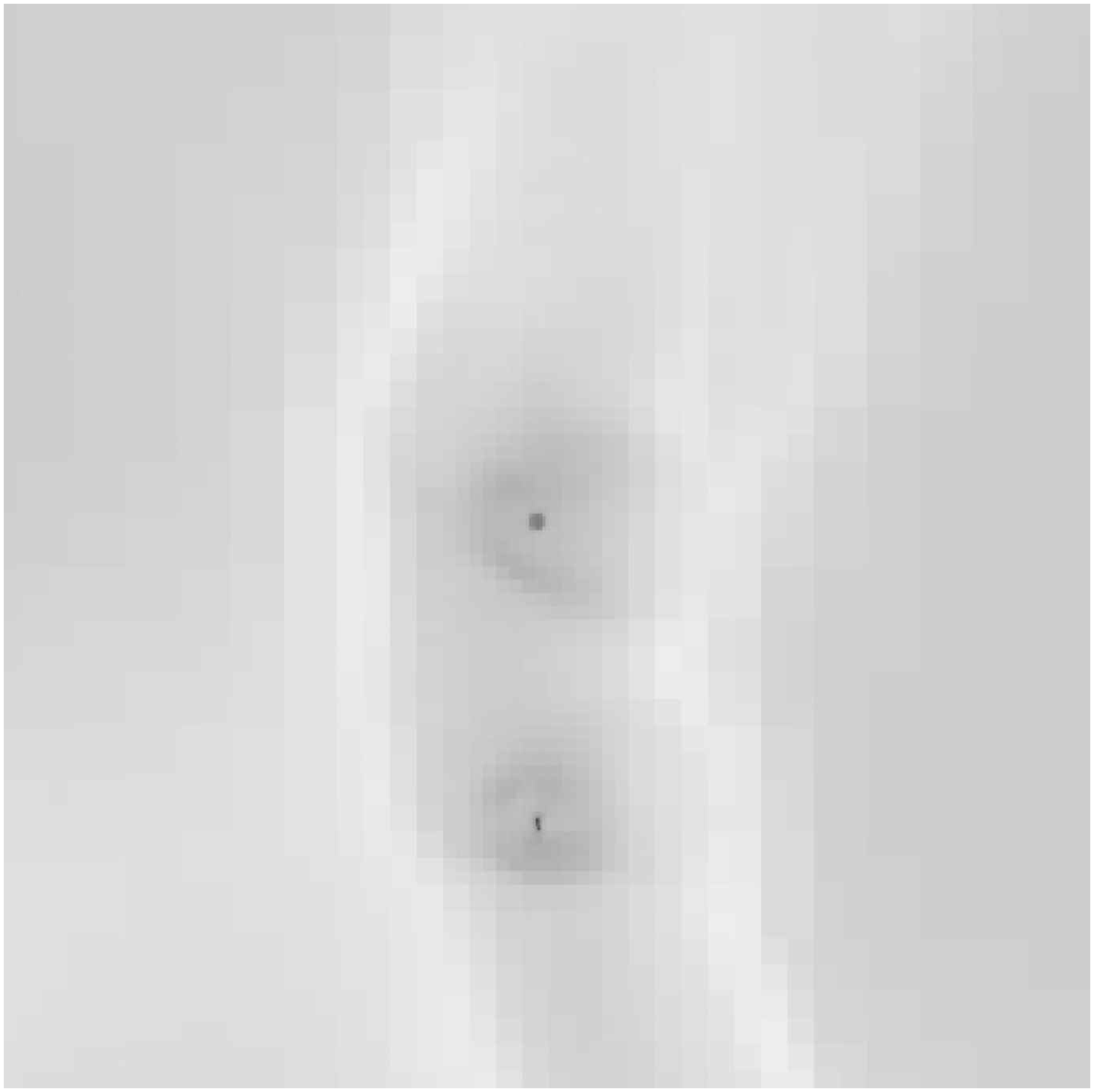}
\includegraphics[width=0.245\textwidth]{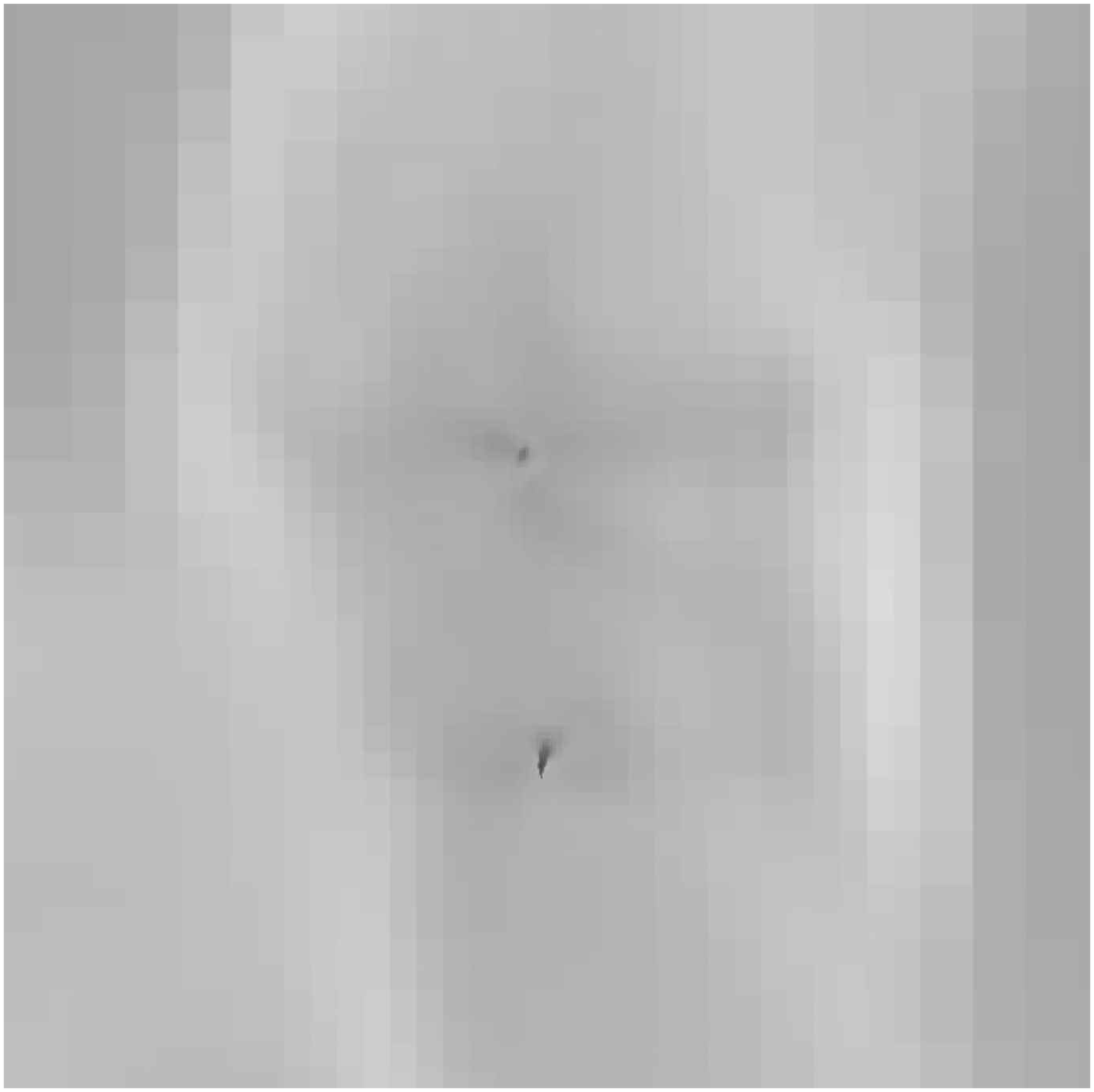}
\includegraphics[width=0.245\textwidth]{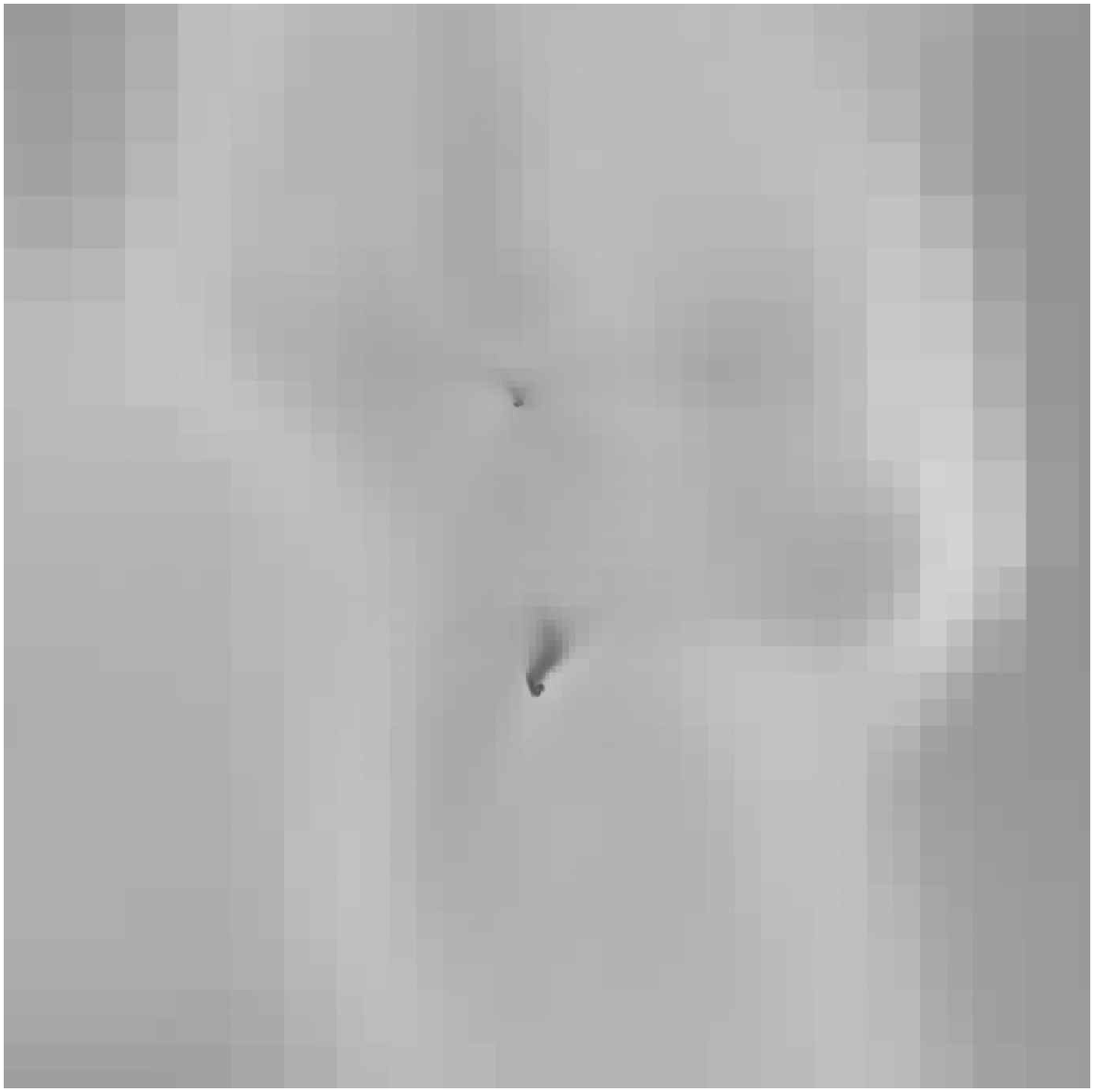}
}
\vspace{-1\baselineskip} \figcaption{Temperature projections of a 20
$h^{-1}$ kpc comoving region surrounding two halos nested in a
filament.  The rows, from top to bottom, correspond to the \nothing,
\flash, \heatinglow, and \heatinghigh\ runs.  The columns, from left
to right, correspond to redshifts of $z=$ 24.62, 24, 22, and 20.  The
scale is logarithmic, with black corresponding to $T < 100$K and white
corresponding to $T = 10^4$ K.
\label{fig:pics}
}
\vspace{-1\baselineskip}
\end{figure*}

In Figure \ref{fig:pics}, we show grey scale temperature projections of
a 20 $h^{-1}$ kpc comoving region surrounding two halos nested in a
filament.  The rows correspond to the \nothing, \flash, \heatinglow,
and \heatinghigh ~runs (from top to bottom).  The columns correspond to
redshifts $z=$ 24.62, 24, 22, and 20 (from left to right).  The scale
is logarithmic, with black corresponding to $T < 100$K and white
corresponding to $T = 10^4$K.  The halo in the lower (upper) part of
each figure grows from $M=7.2\times10^5\Msun$ ($M=6.9\times10^5\Msun$)
at $z=24.62$ to $M=1.4\times10^6\Msun$ ($M=1.2\times10^6\Msun$) at
$z=20$, as measured in the \nothing\ run.

As one would expect, when the UVB is turned on, the gas is quickly
ionized and heated.  Gas which was previously at or close to
hydrostatic equilibrium now has a greatly increased pressure gradient
due to the increase in temperature.  As a result, an outward--moving
shock is formed, as clearly seen in Figure~\ref{fig:pics} for the last
two rows (i.e. the runs which include a UVB with dynamical heating).
Note that this shock is nearly absent in the Flash run.  Subsequently,
the gas in the dense filaments inside the shock is able to cool
through Compton and H$_2$ cooling, and the shock stalls.  The gas
surrounding the halo starts infalling again.  We explore these
processes more quantitatively in \S~\ref{sec:profiles}.

Note that the cores of the halos retain CD gas in all of the runs. In
runs containing a UVB, the low density IGM outside of the filament
still hasn't cooled below $T\sim10^3$K by $z=20$; however, the
filament itself shows evidence of positive feedback in the \flash\ and
\heatinglow\ runs, with lower temperatures at $z\lsim24$ than in the
\nothing\ run.  Furthermore, it is evident that once the UVB is turned
off, the dense filament is able to cool very rapidly, from
$T\sim10^4$K to $T\sim10^3$K in $\Delta z \lsim 0.6$, due to the
increased electron fraction, as we shall see below.

\subsection{Halo Profiles}
\label{sec:profiles}

\begin{figure*}
\vspace{+0\baselineskip}
{
\includegraphics[width=0.33\textwidth]{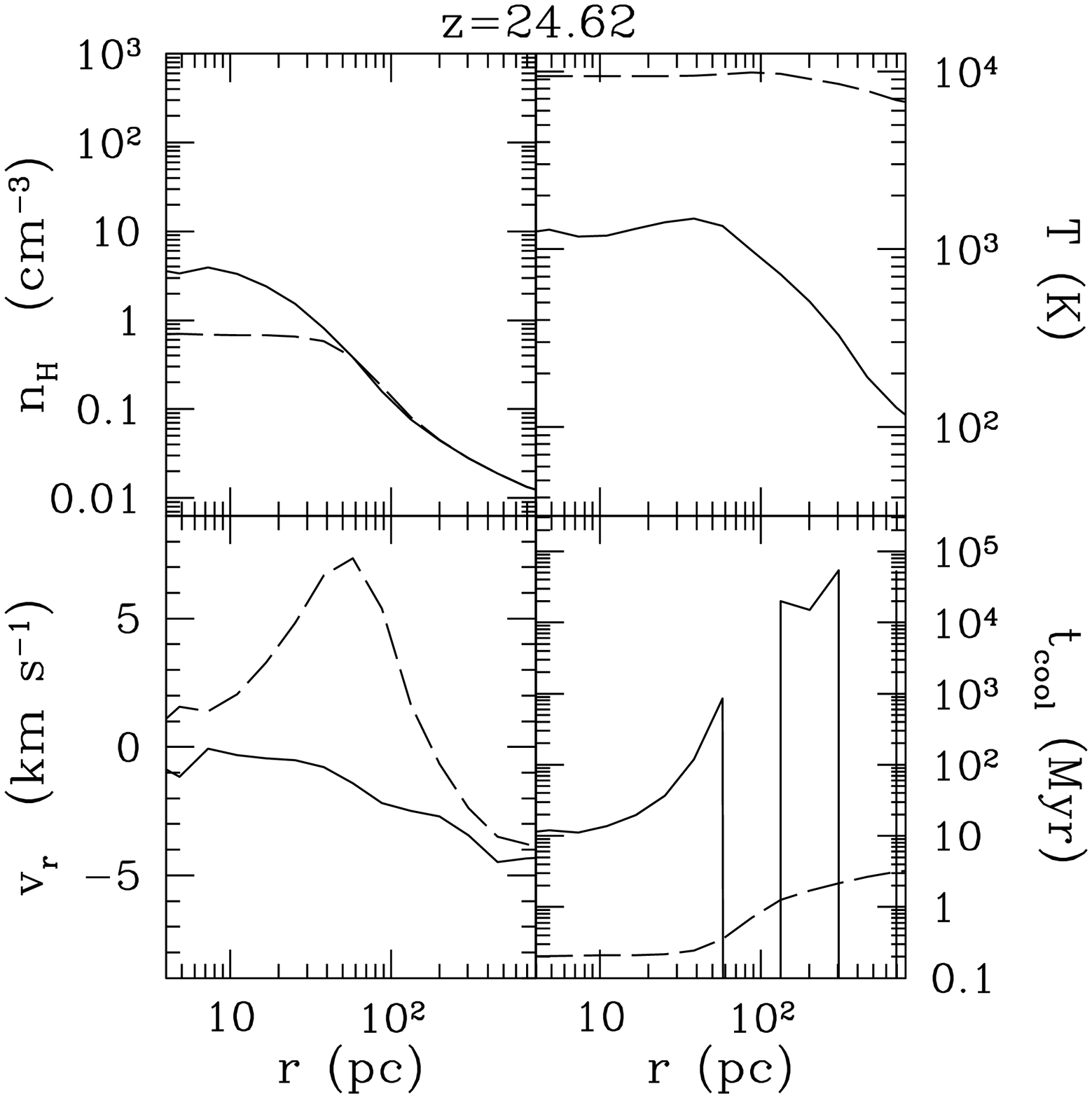}
\includegraphics[width=0.33\textwidth]{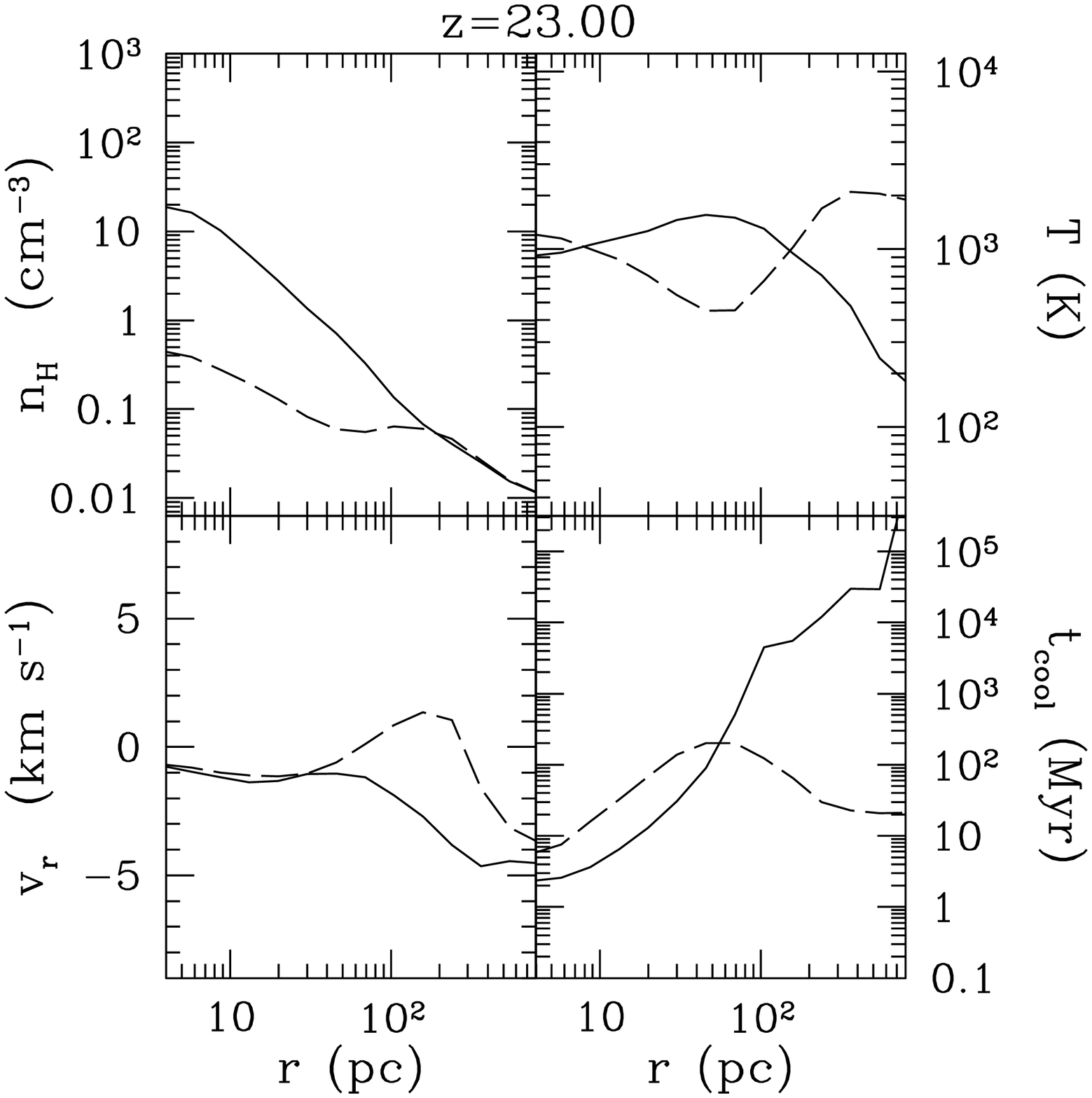}
\includegraphics[width=0.33\textwidth]{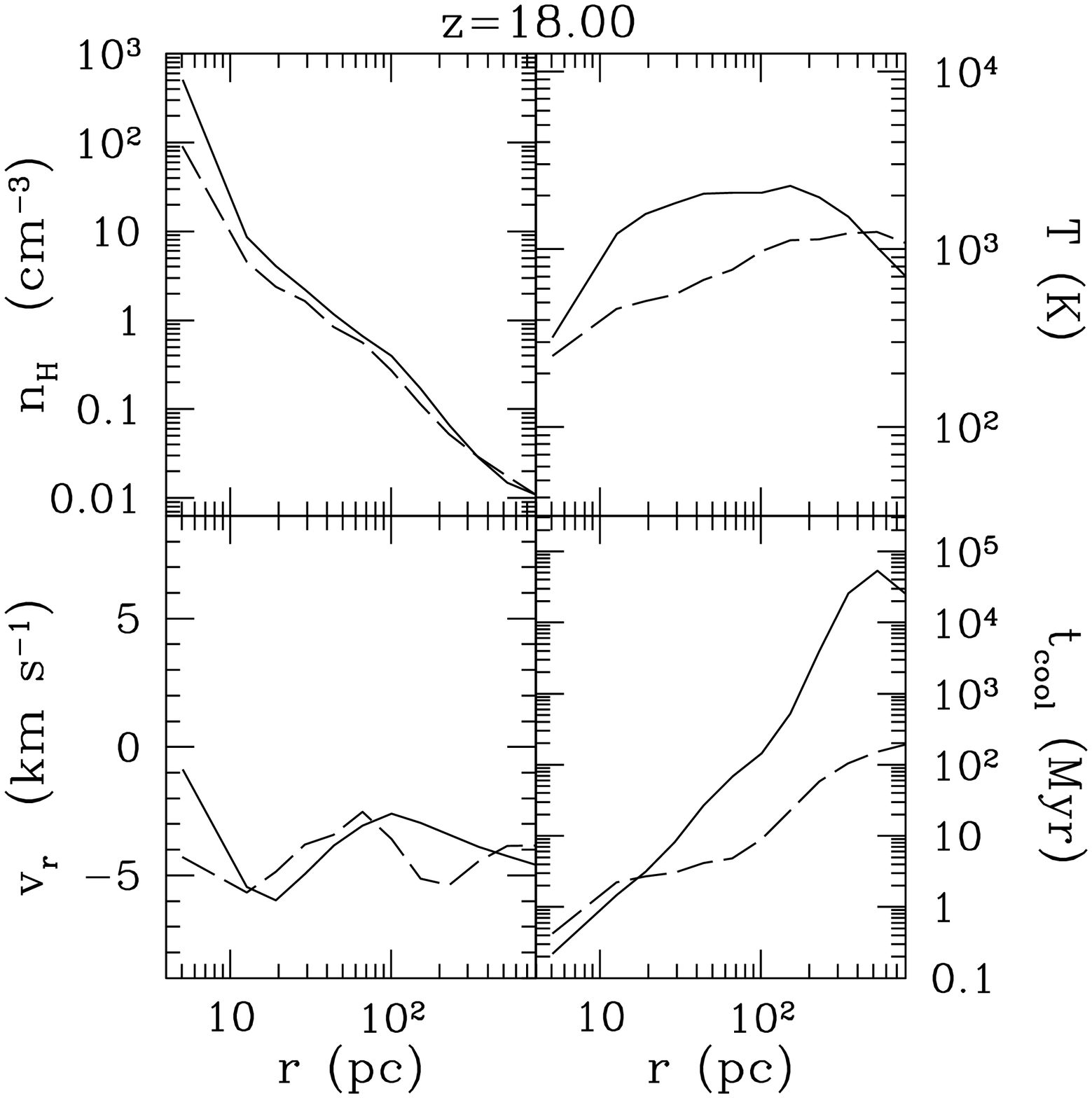}
\vskip0.0pt
}
{
\includegraphics[width=0.33\textwidth]{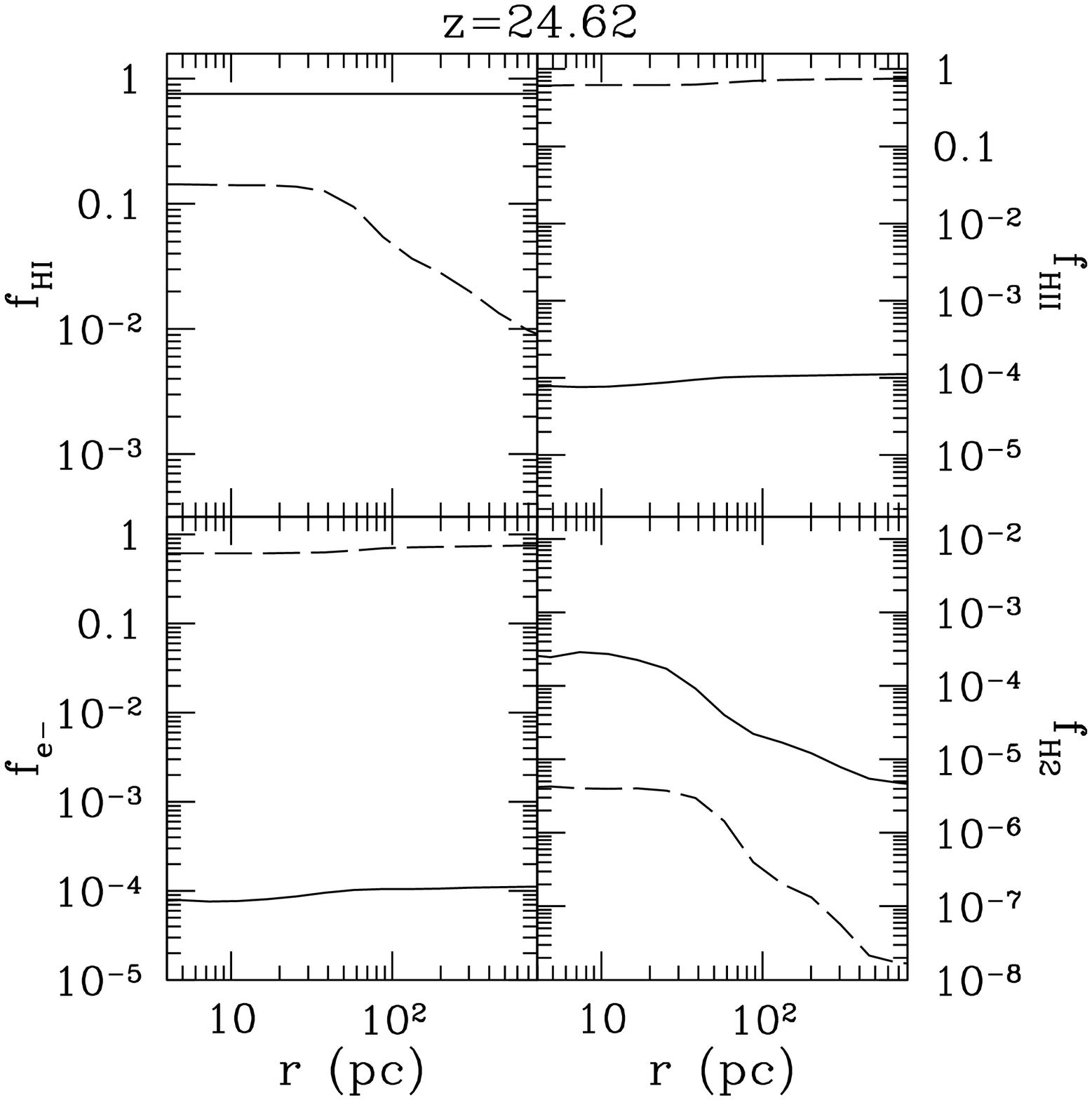}
\includegraphics[width=0.33\textwidth]{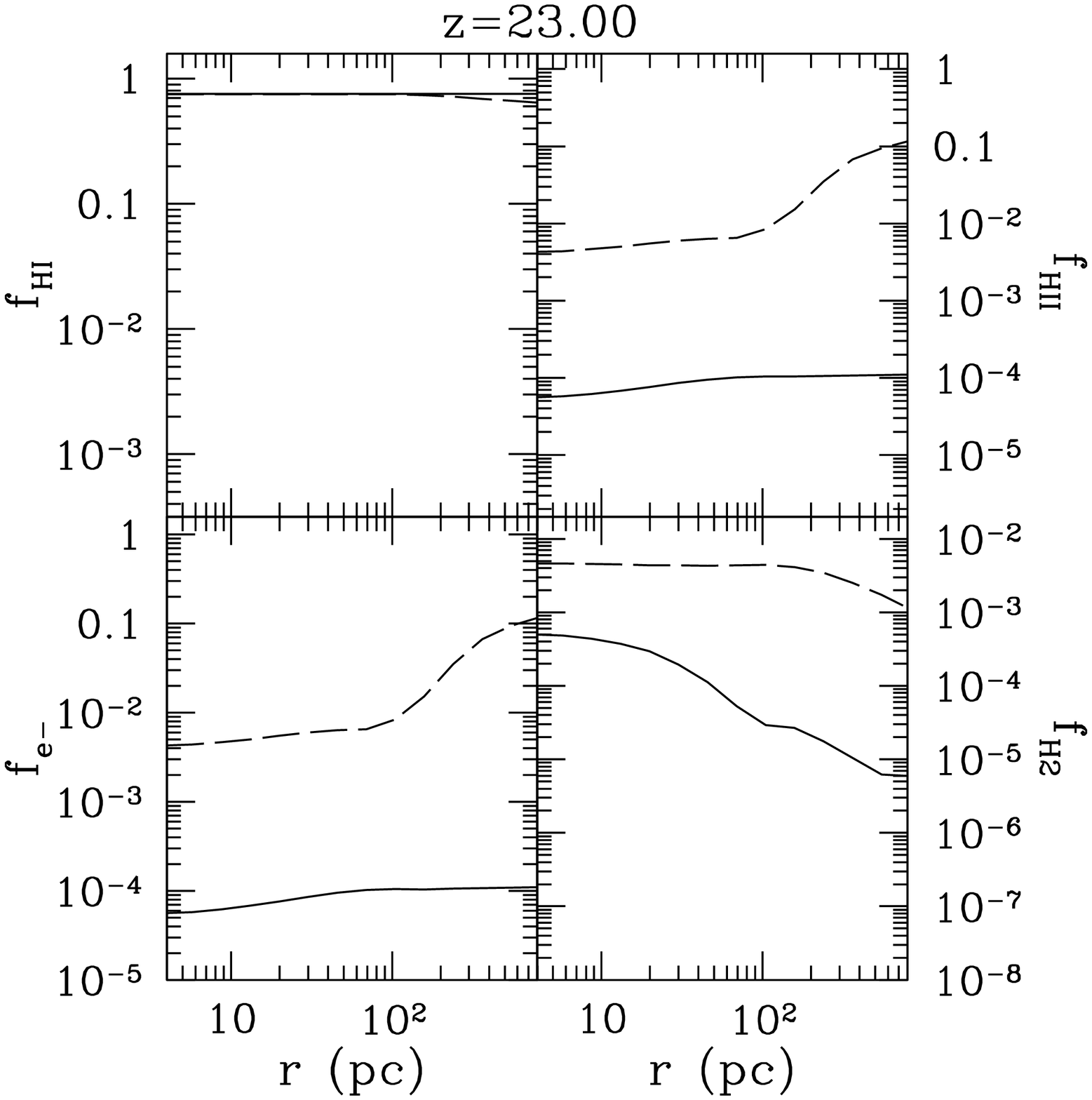}
\includegraphics[width=0.33\textwidth]{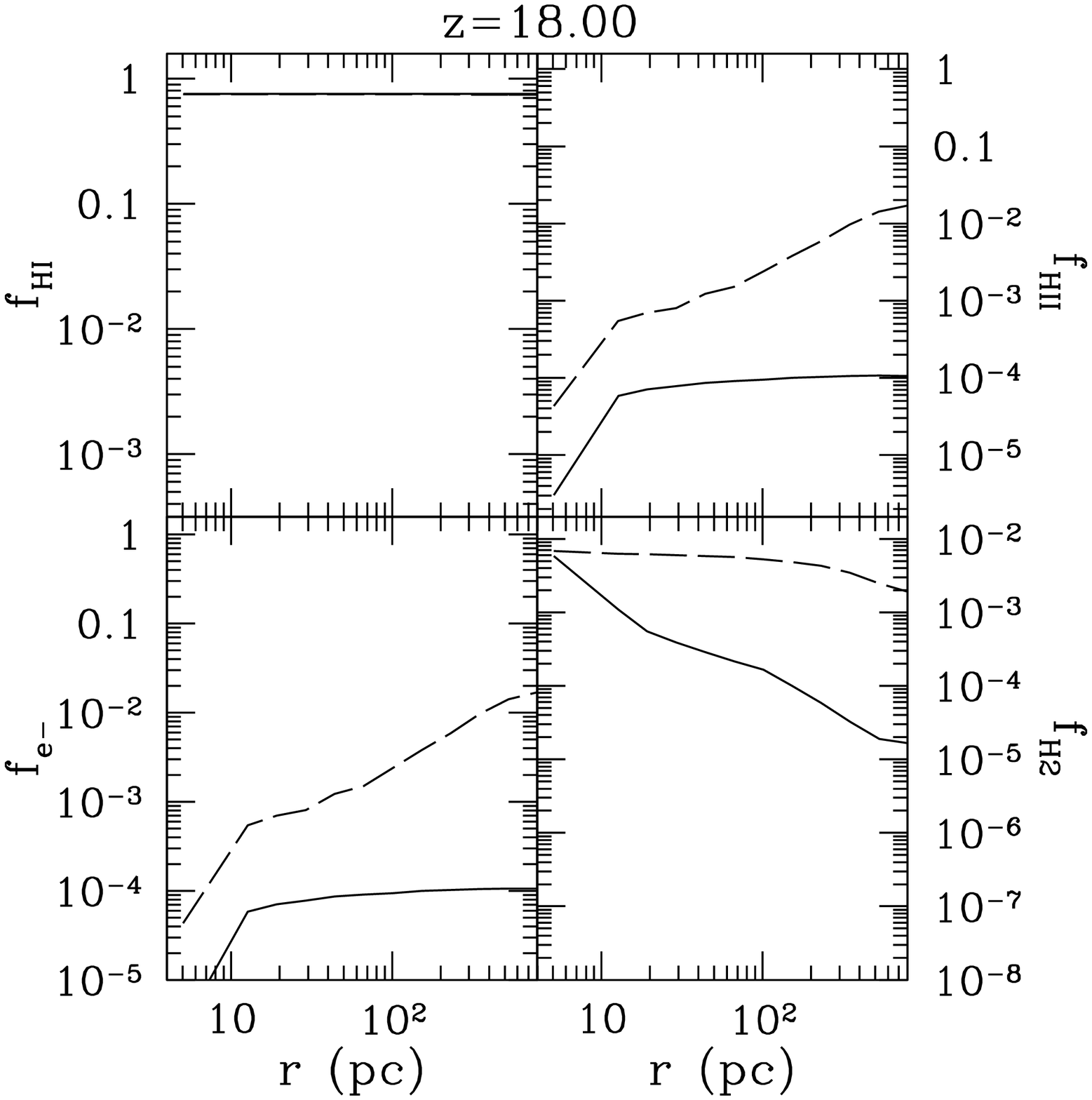}
}
\vspace{-1\baselineskip} \figcaption{Spherically averaged radial
profiles of the same halo in the \nothing\ ({\it solid lines}) and
\heatinghigh\ ({\it dashed lines}) simulation runs.  This halo was
able to first form CD gas at $z=21$ ($z=18$) in the \nothing\
(\heatinghigh) run, respectively.  The left pair of figures show a
snapshot at $z = \zuvboff = 24.62$, the middle pair at $z=23$, and the
right pair at $z=18$.  The virial radius of the halo increases from
$R_{\rm vir} \sim 100$ pc at $z=24.62$, to $R_{\rm vir} \sim 200$ pc
at $z=18$.  All quantities are shown in proper (not comoving) units.
The {\it Upper panels} show the hydrogen density, mass-weighted gas
temperature, gas cooling time, and radial velocity ({\it clockwise
from upper left}).  The {\it Bottom panels} show mass fractions of HI,
HII, H$_2$, and the number fraction of $e^-$ ({\it clockwise from
upper left}).
\label{fig:profiles}
}
\vspace{-1\baselineskip}
\end{figure*}

To get a more quantitative idea of the feedback introduced by a UVB,
in Figure \ref{fig:profiles} we plot spherically averaged radial
profiles for the same individual halo at redshifts $z=$ 24.62, 23, and
18 ({\it left to right}), and in two different runs: \nothing\ and
\heatinghigh\ ({\it solid} and {\it dashed} lines, respectively).
Figures in the top row show hydrogen density, mass-weighted gas
temperature, gas cooling time, and radial velocity ({\it clockwise
from upper left}).  Figures in the bottom row shows mass fractions of
HI, HII, H$_2$, and the number fraction of $e^-$ (more precisely,
$f_e$ is defined as the mass fraction of $e^-$, normalized such that
each $e^-$ is assumed to have the mass of hydrogen) ({\it clockwise
from upper left}).  The halo has a mass of M($z=24.62$) =
$3.42\times10^5\Msun$ and M($z=18$) = $2.37\times10^6\Msun$ [taken
from the \nothing\ run; note that the mass in the \heatinghigh\ run is
somewhat smaller, e.g., M($z=18$) = $2.24\times10^6\Msun$, due to
photo-evaporation and a slight suppression of gas accretion as a
result of the UVB].

 From the profiles, one can see the impact of photo-evaporation in the
\heatinghigh\ simulation run: the radially--outward moving shock
mentioned above, as well as an accompanying decrease in density.  As
soon as the ionizing radiation is turned off, the gas cools from a
temperature of $\sim10^4$ K to $\sim10^3$ K quite rapidly, with the
free electron number fraction (approximately corresponding to the
bottom left panels of the bottom row of Fig. \ref{fig:profiles}),
dropping two orders of magnitude by $z=23$, $\Delta z = 1.62$, since
the UVB was turned off.  Also, the shock starts to dissipate by
$z\sim23$, with most of the gas switching to the infall regime again.
Despite the evident photo-evaporation, the presence of the UVB has
caused over an order of magnitude increase in the H$_2$ fraction, due
to the increased (out--of--equilibrium) number of free electrons soon
after $\zuvboff$.

We note that this halo managed to first form CD gas at $z=21$ in the
\nothing\ case, but the formation of CD gas was delayed in the
\heatinghigh\ case until $z=18$.  This delay can be crudely understood
by looking at three fundamental time-scales: the H$_2$ cooling time,
\begin{equation}
\label{eq:t_H2_ion}
\tHtwo = \frac{1.5 k_B T}{\Lambda_{\rm H_2}} \frac{n_g}{n_{\rm HI} n_ 
{\rm H_2}} ~ ,
\end{equation}
\noindent
the Compton cooling time,
\begin{equation}
\label{eq:t_C}
\tcomp \approx 14 \left( \frac{1+z}{20} \right)^{-4} x_e^{-1} ~ {\rm  
Myr} ~ ,
\end{equation}
\noindent
and the gas recombination time,
\begin{equation}
\label{eq:t_rec}
\trec \approx 39 \left( \frac{1+z}{20} \right)^{-3} \delta^{-1} x_e^ 
{-1} ~ {\rm Myr} ~ .
\end{equation}
\noindent
Here, $k_B$ is the Boltzmann constant, $T$ is the temperature,
$\Lambda_{\rm H_2}$ is the H$_2$ cooling function, $x_e$ is the free
electron number fraction, $n_g$, $n_{\rm HI}$, and $n_{\rm H_2}$ are
the number densities of all baryons and electrons, neutral hydrogen,
and H$_2$, respectively, and $\delta$ is the gas overdensity, $\delta
\equiv n_g/\bar{n}_g - 1$.

The first column of Figure \ref{fig:profiles} shows a snapshot of our
halos immediately prior to turning off the UVB.  Note that the cooling
time in the \heatinghigh\ run is several orders of magnitude lower than
in the \nothing\ run.\footnote{The sharp drops in the cooling time in  
the \nothing\ run
correspond to annuli that include cold, low--density gas below the CMB
temperature ($\sim$10 K), which is heated, rather than cooled, by
Compton scattering.}
At large radii, this is due to Compton cooling, since the UVB
dramatically increases $x_e$ and the Compton cooling time
(eq. \ref{eq:t_C}) scales as $x_e^{-1}$.  Additionally, the
temperature increase to $\sim 10^4$ K allows for far more efficient
line cooling from atomic hydrogen than in the \nothing\ case.  We also
see that the structure of halos is important in accurately predicting
feedback.  Specifically, we note that the central region can behave
quite differently than the outer regions of the halo.\footnote{We do
not include radiative transfer in our analysis, and the high-density
regions ($n_H \geq 1$cm$^{-3}$ ), such as the cores of halos, might be
able to self-shield against UV radiation \citep{ABS05}, decreasing
feedback effects somewhat; see discussion in \S~\ref{sec:nort}.}

Initially, immediately after the radiation field is turned off, the
gas cools due to atomic line cooling.  However, this quickly becomes
inefficient below temperature of about 6000 K.  While the radiation is
on, the H$_2$ abundance is highly suppressed due to Lyman-Werner
dissociation.  However, as the temperature drops, the amount of H$_2$
increases rapidly because of the high electron abundance.  In fact the
molecular hydrogen formation time is shorter than the recombination
time and a large amount of molecular hydrogen is produced, $x_{\rm
H_2} \sim 3 \times 10^{-3}$, irrespective of the density and
temperature (see the discussion in \citealt{OH02} for an explanation
of this freeze-out value).

Due to the relatively high densities throughout the halo ($\delta \sim
10^4$ near the core), as well as the high initial $x_e$, the majority
of the hydrogen recombines shortly after $\zuvboff$ in the
\heatinghigh\ case ($\trec(\zuvboff) \ll 1$Myr).  After a few
recombination times, $n_g \sim n_{\rm HI}$, and so
eq. (\ref{eq:t_H2_ion}) can be simplified to
\begin{eqnarray}
\label{eq:t_H2}
\tHtwo & \approx &
\frac{1.5 k T}{\Lambda} \frac{1}{n_g x_{\rm H_2}} \\
  & \approx & 4 \left( \frac{T}{10^3 K} \right)^{-2.5}
                   \left( \frac{x_{\rm H_2}}{3 \times 10^{-3}} \right) 
^{-1}
                   \left( \frac{n_g}{1 \rm cm^{-3}} \right)^{-1} {\rm  
Myr}~ , \nonumber
\end{eqnarray}
\noindent
where $x_{\rm H_2}$ is the number fraction of H$_2$.  This highly
efficient molecular cooling channel is largely responsible for quickly
driving the temperature down to about 1000 K (although a persistent LW
background can counter this ${\rm H_2}$--enhancement effect; see
below).  Hence, we note that the cooling times near the halo core are
comparable for both runs at $z=23$, with the cooling time in the
\heatinghigh\ case being larger by a factor of a few.

This factor of only a few change in the cooling time is due to the  
remarkable
cancellation of two strong effects, and can be understood by noting
that the UVB induced photo-evaporation reduces $n_g$ by a factor of
$\sim 40$ near the core at $z=23$.  Meanwhile, the UVB boosts $x_{\rm
H_2}$ near the core by a factor of $\sim 10$.  Since Compton cooling
is ineffective at this stage, the cooling time is dominated by the
H$_2$ cooling channel, whose cooling time roughly scales as $\tHtwo
\propto (n_g x_{\rm H_2})^{-1}$, given that the temperature is almost
identical near the core at $z\sim23$, and that the cooling function is
very weakly dependent on $n_{\rm H_2}$ in this regime \citep{GP98}.

 From this crude estimate, one obtains an effective ``delay'' in the
formation of CD gas in the \heatinghigh\ run with respect to the
\nothing\ run:
\begin{equation}
\label{eq:delay}
\fdelay \sim
\frac{\tHtwo^{\rm\heatinghigh}}{\tHtwo^{\rm\nothing}} \sim
\left( \frac{n^{\rm\nothing}_g}{n^{\rm\heatinghigh}_g} \right)
\left( \frac{x^{\rm\heatinghigh}_{\rm H_2}}{x^{\rm\nothing}_{\rm  
H_2}} \right)^{-1} \sim
\frac{40}{10} \sim 4.
\end{equation}
\noindent
This is in excellent agreement with the delay observed in the pair of
simulation runs, where the halo obtains CD gas $\sim$ 20 (60) Myr
after our data point at $z=23$ in the \nothing\ (\heatinghigh) case,
yielding a delay of a factor of $\fdelay \sim 3$.

\subsection{$H_2$ production}
\label{sec:Htwo}

\begin{figure}
\vspace{+0\baselineskip}
\myputfigure{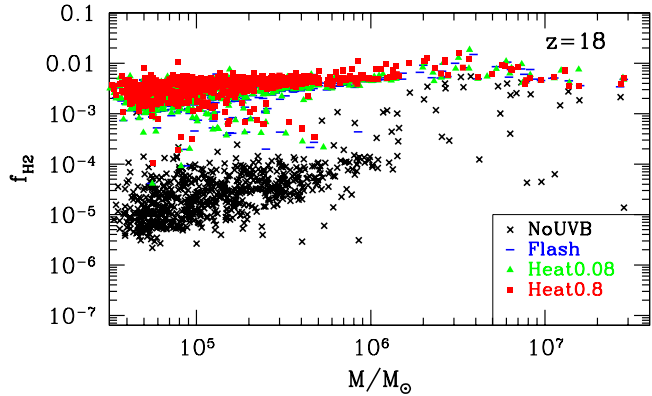}{3.3}{0.5}{.}{0.}
\vspace{-1\baselineskip} \figcaption{ Mass fractions of H$_2$ at
$z=18$.  Results are shown from the \nothing\ ({\it black crosses}),
\flash\ ({\it blue dashes}), \heatinglow\ ({\it green triangles}), and
\heatinghigh\ ({\it red squares}) simulation runs.
\label{fig:hii}}
\vspace{-1\baselineskip}
\end{figure}

As mentioned above, molecular hydrogen can provide a dominant cooling
mechanism, especially in high density regions.  To study the impact of
the UVB on the formation of H$_2$, we plot the H$_2$ mass fractions as
functions of halo mass in Figure \ref{fig:hii}, at the lowest redshift
of our simulations, $z=18$.  Results are shown from the \nothing\
({\it black crosses}), \flash\ ({\it blue dashes}), \heatinglow\ ({\it
green triangles}), and \heatinghigh\ ({\it red squares}) simulation
runs.

We note that in all of our runs which include a UVB, the molecular
hydrogen fraction converges to a value which is nearly independent of
the strength and duration of the UVB.  As seen in Figure
\ref{fig:hii}, by the end of our simulations, most halos which have
been exposed to a UVB in their past have settled on a value of a few
$\times 10^{-3}$ for the mass fraction of H$_2$.  This freeze--out
value is due to the fact that the number density of H$_2$ follows its
equilibrium value until about 3000 K, below which recombination
proceeds more quickly than H$_2$ formation and the fraction freezes
out at this point \citep{OH02}.  This number is fairly independent of
mass, though there is a weak trend towards higher mass fractions for
higher mass halos, up to mass fractions of $\sim 10^{-2}$ for
$M\sim4\times10^6\Msun$.  Note also that there is some evidence for a
non-monotonic evolution of the H$_2$ fraction, with mass fractions
falling back down to $\sim 4\times10^{-3}$ by $M\sim3\times10^7\Msun$.

In contrast, molecular hydrogen in halos which have not been exposed
to a UVB ({\it black crosses} in Figure \ref{fig:hii}) is distinctly
sparser (over two orders of magnitude for $M\lsim10^6\Msun$) than in
our other runs.  Also, there is a stronger evolution with respect to
mass, as well as more scatter (which makes sense, since the ${\rm
H_2}$ abundance in this case is not a result of a freeze--out process,
and depends strongly on local density and temperature).

An interesting conclusion can be drawn from the similarity of H$_2$
fractions in our runs which include a UVB.  Namely, if our analysis of
the dominant cooling processes in \S~\ref{sec:profiles} is accurate,
the differences between the CD gas fractions among our UVB runs is
predominately due to disparate effectiveness of photo-evaporation.  In
other words, as the positive feedback (i.e. the increase of the
$x_{\rm H_2}$ term in eq.~\ref{eq:t_H2}) is nearly independent of the
strength and duration of the UVB in our runs, only the amount of
negative feedback (decrease in $n_g$) causes variations in the delay
in the formation of CD gas (see a more detailed discussion in
\S~\ref{sec:lw} below).  We have also verified for several halos that
this similarity in the total H$_2$ fraction extends to the radial
profiles of H$_2$.

\subsection{Ensemble Evolution of the Cold, Dense Gas Fractions}
\label{sec:fcd}

\begin{figure*}
\plottwo{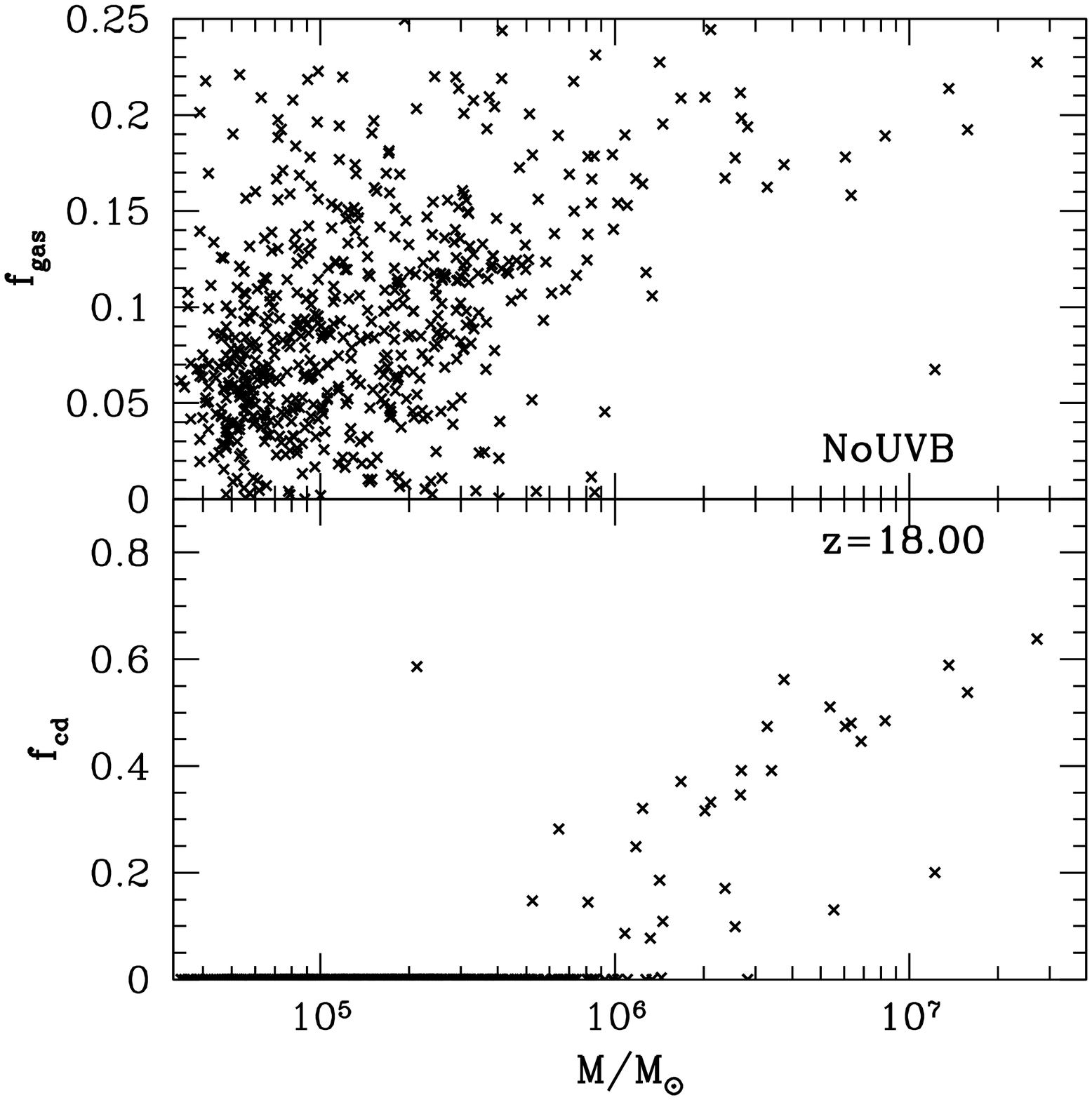}{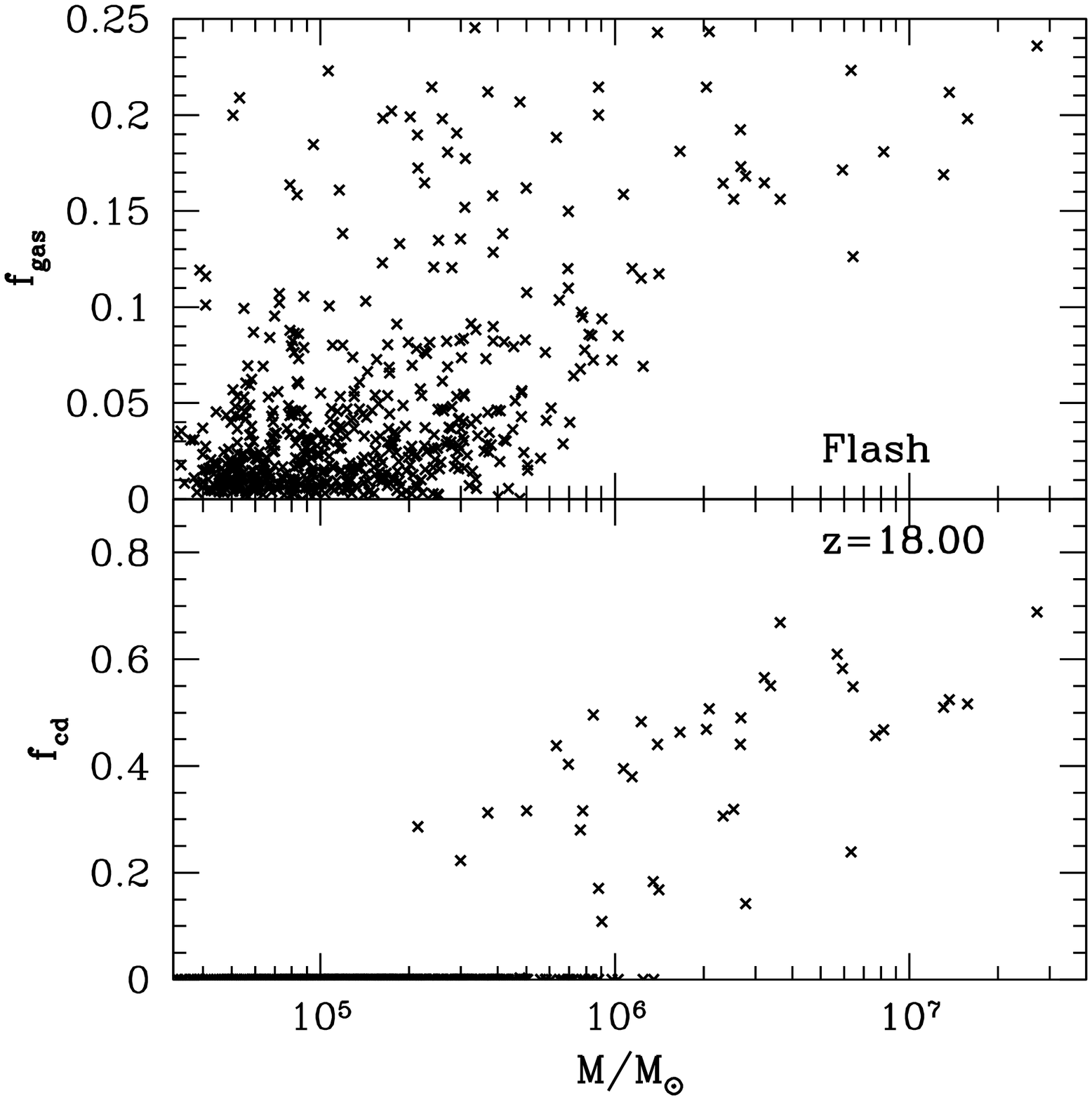}
\plottwo{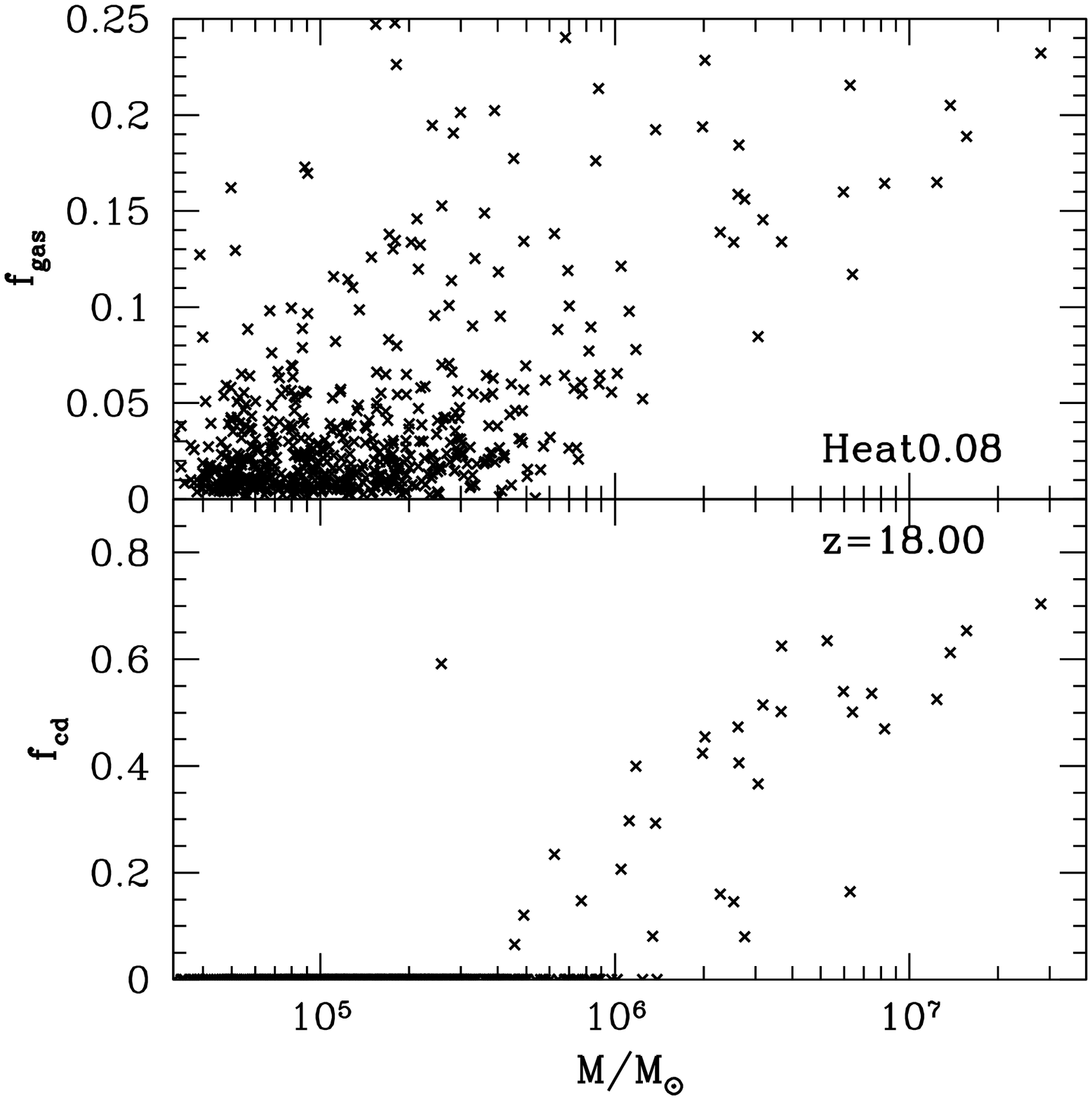}{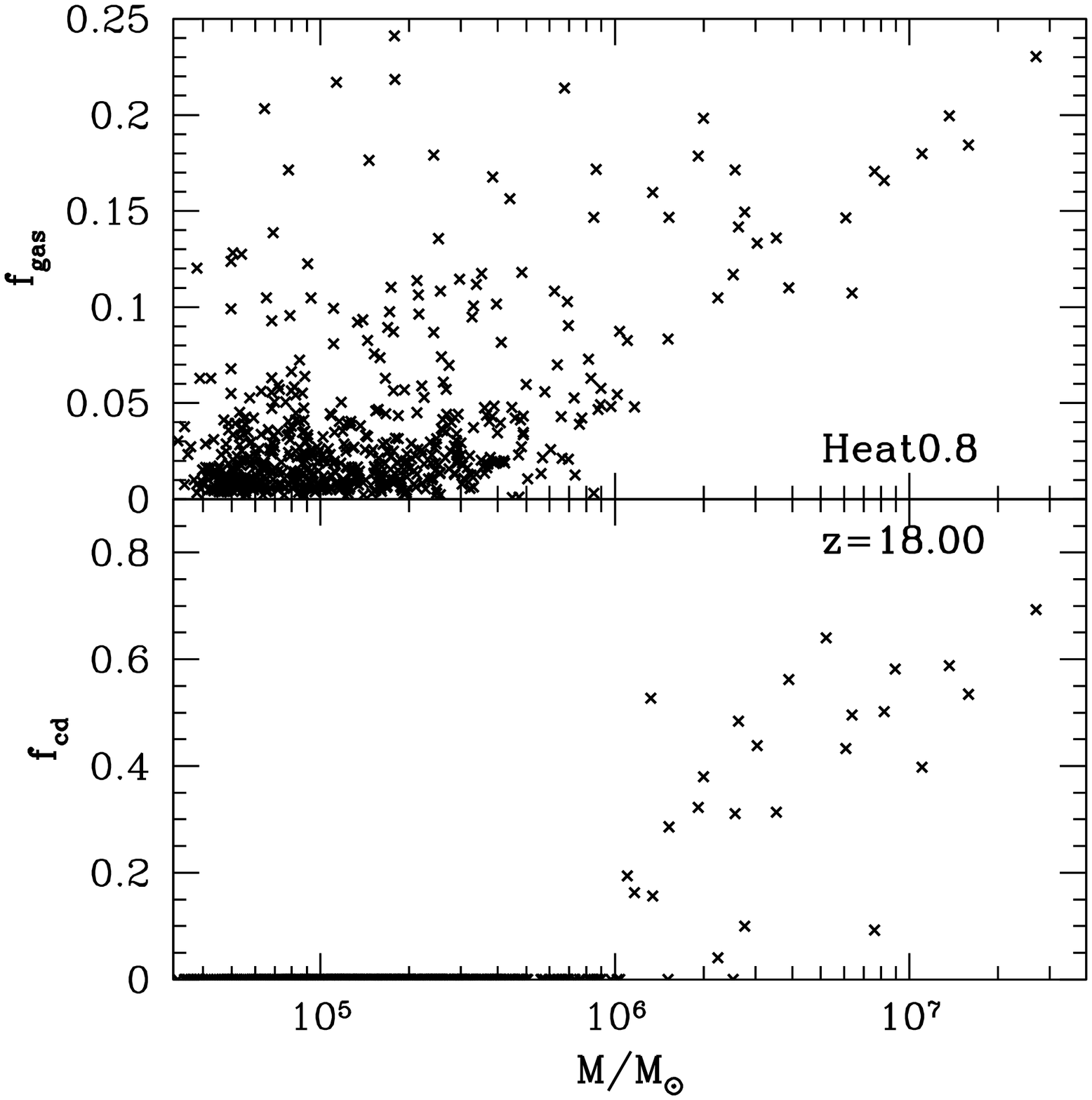}
\figcaption{ Total gas fractions ({\it top panels}) and cold, dense
gas fractions ({\it bottom panels}) as a function of total halo mass
at redshift $z=18$, the lowest redshift output of our simulations.
The four panels correspond to the \nothing\ ({\it top left}), \flash\
({\it top right}), \heatinglow\ ({\it bottom left}), and \heatinghigh\
({\it bottom right}) simulation runs.
\label{fig:gas_fract}
}
\vspace{-1\baselineskip}
\end{figure*}

As mentioned above, a quantity of particular importance in studying
the capacity of a halo at hosting stars is its cold, dense (CD) gas fraction,
$\fcd$, defined above.  Here we show the general trends for the
evolution of this quantity for the population of halos in our
simulations, as we vary the UVB.

In Figure \ref{fig:gas_fract}, we plot the total gas fractions ({\it
upper panels}) and CD fractions ({\it lower panels}) as a function of
total halo mass at redshift $z=18$, the lowest redshift output of our
simulations.  The figures correspond to the \nothing\ ({\it top
left}), \flash\ ({\it top right}), \heatinglow\ ({\it bottom left}),
and \heatinghigh\ ({\it bottom right}) simulation runs.

Note that while the total gas fractions of small halos ($M\lsim$ few
$\times 10^5 \Msun$) that have been exposed to a UVB are suppressed
with respect to the \nothing\ case, there is little immediate visual
evidence of either negative or positive feedback for halos large
enough to host CD gas ($M\gsim$ few $\times 10^5 \Msun$).  The \flash\
CD gas fractions show evidence of positive feedback in the mass range
$\sim ~ 2 \times 10^5$ -- $10^6$ $\Msun$, while the \heatinghigh\ run
shows evidence of strong negative feedback in the same mass range,
with no halos hosting CD gas at $M < 10^6 \Msun$.  These small halos
are the ones which would be most affected by photo--evaporation.  This
lends further credibility to our assertion above that {\it positive}
feedback in runs which include a UVB is fairly independent of the UVB
strength, and hence the total feedback is set by photo--evaporation
effects (i.e. {\it negative} feedback).  The \heatinglow\ run has a
near zero balance of positive and negative feedback.

In order to better quantify the amount of suppression of CD gas in our
models incorporating a UVB, as well as the evolution of such a
suppression with redshift, we define the cumulative, fractional
suppression of the halo number as
\begin{equation}
\label{eq:delN}
\delN \equiv \frac{\Nrun(z) - \Nrun(\zuvbon)}{N_{\rm cd}^{\rm
\nothing}(z) - N_{\rm cd}^{\rm \nothing}(\zuvbon)} - 1 ~ ,
\end{equation}
where $N_{\rm cd}^{\rm \nothing}(z)$ and $\Nrun(z)$ are the total
number of halos with CD gas at redshift $z$ in the \nothing\ run and
some given run $i$, respectively.  This expression is well--defined
for $N_{\rm cd}^{\rm \nothing}(z)$ $>$ $N_{\rm cd}^{\rm
\nothing}(\zuvbon)$; for $N_{\rm cd}^{\rm \nothing}(z)$ = $N_{\rm
cd}^{\rm \nothing}(\zuvbon)$, we set $\delN \equiv 0$.  Note that by
definition, $N_{\rm cd}^{\rm \nothing}(\zuvbon)$ = $\Nrun(\zuvbon)$.

Similarly, we define the cumulative, fractional suppression of the CD
gas mass as
\begin{equation}
\label{eq:delM}
\delM \equiv \frac{\Mrun(z) - \Mrun(\zuvbon)}{M_{\rm cd}^{\rm
\nothing}(z) - M_{\rm cd}^{\rm \nothing}(\zuvbon)} - 1 ~ ,
\end{equation}
where $M_{\rm cd}^{\rm \nothing}(z)$ and $\Mrun(z)$ are the total mass
of CD gas at redshift $z$ in the \nothing\ run and some given run $i$,
respectively.  The total CD gas mass is obtained by merely summing the
CD gas masses for all of the halos in the simulation.  As for
equation~(\ref{eq:delN}), this expression is well--defined for $M_{\rm
cd}^{\rm \nothing}(z)$ $>$ $M_{\rm cd}^{\rm \nothing}(\zuvbon)$, and
for $M_{\rm cd}^{\rm \nothing}(z)$ = $M_{\rm cd}^{\rm
\nothing}(\zuvbon)$, we set $\delM \equiv 0$.

\begin{figure}
\myputfigure{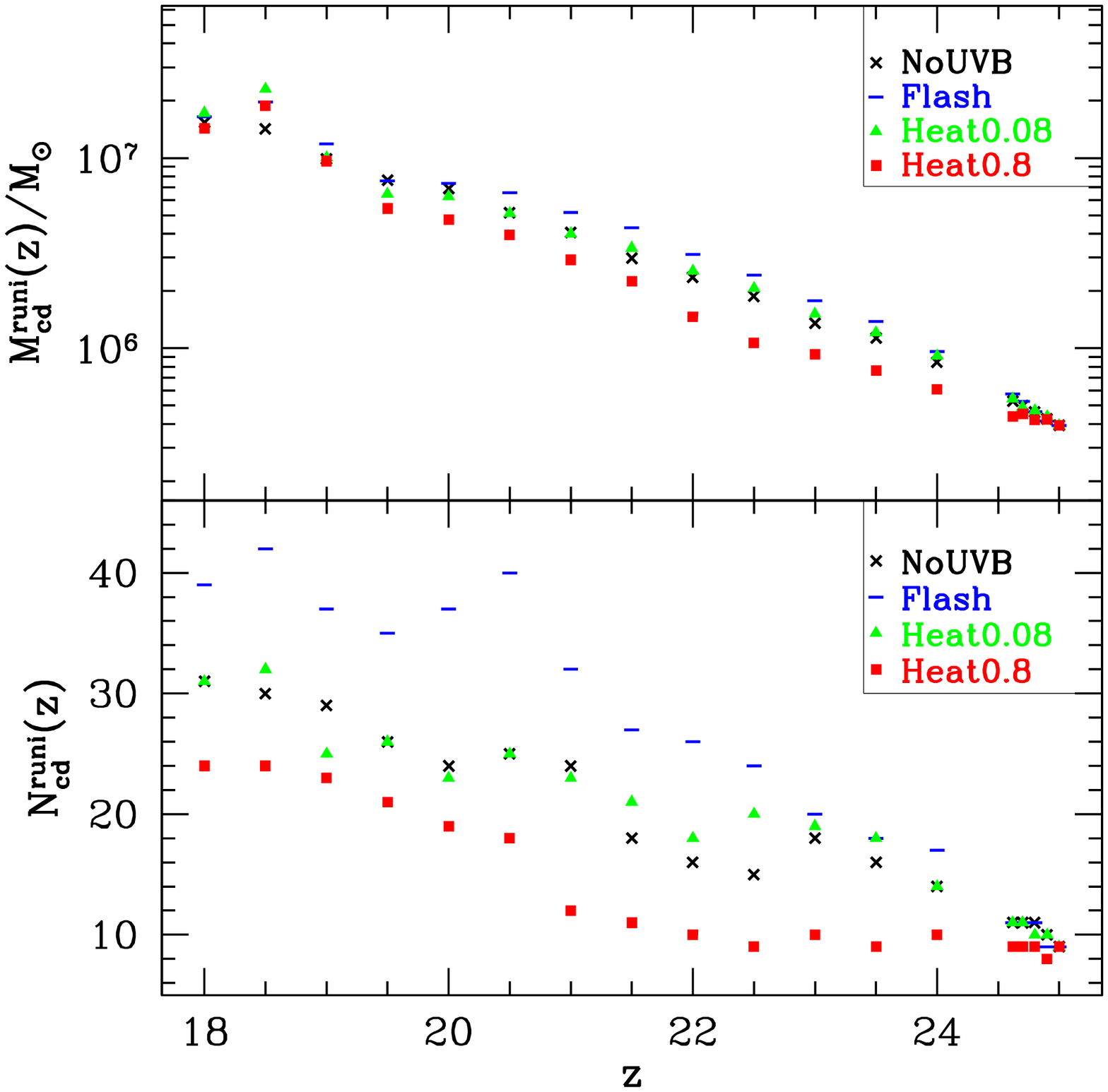}{3.3}{0.5}{.}{0.}  \figcaption{Values of
$\Mrun(z)$ ({\it top panel}) and $\Nrun(z)$ ({\it bottom panel}) as
defined in equations (\ref{eq:delM}) and (\ref{eq:delN}).  The results
are displayed for the \nothing\ ({\it crosses}), \flash\ ({\it
dashes}), \heatinglow\ ({\it triangles}), and \heatinghigh\ ({\it
squares}) simulation runs.
\label{fig:NM}}
\vspace{-1\baselineskip}
\end{figure}

Equations (\ref{eq:delN}) and (\ref{eq:delM}) provide an estimate of
how the CD gas has been affected by the presence of a UVB, {\it
following} the turn-on redshift of the UVB, $\zuvbon$ (the values at
$\zuvbon$ are subtracted in order to provide a more sensitive measure
of {\it relative} changes of CD gas).  As defined above, $\delN = 0$
and $\delM = 0$ if the UVB has no effect.  If the effect of a UVB is
positive, resulting in positive feedback, $\delN$ and $\delM$ would be
positive.  If the effect of the UVB is negative, $\delN$ and $\delM$
would be negative.

\begin{figure}
\myputfigure{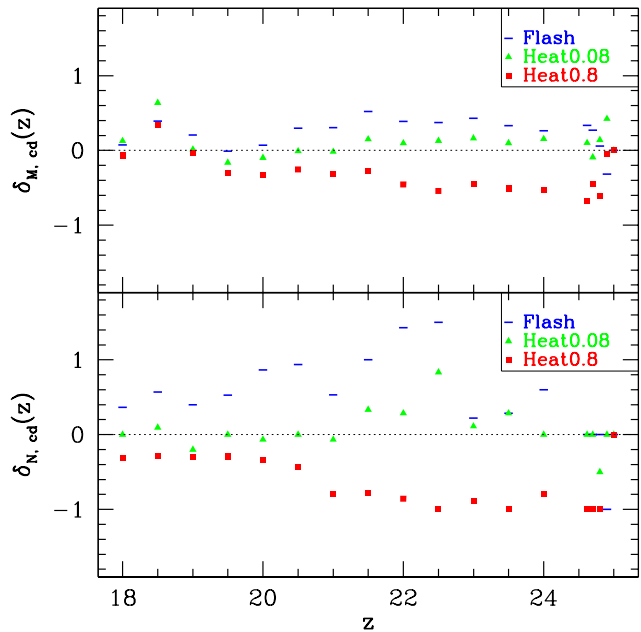}{3.3}{0.5}{.}{0.}  \figcaption{Values of
$\delM$ ({\it top panel}) and $\delN$ ({\it bottom panel}) as defined
in equations (\ref{eq:delM}) and (\ref{eq:delN}).  The results are
derived from Figure~\ref{fig:NM} and displayed for the \flash\ ({\it
dashes}), \heatinglow\ ({\it triangles}), and \heatinghigh\ ({\it
squares}) simulation runs.
\label{fig:delta}}
\vspace{-1\baselineskip}
\end{figure}

In Figure \ref{fig:NM}, we plot the values of $\Mrun(z)$ ({\it top
panel}) and $\Nrun(z)$ ({\it bottom panel}) in our four main
simulation runs: ${\rm run}i$ = \nothing\ ({\it crosses}), \flash\
({\it dashes}), \heatinglow\ ({\it triangles}), and \heatinghigh\
({\it squares}).  The corresponding values of $\delM$ and $\delN$ are
plotted in Figure \ref{fig:delta} in the top and bottom panels,
respectively.  The results are displayed for the \flash\ ({\it
dashes}), \heatinglow\ ({\it triangles}), and \heatinghigh\ ({\it
squares}) simulation runs.  Although some of the notable fractional
changes shown in Figure \ref{fig:delta} might appear statistically
insignificant due to the small number statistics inferred from Figure
\ref{fig:NM}, it should be noted that these runs are not uncorrelated
experiments.  In other words, each of our runs in Table \ref{tbl:runs}
is seeded with the same initial conditions, and so small relative
changes compared to the \nothing\ run are significant (i.e. the errors
are not Poisson).

One can infer from Figures \ref{fig:NM} and \ref{fig:delta} that the
\heatinghigh\ run shows evidence of strong negative feedback down to
$z\sim 20$, with values approaching the \nothing\ run by the end of
our simulation ($z=18$). Conversely, the \flash\ run exhibits strong
positive feedback down to $z\sim20$, and again approaches the
\nothing\ run by the end of our simulation.  In the middle is the
\heatinglow\ run, which shows very little difference compared to the
\nothing\ run (initially there is some evidence of mild positive
feedback down to a redshift of $z=21$, but at redshifts below that,
little evidence remains of a UVB ever being present).

It is also interesting to note that while the halo number in the
\heatinghigh\ run shows negative feedback down to $z=18$ ({\it bottom
panels of Figures \ref{fig:NM} and \ref{fig:delta}}), the total mass
of CD gas ({\it top panels of Figures \ref{fig:NM} and
\ref{fig:delta}}) shows no such feedback at $z\lsim19$.  The
explanation for this apparent contradiction is that the total mass of
CD gas is dominated by the largest halos (both because these halos are
more massive and because the fraction of CD gas increases with mass), and
as Figure~\ref{fig:gas_fract} shows, these large halos are largely
unaffected by the ionizing radiation.  Conversely, the elimination of
the CD gas from the lowest--mass halos even at $z=18$ is a genuine
effect (as is clearly visible in the lower right panel in
Fig~\ref{fig:gas_fract}), but these halos do not contribute
significantly to the total CD mass summed over all halos.

Figure \ref{fig:delta} agrees well with the qualitative inferences
drawn above.  Furthermore, it explicitly shows that the critical UVB
flux cutoff in our simulation between inducing a net negative and net
positive feedback is $J_{\rm UV}\sim 0.1$.  Halos which have been
exposed to a fainter UVB exhibit positive feedback, whereas halos
which have been exposed to a brighter UVB exhibit negative feedback.
However, it is also important to note that any such feedback is
temporary, as all of our runs begin to converge by the end of our
simulations at $z=18$.  The exception is that the Heat0.8 run shows
persistent suppression of the smallest halos (with $M<10^6 \Msun$) all
the way down to $z=18$.

\subsection{Relating initial densities at $\zuvbon$ to subsequent  
suppression of cold, dense gas}
\label{sec:densities}

Here we attempt to generalize and physically motivate some of the
results from the previous section.  In particular, we have already
seen that feedback depends on $J_{\rm UV}$ and $M_{\rm halo}$. Here we
examine whether a halo's capacity for forming CD gas depends strongly
on the properties of its progenitor region at the time of the
UV--illumination ($\zuvbon$).  Specifically, we expect those
progenitor regions which are less dense at $\zuvbon$, and hence at an
earlier evolutionary stage, to be more susceptible to negative
photo--heating and photo--evaporation feedback than more dense
regions.  This is because the ${\rm H_2}$ photo--dissociation rate
scales with the density, whereas ${\rm H_2}$--forming reaction rates
scale with the square of the density; as a result, photo--dissociation
becomes comparatively more important at low densities \citep{OH03}.
Below we focus on the \heatinghigh\ run, as it exhibits the strongest
negative feedback.

We divide the set of halos with CD gas at redshift $z$ in our
\nothing\ run into two groups: those that also {\it have} CD gas in
the \heatinghigh\ run (group 1), and those that {\it do not} also have
CD gas in the \heatinghigh\ run (group 2).  From Figure
\ref{fig:gas_fract}, one can note that at $z=18$ it is possible to
define a rough mass scale that separates these two groups; namely
halos with masses $\gsim 10^6 \Msun$ {\it do not} have their CD gas
suppressed (group 1), and halos with masses $\lsim 10^6 \Msun$ {\it
do} have their CD gas suppressed (group 2).  As stated above, we
hypothesize that the physical distinction between the two sets occurs
due to their differences at the redshift they were exposed to the UVB,
$\zuvbon$.  Specifically, we compare the mass-weighted, average
densities of progenitor regions at $\zuvbon$ = 25, which are to become
our halos from groups 1 or 2 at some later $z$.  We do this by tracing
back all dark matter particles comprising each halo at $z$ to their
positions at $z=\zuvbon$, and then obtaining the average gas density
at that position.

As there are too few halos to accurately construct the group 1 and
group 2 mean density distribution functions (see bottom panel of
Fig. \ref{fig:NM}), we present their properties via a density cutoff.
We adopt a simple criterion to define a density cutoff, $\rhocut$,
between the two groups.  We chose $\rhocut$ so that the sum of the
fraction of group (1) points below $\rhocut$ and the fraction of group
(2) points above $\rhocut$ is minimized.  Specifically, this
fractional sum used as a proxy for the disjointness of the two
distributions is defined as
\begin{equation}
\label{eq:ffit}
\ffit \equiv {\rm MIN} \left[ f_1(< \rhocut) + f_2(>\rhocut) \right]
\end{equation}
where $f_1(<\rhocut)$ is the fraction of halos in group 1 which have
mean densities {\it less} than the cutoff density, and $f_2(>\rhocut)$
is the fraction of halos in group 1 which have mean densities {\it
greater} than the cutoff density.  In essence, each term is the
fraction of ``misclassified" halos, and so we want to select $\rhocut$
such that $\ffit$ is minimized.  The sum as defined in
equation (\ref{eq:ffit}) ranges from 0 to 2, while our figure of merit
ranges from $\ffit = 0$ for two completely disjoint distributions to
$\ffit = 1$ for the case where group 1 and group 2 are drawn from the
same underlying distribution (not taking into account Poisson errors).

\begin{figure}
\vspace{+0\baselineskip}
\myputfigure{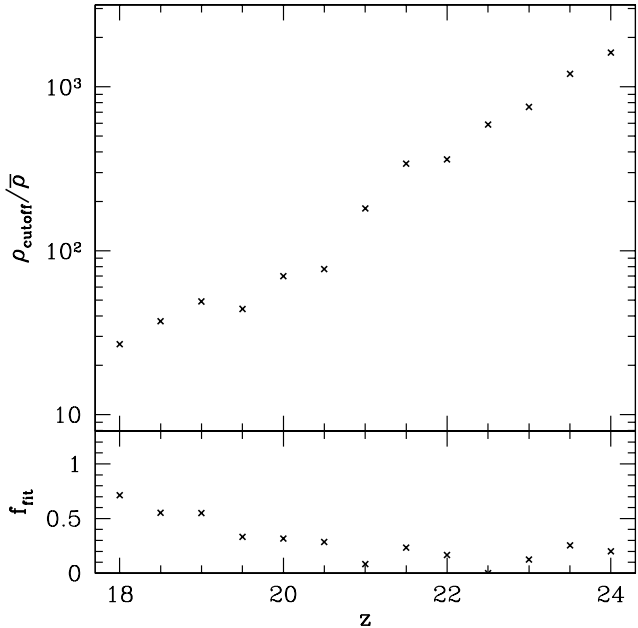}{3.3}{0.5}{.}{0.}
\vspace{-1\baselineskip} \figcaption{The critical density $\rhocut$,  
{\it upper panel}, for the progenitor gas at $\zuvbon$ =
25 of halos that collapse at some later redshift $z$, roughly divides
halos experiencing negative vs. positive feedback at redshift $z$
(i.e. halos that were more/less dense than $\rhocut$ at the time of
illumination will experience positive/negative feedback).  $\rhocut$
is a mass-weighted mean density of the progenitors pieces of the halo,
shown in units of the average comoving density, $\bar{\rho}$ in our
\heatinghigh\ run.  The values of $\ffit$, shown in the {\it Lower
panel}, show a rough figure of merit for how well the fixed density
$\rhocut$ separates halos into two disjoint categories ($\ffit \ll 1$
indicates a clear separation).  See text and equation (\ref{eq:ffit})
for definitions and discussion.
\label{fig:densities}}
\vspace{-1\baselineskip}
\end{figure}

We plot values for the density cutoff (in units of the average
comoving density, $\bar{\rho}$) as a function of redshift in the top
panel of Figure \ref{fig:densities}.  Our disjointness figure of merit
is plotted in the bottom panel of Figure \ref{fig:densities}.  Note
that for most redshifts, $\ffit \ll 1$, meaning that the group 1 and
group 2 density distributions are quite disjoint, and that the density
cutoff, $\rhocut$, has a well-defined value.  One can get a sense of
the Poissonian errors associated with $\rhocut$ by looking at the
bottom panel of Figure \ref{fig:NM}, since ${\rm N_{group 1}} \sim
N_{\rm cd}^{\rm \heatinghigh}\sim 10-20$ and ${\rm N_{group 2}} \sim
N_{\rm cd}^{\rm \nothing} - N_{\rm cd}^{\rm \heatinghigh}\sim 10$.
One should also note that $\ffit$ increases with decreasing $\rhocut$,
which is to be expected as the density distributions can not have
negative values, so both distributions start being ``packed'' together
as they approach zero.  In other words, there is an intrinsic
``noise'' consisting of small environmental fluctuations (halo
location, peculiar velocity, etc.), and this noise becomes more
noticeable as $\rhocut \rightarrow 0$.  In practice, it is difficult
to disentangle this effect from an actual merging of the two
distributions.

As expected, halos with less dense progenitor regions at $\zuvbon$ are
more susceptible to negative feedback.  It is quite interesting to
note that our density cutoff decreases exponentially with redshift,
implying that an increasing fraction of the photo--heated mass will
fall in the ``borderline'' region between negative and positive
feedback. This provides further support for our earlier claim that the
fossil HII region ``forgets'' the UVB as time passes.  In other words,
a strong UVB serves to merely delay the gas from cooling and
collapsing; the gas eventually manages to cool, aided by an enhanced
$\rm H_2$ fraction and enhanced infall (see Figure \ref{fig:profiles},
and associated discussion).  The length of this delay is a strong
function of the density of halo progenitor regions at $\zuvbon$, as
one would expect from our analysis in \S~\ref{sec:profiles}.

It is numerically impractical to run our simulations to redshifts much
lower than $z\lsim18$, due to the rapidly increasing collapsed
fraction in our refined region (see Fig. \ref{fig:f_col}).  On the
other hand, it would be interesting to know what eventually happens to
{\it most of the mass} of our refined region. A step towards answering
this question is to find out whether the majority of the mass of our
refined region at $z=\zuvbon$ is located in overdensities below or
above the lowest redshift $\rhocut$ value shown in Figure
\ref{fig:densities}.  With this motivation, we obtained a
mass-weighted density distribution over randomly generated positions
inside our refined region.  We select a radius surrounding each
position, such that a given number of DM particles lie within that
radius (the number of particles is chosen to correspond to the halo
masses capable of hosting CD gas).  We then obtain a mass-weighted
average density by averaging over the gas densities at the location of
each DM particle inside our chosen radius.

\begin{figure}
\myputfigure{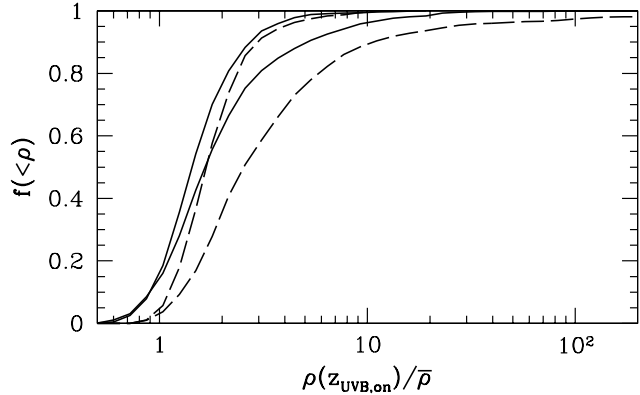}{3.3}{0.5}{.}{0.}
\vspace{-1\baselineskip} \figcaption{Mass-weighted, cumulative
density distributions for regions of $M\sim 8.9\times10^5\Msun$ ({\it
solid curves}) and $M\sim 8.9\times10^6\Msun$ ({\it dashed curves}).
Two redshift values are presented: $\zuvbon$ = 33 and 25, from left to
right.  Note that $\rho$ is the comoving gas density.
\label{fig:den_hist}}
\vspace{-1\baselineskip}
\end{figure}

In Figure \ref{fig:den_hist}, we plot the cumulative density
distributions (fraction of regions with mass-weighted density less
than $\rho$) thus generated at $z$ = 33 and 25, from left to right,
and for regions of mass scale $M\sim 8.9\times10^5\Msun$ ($\sim10^3$
DM particles) and $M\sim 8.9\times10^6\Msun$ ($\sim10^4$ DM particles)
with solid and dashed lines, respectively.  Understandably, the larger
mass scales shift the mean density towards larger values, due to the
increased likelihood of averaging over dense patches.  Also, we see
that for regions of equal mass scales, the higher redshift
counterparts have a lower mean density, partly due to a smaller
clumping factor and partly due to to the fact that we plot comoving
density, which increases with decreasing redshift.

Figure \ref{fig:den_hist} shows that the majority of the mass ($\gsim
90\%$) of our refined region is located in regions with mean densities
lower than $\sim10\bar{\rho}$, the lowest cutoff obtained by our
analysis (see Figure \ref{fig:densities}).  Hence, we cannot rule out
the possibility of significant negative feedback at lower redshifts,
not probed by our simulation.  Nevertheless, we regard this as unlikely,
for two reasons.  First, halos will be centered around
overdensities, not random points, and subsequent growth of the halo's
mass need not be spherically symmetric; these effects will bias the
relevant density distribution to higher values than shown in Figure
\ref{fig:den_hist}.  Second, it is likely that most halos massive
enough to host CD gas, which form in our biased region at $z<18$,
already had a dense progenitor core at $z=\zuvbon$.  Indeed, we find
that all halos hosting CD gas at $z=18$ in the \nothing\ run had {\it
some} dense progenitor gas ($\rho \gsim 100 \bar{\rho}$) at
$\zuvbon=25$.  Subsequent growth could be dominated by gas accretion
onto these dense regions, and merging with other halos (rather than
forming fresh halos entirely from low--density gas).  Nevertheless, we
emphasize that we chose to simulate a biased volume (favoring early
collapse), and the above arguments will be less valid in a more
typical patch of the IGM (with lower densities).

\subsection{Early UVB Run}
\label{sec:early_heating}

Ideally, one would want to explore all of parameter space by varying
$J_{\rm UV}$, $\zuvbon$ and $\zuvboff$, with different realizations of
the density field.  Unfortunately, given computational limitations,
this is impossible.  However, in order to confirm the trends we
present above, we run another simulation, \earlyheating, in which we
turn on a UVB, with an amplitude of $J_{\rm UV}=0.8$, at $\zuvbon=33$,
and turn it off at $\zuvboff=32.23$.  We then repeat the analysis in
\S~\ref{sec:densities}.  The corresponding figures, Figures
\ref{fig:early_NM} and \ref{fig:early_delta}, are presented below.

Figure \ref{fig:early_delta} shows that we once again find strong
negative feedback down to $z\sim23$.  For $z<23$, we see virtually no
evidence of any feedback, lending further credibility to our
interpretation above, that our other runs ``forget'' the episode of UV
heating, and start converging to the \nothing\ run by the end of our
simulations ($z=18$).

\begin{figure}
\vspace{+0\baselineskip}
\myputfigure{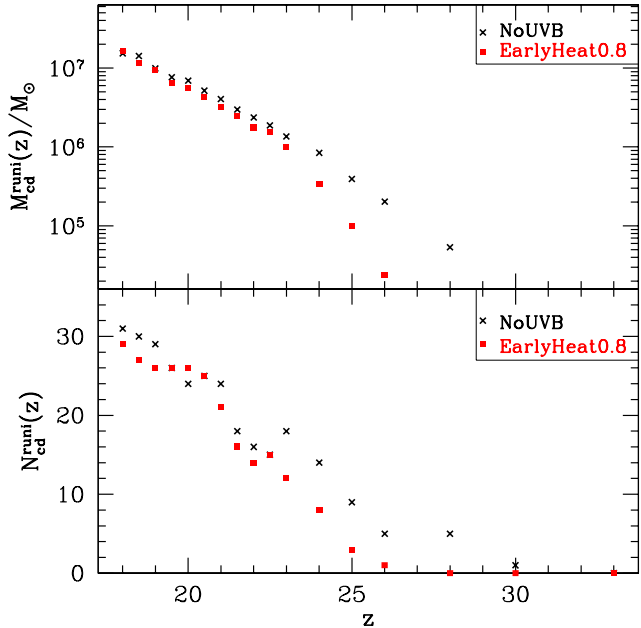}{3.3}{0.5}{.}{0.}
\vspace{-1\baselineskip}\figcaption{ Values of $\Mrun(z)$ ({\it top
panel}) and $\Nrun(z)$ ({\it bottom panel}) as defined in equations
(\ref{eq:delM}) and (\ref{eq:delN}).  The results are displayed for
the \nothing\ ({\it crosses}) and \earlyheating\ ({\it squares})
simulation runs.
\label{fig:early_NM}}
\vspace{-1\baselineskip}
\end{figure}

\begin{figure}
\vspace{+0\baselineskip}
\myputfigure{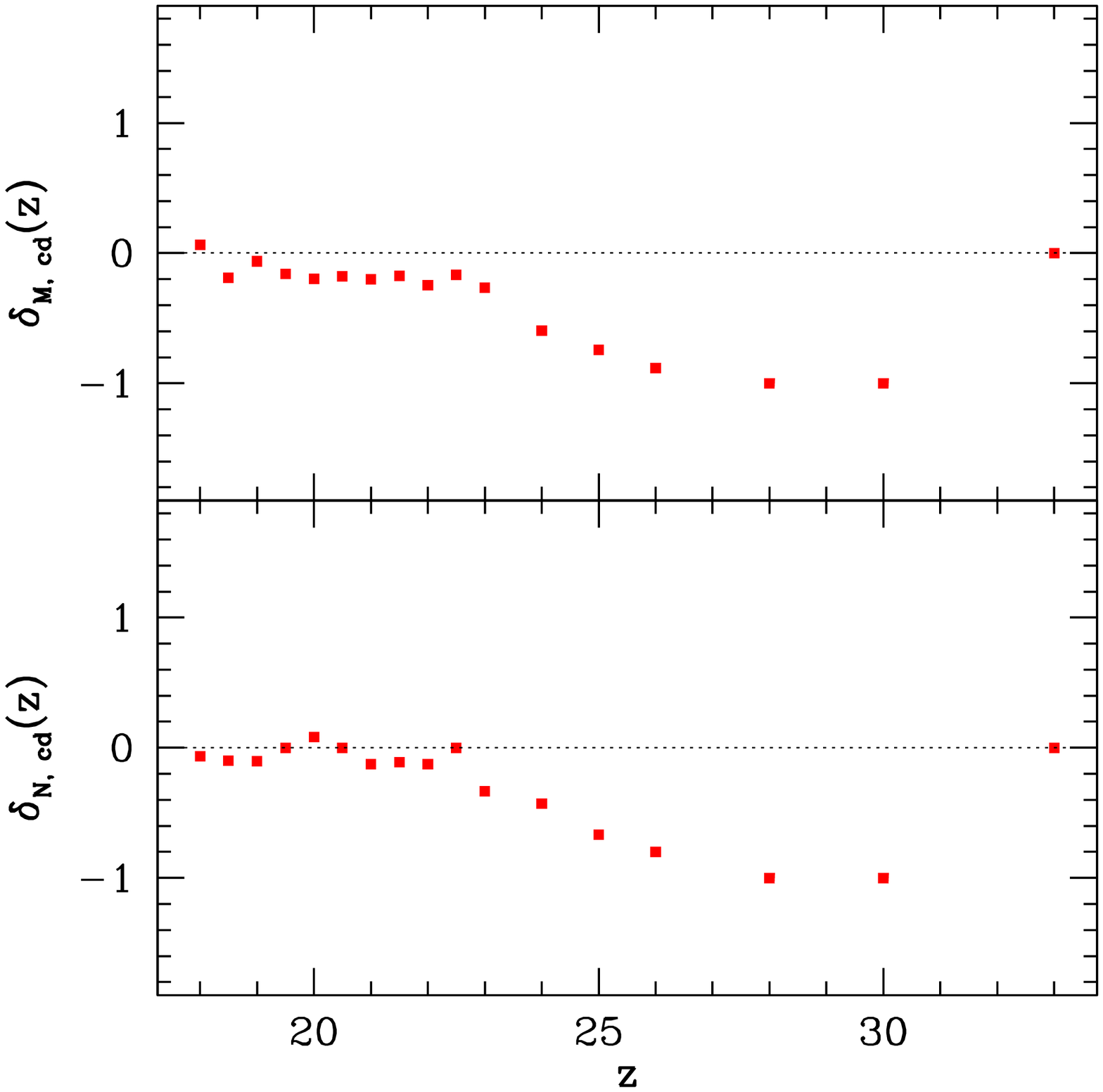}{3.3}{0.5}{.}{0.}
\vspace{-1\baselineskip}\figcaption{Values of $\delM$ ({\it top
panel}) and $\delN$ ({\it bottom panel}) as defined in equations
(\ref{eq:delM}) and (\ref{eq:delN}), shown here in the \earlyheating\
run.
\label{fig:early_delta}}
\vspace{-1\baselineskip}
\end{figure}

\begin{figure}
\vspace{+0\baselineskip}
\myputfigure{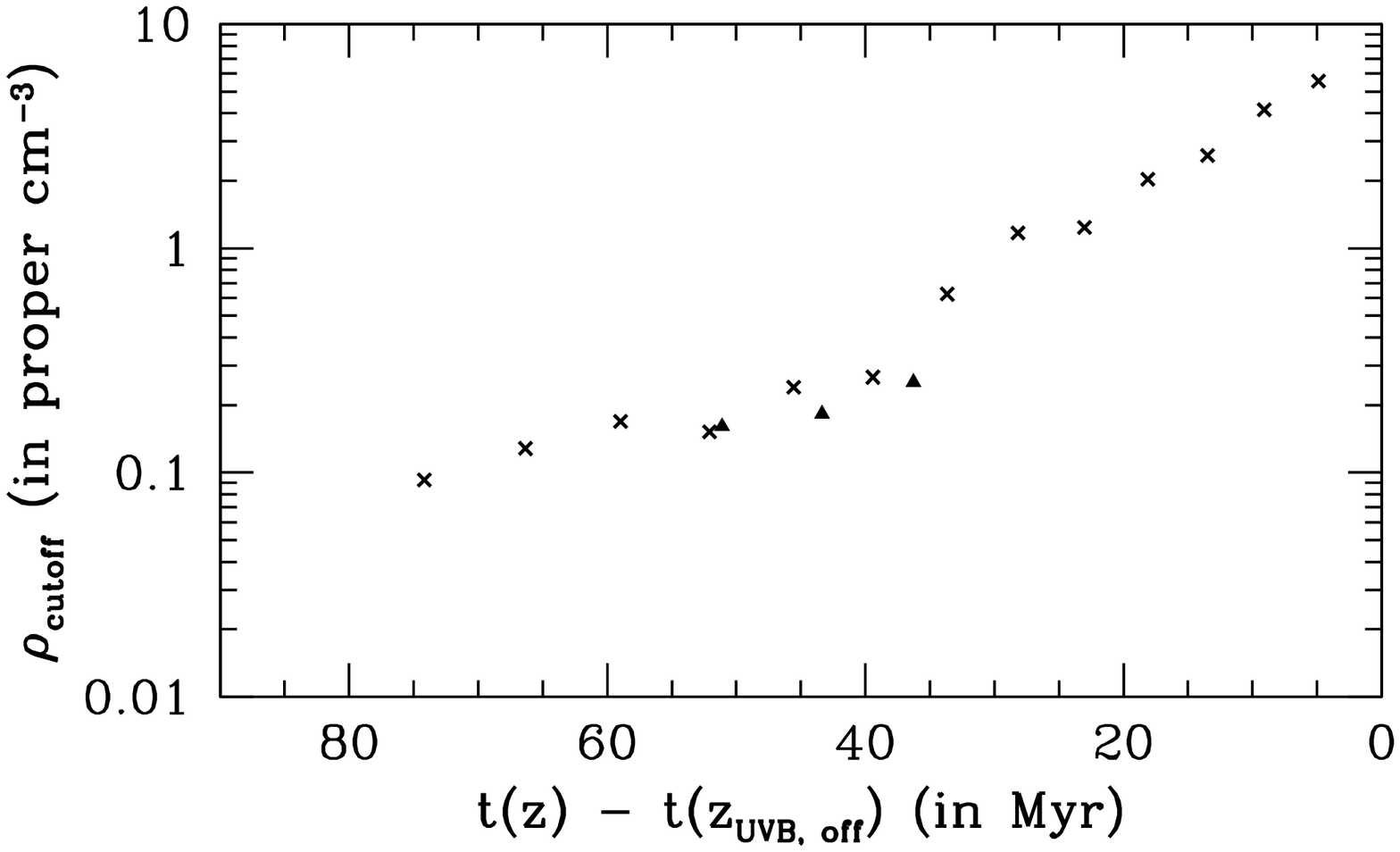}{3.3}{0.5}{.}{0.}
\vspace{-1\baselineskip} \figcaption{Values of $\rhocut$ (in proper
cm$^{-3}$) as a function of time elapsed since $\zuvboff$.  Crosses
correspond to our \heatinghigh\ run (i.e. $\zuvbon=25$); triangles
correspond to our \earlyheating\ run (i.e. $\zuvbon=33$).
\label{fig:t_both}}
\vspace{-1\baselineskip}
\end{figure}

In Figure \ref{fig:t_both}, we plot the density cutoff, $\rhocut$,
defined in \S~\ref{sec:densities}, for both the \heatinghigh\ ({\it
crosses}) and \earlyheating\ ({\it triangles}) runs.  For the sake of
a direct comparison, this time we use physical units both for
$\rhocut$, (proper cm$^{-3}$), and for the time elapsed since the UVB
turn-off (Myr).  Unfortunately, the drawback to having a simulation
run with such an early heating episode is that there are fewer halos
to analyze at earlier epochs.  Specifically, in the epoch with evident
negative feedback ($z\gsim23$), as seen below, there are only three
redshift outputs containing {\it both} halos exhibiting suppression
{\it and} halos not exhibiting suppression of CD gas (groups 2 and 1,
respectively, defined in \S~\ref{sec:densities}).  While it is
difficult to draw strong conclusions from Figure \ref{fig:t_both}, the
density cutoff values do appear similar in the two runs.

We examined the radial profiles of the same halo pictured in Figure
\ref{fig:profiles}, to verify that we can apply the same cooling
arguments as discussed in \S~\ref{sec:profiles}.  We compare the
\nothing\ and \earlyheating\ runs at $z=30$, shortly after
$\zuvboff=33$.  As in Figure~\ref{fig:profiles}, this redshift
corresponds roughly to the regime where the induced shock begins to
dissipate, and the gas starts falling back into the core.  In all of
the runs, the temperature drops to $T\sim \Tvir \sim 10^3$K near the
core very soon after $\zuvboff$.  As in \S~\ref{sec:profiles}, we
characterize the delay in formation of CD gas with
(c.f. eq. (\ref{eq:delay}))
\begin{equation}
\label{eq:earlydelay}
\fdelay \sim \left( \frac{n^{\rm\nothing}_g}{n^{\rm\heatinghigh}_g}
\right) \left( \frac{x^{\rm\heatinghigh}_{\rm
H_2}}{x^{\rm\nothing}_{\rm H_2}} \right)^{-1} \sim \frac{25}{20} \sim
1.
\end{equation}
\noindent
From our simple cooling argument, we predict a nearly negligible delay
for this halo.  Indeed, the halo ends up forming CD gas at $z=20.5$ in
the \earlyheating\ run, and at $z=21$ in the \nothing run, showing a
very small delay.

Despite the halo's exposure to the UVB earlier in its evolution and
subsequent lower gas density, the total negative feedback is reduced
compared to the \heatinghigh\ run.  Compared to the \heatinghigh\ run,
the negative feedback, when expressed as the photo-evaporation term in
the above equation $( n^{\rm\nothing}_g /n^{\rm\heatinghigh}_g)$ is
smaller by a factor of $\sim 2$, and the positive feedback, when
expresses as the H$_2$ fraction term in the above equation $(
x^{\rm\heatinghigh}_{\rm H_2}/x^{\rm\nothing}_{\rm H_2})$ is larger by
a factor of $\sim 2$.  These changes are explained by more efficient
Compton cooling (which more effectively eliminates the impact of the
photo--heating), and the lower gas density (which leads to a lower
value for the ${\rm H_2}$ fraction in the \nothing\ run, and hence a
larger relative enhancement in the \heatinghigh\ run), respectively
\citep{OH03}.

\subsection{The impact of not including radiative transfer}
\label{sec:nort}

Our simulations treat photo-ionization in the optically thin limit and
so do not include radiative transfer effects.  This results in two
differences compared to a self-consistent treatment.

The most obvious effect is that all of our halos are ionized
simultaneously, while in reality halos are ionized by very nearby
stars with distances less than the few kpc radius of typical HII
regions \citep{WAN04, KYSU04}.  Nevertheless, we argue that the
primary effect of this is to vary the flux felt by the halo and we
explore a range of reasonable fluxes in our simulations.  The
exception is if the halo is so close that it is enveloped within the
shock generated by the gas expelled from the halo hosting the star
that produces the ionizing radiation.  However, typically this shocked
region occupies a volume of less than 1\% of the ionized region (e.g.,
\citealt{WAN04}).

The second, and more important, effect of radiative transfer will be
to shield the high density cores of our minihalos.  If the cores are
not ionized, then both the positive and negative feedback effect will
clearly not occur in the neutral gas.  \citet{ABS05} estimate that
self-shielding will set in at densities around a few particles per
cm$^{-3}$ (depending on the strength of the flux and the size of the
halo).  This value is approximately the density we find in the cores
of our simulated halos (e.g., Figure 2), and so we conclude that
radiative transfer effects may play an important role in the cores of
our halos.  We note that at these densities, we typically find very
little negative feedback anyway because of the short cooling times in
the ionized gas.  Most of the negative feedback we observe arises due
to the photo-heating of low-density gas which is then later accreted
onto halos.  This means that we do not expect our results to be
strongly affected by the missing radiative transfer effects.  The
effect, where important, will be to decrease the amount of feedback,
making our statements about feedback upper limits on the amplitude of
the expected feedback.  Finally we note that, given time, the halos
will be evaporated and eventually ionized despite the high densities
in the core; however, this photo-evaporation time will be typically
longer than 3 Myr \citep{HAM01, SIR04, ISR05}.

\section{The Impact of a Lyman Werner Background}
\label{sec:lw}

While up to now we have ignored a possible Lyman--Werner background,
such a background is likely to be present early on, and could have a
strong impact on the ${\rm H_2}$ chemistry and gas cooling.  In
particular, the IGM is nearly optically thin, or quickly becomes so,
at frequencies below 13.6eV \citep{HAR00}.  For reference, we note
that one photon per hydrogen atom (the minimum UV background required
for reionizing the IGM) would translate to a background intensity of
$J_{\rm LW}\sim 20 [(1+z)/21]^3$. Background levels 2--4 orders of
magnitude below this value will be established well before
reionization, and have the potential to already photodissociate ${\rm
H_2}$ molecules at these early epochs \citep{HRL97,
MBA01}. Furthermore, and of more direct interest in the present study,
\citet{OH03} have argued that the presence of an entropy floor,
generated by the UV heating, reduces gas densities, and makes the
${\rm H_2}$ molecules in collapsing halos more vulnerable to a LW
background.  Motivated by the above, in this section, we study the
impact of a LW background on our results.  We start by a brief
discussion of our results without a LW background
(\S~\ref{subsec:disc}), use these results to build up some
expectations for the impact of the LW background
(\S~\ref{subsec:LWtoy}), and then present the results of simulation
runs with LW backgrounds (\S~\ref{subsec:LWsims}).

\begin{figure}
\vspace{+0\baselineskip}
\myputfigure{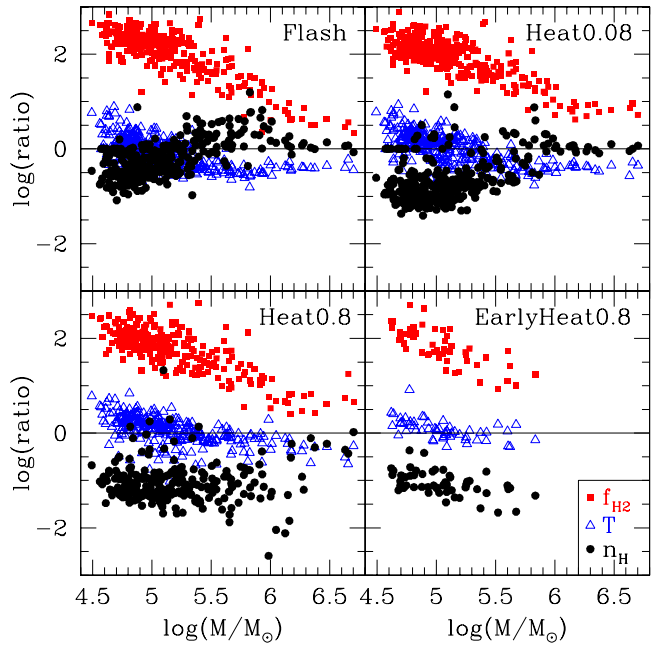}{3.3}{0.5}{.}{0.}
\vspace{-1\baselineskip} \figcaption{The figure illustrates the impact
of the UV heating in the four different runs (Flash, Heat0.08,
Heat0.8, EarlyHeat0.8, as labeled in each panel).  For each halo, we
show, at $z=23$ (or $z=30$, in the early heating run), the ratio of the
${\rm H_2}$ fractions (red filled squares), the temperatures (blue
empty triangle) and of the average gas density within the central 15pc
(red filled circles) in the runs with UV heating, compared to the runs
with no UVB. Note that the ${\rm H_2}$ fraction tends to increase by
the same factor, regardless of the nature of the heating. As a result,
the sign of the overall feedback (negative or positive) is determined
primarily by the changes in the gas density, which does depend
strongly on the type and amount of heating. \label{fig:fossilf4}}
\vspace{-1\baselineskip}
\end{figure}

\subsection{Discussion of Results without the LW Background}
\label{subsec:disc}

As already discussed above, the UV heating produces two prominent
effects: it boosts the ${\rm H_2}$ fraction and it decreases the gas
density.  In the case of the individual halo studied in
Figure~\ref{fig:profiles} and described in \S~\ref{sec:profiles}, and
also in the case of the analogous halo in the early heating run,
described in \S~\ref{sec:early_heating} above, we have seen that the
overall impact of the UV heating is a delay in the development of cold
dense gas; this delay can be understood by the increase in the cooling
time, given by the product of the two effects above.

In order to understand the net effect of the UV heating on the overall
halo population, in Figure~\ref{fig:fossilf4}, we show the ratio of
the ${\rm H_2}$ fractions (red filled squares), of the temperatures
(blue empty triangle) and of the mean gas density within the central
15pc (red filled circles) in all four of our runs with UV heating.
Each quantity is computed for every halo present at $z=23$ (or at
$z=30$ in the early heating run), in the runs with UV heating, and the
ratio refers to this value, divided by the same quantity in the run
without a UV background.

The figure clearly shows that the ${\rm H_2}$ fractions are enhanced
in a similar fashion in all of the runs (by factors ranging from several 
hundred at low mass, to a few at high mass).
This is indeed expected: while the H$_2$ abundance is nearly 
independent of halo mass in runs with a UVB, it is a strongly 
increasing function of mass in the NoUVB case (see Figure 3).
Because the ``freeze--out'' value of the ${\rm H_2}$
abundance, $f_{\rm H_2}\approx 2\times 10^{-3}$ is
insensitive to the background flux or duration, or to the gas density
\citep{OH02}, the enhancement factor over the NoUVB run is always the same.

Contrary to the ``universal'' effect on the ${\rm H_2}$ fraction, the
impact of the UV background on the gas density depends strongly on the
nature of the heating.  Not surprisingly, the flash heating case shows
the weakest gas dilution; heating the gas for an extended period, at
increasing flux levels, causes larger dilutions.  Note that the impact
on the density tends to diminish for more massive halos. This is
partly because a fixed amount of heating/energy input corresponds to a
smaller fraction of the halo's total binding energy. In addition, the
${\rm H_2}$--cooling time is shorter than $10^7$ years in halos with
$M\gsim 10^{5.8}~{\rm M_\odot}$ and the UV--heated gas is able to cool
prior to $z=23$.  This latter effect is also directly evident in the
gas temperature ratios, which decrease towards larger halos (and
decrease below unity).

The above trends account for the basic results shown in
Figure~\ref{fig:delta}. Note that this figure shows only those halos
that develop cold dense gas; i.e. those with $M\gsim 10^{5.5}~{\rm
M_\odot}$.  In the Flash--ionization case, the ${\rm H_2}$ fraction
enhancement in these halos dominates, and results in a positive
overall feedback. In the Heat0.08 case, the effects on the ${\rm H_2}$
fraction and on the gas density nearly cancel each other and the net
result is that the UV heating has almost no impact.  In the Heat0.8
case, the dilution of the gas density dominates, and results in a
delay in the cooling time, and in the development of the cold dense
gas, by a factor of $1-10$.

\subsection{The Impact of a Lyman Werner Background - Expectations}
\label{subsec:LWtoy}

The above trends suggest that the UV heating can render the halos more
susceptible to the negative effect of a LW background.  As argued in
\citet{OH03}, the ${\rm H_2}$ photodissociation rate depends linearly
on the gas density, while the rate of ${\rm H_2}$--forming two--body
collisions scales with the square of the density; hence density
dilution makes ${\rm H_2}$ photodissociation comparatively more
important.

\begin{figure}
\vspace{+0\baselineskip}
\myputfigure{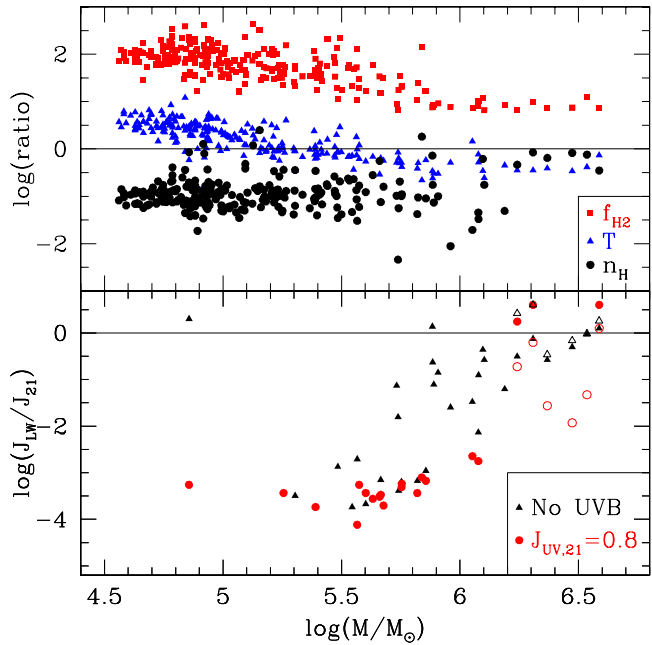}{3.3}{0.5}{.}{0.}
\vspace{-1\baselineskip} \figcaption{{\it Upper panel:} The ratio of the ${\rm H_2}$ fractions, temperatures, and average gas density between the Heat0.8 and NoUVB runs, at $z=24$ (following the notation in Figure~\ref{fig:fossilf4}).  
{\it Lower panel:} The critical value of the background LW flux, $J_{\rm LW}$ (in units of $10^{-21}{\rm ergs~s^{-1}~cm^{-2}~Hz^{-1}~sr^{-1}}$) such that the gas temperature cools to $300$K by redshift $z=18$.  Only those halos whose gas manages to cool by $z=18$ are shown.  Values for the \nothing\ run are represented by filled black triangles; values for the \heatinghigh\ run are represented by filled red circles.  Empty symbols denote the critical value of $J_{\rm LW}$, such that the gas temperature decreases by {\it half} between redshifts $z=24$ and $z=18$ (see discussion in text).
\label{fig:fossilf1}}
\vspace{-1\baselineskip}
\end{figure}

\begin{figure}
\vspace{+0\baselineskip}
\myputfigure{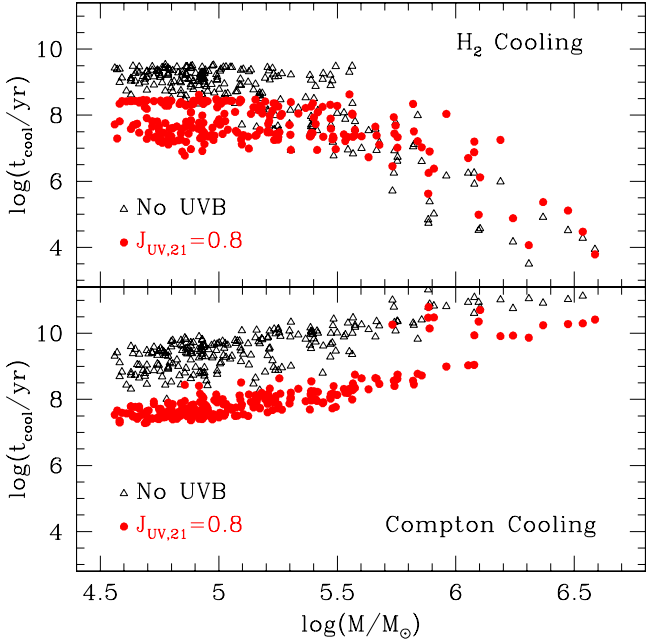}{3.3}{0.5}{.}{0.}
\vspace{-1\baselineskip} \figcaption{{\it Upper panel:} ${\rm H_2}$--cooling
time for each halo in the Heat0.8 (red, filled circles) and NoUVB
(black, empty triangles) runs, at redshift $z=24$.  
{\it  Lower panel:}
Compton--cooling timescale for the same halos. Note that the ${\rm
H_2}$--cooling time is shorter than the Hubble time in halos with
$M\gsim 10^{5.5}~{\rm M_\odot}$.
\label{fig:fossilf3}}
\vspace{-1\baselineskip}
\end{figure}

\begin{figure}
\vspace{+0\baselineskip}
\myputfigure{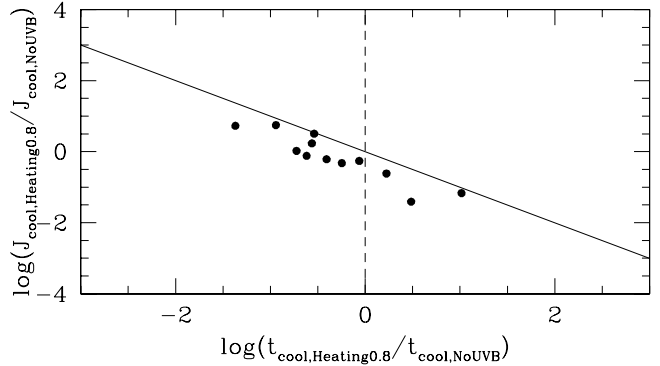}{3.3}{0.5}{.}{0.}
\vspace{-1\baselineskip} \figcaption{The figure shows the ratio of the
amplitude of critical LW background (defined to prevent gas cooling
between $z=24$ and $z=18$) between the Heat0.8 and NoUVB runs, as a
function of the ratio of the ${\rm H_2}$--cooling times.  The scaling
is close to the $J_{\rm LW} \propto t_{\rm cool}^{-1}$ expected from
equating the ${\rm H_2}$--cooling and ${\rm H_2}$--photodissociation
timescales. \label{fig:fossilf2}}
\vspace{-1\baselineskip}
\end{figure}

In order to investigate the impact of a LW background on the amount of
cold dense gas, we performed a set of six additional simulation runs.
Before describing these runs, however, we use the no--LW runs with UV
heating (Heat0.8) and without heating to develop some expectations.
These are shown in Figures~\ref{fig:fossilf1}--\ref{fig:fossilf2}, as
follows.

In Figure~\ref{fig:fossilf1}, in the upper panel, we show the ratio of
the ${\rm H_2}$ fractions (red filled squares), of the temperatures
(blue empty triangles) and of the mean gas density within the central
15pc (red filled circles).  The ratios are computed in the Heat0.8 and
NoUVB runs, as in Figure~\ref{fig:fossilf4}, but we here use $z=24$,
rather than $z=23$.  This choice is made to allow for some Compton
cooling, but to minimize the ${\rm H_2}$--cooling that occurs after
the heating is turned off (the latter may not occur if a LW background
is always on).  In Figure~\ref{fig:fossilf3}, we explicitly show the
${\rm H_2}$--cooling and Compton--cooling times for each halo in the
Heat0.8 and NoUVB runs, at $z=24$.  Note that the ${\rm H_2}$--cooling
time is shorter than the Hubble time in halos with $M\gsim
10^{5.8}~{\rm M_\odot}$.

In the bottom panel of Figure~\ref{fig:fossilf1}, we compute the
coupled chemical and thermal evolution at fixed density, and compute,
for each halo, the critical value of the background LW flux, $J_{\rm
LW}$ (in units of $10^{-21}{\rm ergs~s^{-1}~cm^{-2}~Hz^{-1}~sr^{-1}}$)
such that the gas temperature cools to $300$K by redshift $z=18$.
This will represent a proxy for a critical value for the LW
background, above which the halo is prevented from developing cold
dense gas in the simulation prior to $z=18$.  The choice of the
temperature, 300 K, matters relatively little for the low--mass
halos. On the other hand, we find that the critical $J_{\rm LW}$ we
derive for the larger ($M\gsim 10^6~{\rm M_\odot}$) halos is more
sensitive to this choice; in particular, these halos have high ($\gsim
1000$K) virial temperatures, and typically never cool down to 300K,
even in the absence of a LW background.  Hence, for these halos, we
show (in empty symbols), the critical value of the background LW flux,
$J_{\rm LW}$, such that the gas temperature decreases by {\it half}
between redshifts $z=24$ and $z=18$.  The bottom panel in
Figure~\ref{fig:fossilf1} shows that the critical $J_{\rm LW}$ is
between $10^{-4}$ and $10^0$, with a relatively large scatter at fixed
halo mass.  There is, nevertheless, a clear impact of the heating by
the UV background, which reduces the critical LW flux by about an
order of magnitude (shown as a vertical offset between circles and
triangles).  Note that the majority of halos (especially at low
masses) never develop cold dense gas, and are not shown in the bottom
panel.

In Figure~\ref{fig:fossilf2}, we show the ratio of the critical LW
background as a function of the ratio of the ${\rm H_2}$--cooling time
(which scales approximately as $t_{\rm cool}\propto T/[n_g f_{\rm
H2}]$) at $z=24$.  Note that there were only 12 halos for which the
critical LW background was finite in both the Heat0.8 and NoUVB runs
(this excludes the majority of halos, which do not form cold dense gas
even if $J_{\rm LW}=0$, and also those handful of halos that have
already formed cold dense gas prior to $z=24$, in either run).  As a
result, the range shown by this plot is not necessarily
representative.  Nevertheless, the figure shows a clear trend: the
critical LW background scales nearly as the inverse of the cooling
time. This can be understood easily: in order to prevent the gas from
cooling, the ${\rm H_2}$--dissociation time, $t_{\rm dissoc}\approx
2\times 10^7 (J_{\rm LW}/10^{-3})$ yr, must be comparable or shorter
than the ${\rm H_2}$--cooling time.  The critical LW flux falls
somewhat below the value predicted by this scaling, because at higher
LW fluxes, the ${\rm H_2}$ abundance starts saturating as it
approaches its equilibrium value (rather than decreasing linearly with
time under the influence of the background).

\subsection{The Impact of a Lyman Werner Background - Simulation  
Results}
\label{subsec:LWsims}

To better quantify the feedback effects of UV heating combined with a
persistent LW background, we performed six additional simulations, in
which the LW background was left on after the UVB was turned off (at
$z=24.62$).  Our background flux is constant throughout the narrow LW
frequency band (11.18--13.6eV).  We normalize the specific intensity
at the mean photon energy of 12.87 eV, in units of $10^ {-21}{\rm
ergs~s^{-1}~cm^{-2}~Hz^{-1}~sr^{-1}}$.  We include three values of the
LW background: $J_{\rm LW}=$ 0.001, 0.01, and 0.1.  Each of these
three LW backgrounds is applied to both our \nothing\ and
\heatinghigh\ runs at $z=24.62$, and is subsequently left on.

\begin{figure}
\vspace{+0\baselineskip}
\myputfigure{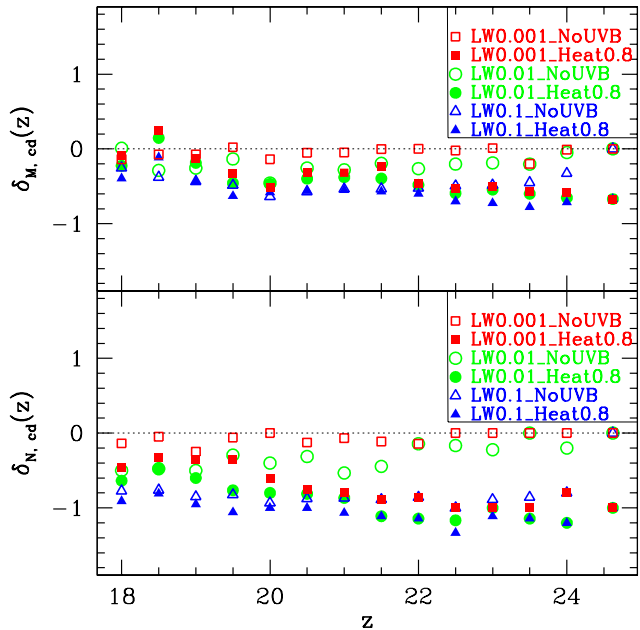}{3.3}{0.5}{.}{0.}
\vspace{-1\baselineskip}\figcaption{This figure shows the suppression
of cold dense gas in halos in simulations runs that include a
persistent LW background. The LW background had specific intensities
of $J_{\rm LW}=0.001, 0.01$, or $0.1$ (normalized at 12.87 eV, in
units of $10^ {-21}{\rm ergs~s^{-1}~cm^{-2}~Hz^{-1}~sr^{-1}}$). Each
of these three LW backgrounds is applied to both our \nothing\ and
\heatinghigh\ runs at $z=24.62$, and is subsequently left on.  Values
of $\delM$ ({\it top panel}) and $\delN$ ({\it bottom panel}) are
show, as defined in equations (\ref{eq:delM}) and (\ref{eq:delN}).
\label{fig:LWruns}}
\vspace{-1\baselineskip}
\end{figure}

In Figure \ref{fig:LWruns}, we show the resulting CD gas suppression,
as defined in eq. (\ref{eq:delN}) and (\ref{eq:delM}).  Empty symbols
refer to runs with no heating (\nothing), while filled symbols refer
to runs with heating (\heatinghigh).  In both cases, squares, circles,
and triangles denote simulation runs with increasing LW backgrounds
($J_{\rm LW}$ = 0.001, 0.01, 0.1, respectively).  Note that we obtain
values of $\delN < -1$ in Figure \ref {fig:LWruns}; this is due to the
fact that the CD gas disappears from some of the low mass halos in the
presence of strong LW fluxes (c.f the normalization of
eq. (\ref{eq:delN}), which tracks relative changes since
$z=\zuvbon=25$).

The results of the simulation runs in Figure \ref{fig:LWruns} agree
fairly well with the semi-analytical arguments in
\S~\ref{subsec:LWtoy} above.  Namely, the value of the LW background
at which significant suppression occurs by $z=18$ in the \nothing\
runs is found to be approximately $J_{\rm LW} \sim 0.01$.  In the
Heat0.8 run, there is significant suppression at $z=18$ already for
$J_{\rm LW} \sim 0.001$.  While the bulk of this suppression is due to
the UV heating alone (and not the LW background; cf. Figure
\ref{fig:delta}), the LW background does prevent three additional
halos from cooling their gas prior to $z=18$.  This is consistent with
the expectation that the UV heating lowers the value of $J_{\rm LW}$
required for appreciable negative feedback, by a factor of $\sim$ 10.

More generally, our results reveal that for $J_{\rm LW}\lsim 0.01$,
negative feedback is dominated by UV heating, while for $J_{\rm
LW}\gsim 0.01$, negative feedback is dominated by the LW background.
Near the threshold value of $J_{\rm LW}\sim 0.01$, negative feedback
transitions from being UV heating dominated ($\lsim 100$Myr after
$\zuvboff$) to being LW background dominated ($\gsim 100$Myr after
$\zuvboff$).  This ``transition'' behavior can be understood as a
combined result of two effects: the UV heating is turned off, and its
impact is transient, as discussed above, while the critical LW
background scales roughly with the inverse of the density
\citep{HAR00, OH03} and hence a fixed LW background will have a larger
impact at lower densities or decreasing redshifts.

\section{Conclusions}
\label{sec:conc}

We used three-dimensional hydrodynamic simulations to investigate the
effects of a transient ultraviolet (UV) flux on the collapse and
cooling of pregalactic clouds, with masses in the range $10^5$ --
$10^7~\Msun$, at high redshifts ($z\gsim18$).  Although in the
scenario we envision, the radiation is due to nearby PopIII star
formation, in order to study its effect in a statistical way, we
adopted a spatially constant but short-lived photo-ionizing background
throughout the simulation box.  This was done to mimic the effect of a
$\sim 100$ solar mass star forming at $z = 25$ and shining for 3 Myr.
Of course, in reality, the closest star can be located at a range of
distances and so we effectively covered this range by varying the
strength of the background.  The effect of the ionizing background
will be strongest on relatively low density gas which is in the
process of assembling to form halos at later times.  The sign of the
effect has been uncertain with suggestions of positive feedback due to
enhanced H$_2$ formation, and negative feedback due to the increased
entropy of gas in the relic HII region.  In addition, we studied the
combined effects of this transient UV flux and a persistent
Lyman--Werner (LW) background (at photon energies below 13.6eV) from
distant sources.

In the absence of a LW background, we
find that a critical specific intensity of $J_{\rm UV} \sim 0.1 \times
10^{-21}{\rm ergs~s^{-1}~cm^{-2}~Hz^{-1}~sr^{-1}}$ demarcates the
transition from net negative to positive feedback for the halo
population. A weaker UV flux stimulates subsequent star formation
inside the fossil HII regions, by enhancing the ${\rm H_2}$ molecule
abundance. A stronger UV flux significantly delays star--formation by
reducing the gas density, and increasing the cooling time at the
centers of collapsing halos.  At a fixed $J_{\rm UV}$, the sign of the
feedback also depends strongly on the density of the gas at the time
of UV illumination.  In either case, we find that once the UV flux is
turned off, its impact starts to diminish after $\sim30\%$ of the
Hubble time.  

In the more realistic case when a LW background is
present (in addition to the ionizing source), with $J_{\rm LW} \gsim 0.01 \times 10^{-21}{\rm
ergs~s^{-1}~cm^{-2}~Hz^{-1}~sr^{-1}}$, strong suppression persists
down to the lowest redshift ($z=18$) in our simulations.  Finally, we
find evidence that heating and photoevaporation by the transient UV
flux renders the $\sim 10^6~{\rm M_\odot}$ halos inside fossil HII
regions more vulnerable to subsequent ${\rm H_2}$ photo--dissociation
by a LW background.

The results of this study show that the combined negative feedback of
a transient UV and a persistent LW background is effective at high
redshift in suppressing star--formation in the low--mass halos; this
suppression will help in delaying the reionization epoch to $z=6-10$
as inferred from SDSS quasar spectra and from CMB polarization
anisotropy measurements in the 3--yr {\it WMAP} data.

\acknowledgments{We thank Peng Oh for many stimulating and helpful
discussions.  AM acknowledges support by NASA through the
GSRP grant NNG05GO97H.  GB acknowledges support through NSF grants
AST-0507161 and AST-0547823.  ZH acknowledges partial support by NASA
through grants NNG04GI88G and NNG05GF14G, by the NSF through grants
AST-0307291 and AST-0307200, and by the Hungarian Ministry of
Education through a Gy\"orgy B\'ek\'esy Fellowship.  This work was
also supported in part by the National Center for Supercomputing
Applications under grant MCA04N012P.}

\bibliographystyle{apj}
\bibliography{apj-jour,ms}

\begin{thebibliography}{56}
\expandafter\ifx\csname natexlab\endcsname\relax\def\natexlab#1{#1}\fi

\bibitem[{{Abel} {et~al.}(2002){Abel}, {Bryan}, \& {Norman}}]{ABN02}
{Abel}, T., {Bryan}, G.~L., \& {Norman}, M.~L. 2002, Science, 295, 93

\bibitem[{{Alvarez} {et~al.}(2006){Alvarez}, {Bromm}, \& {Shapiro}}]{ABS05}
{Alvarez}, M.~A., {Bromm}, V., \& {Shapiro}, P.~R. 2006, \apj, 639, 621

\bibitem[{{Anninos} \& {Norman}(1996)}]{AN96}
{Anninos}, P., \& {Norman}, M.~L. 1996, \apj, 460, 556

\bibitem[{{Anninos} {et~al.}(1997){Anninos}, {Zhang}, {Abel}, \&
  {Norman}}]{Anninos97}
{Anninos}, P., {Zhang}, Y., {Abel}, T., \& {Norman}, M.~L. 1997, New Astronomy,
  2, 209

\bibitem[{{Barkana} \& {Loeb}(1999)}]{BL99}
{Barkana}, R., \& {Loeb}, A. 1999, \apj, 523, 54

\bibitem[{{Barkana} \& {Loeb}(2004)}]{BL04}
---. 2004, \apj, 609, 474

\bibitem[{{Bond} {et~al.}(1991){Bond}, {Cole}, {Efstathiou}, \&
  {Kaiser}}]{Bond91}
{Bond}, J.~R., {Cole}, S., {Efstathiou}, G., \& {Kaiser}, N. 1991, \apj, 379,
  440

\bibitem[{{Bromm} {et~al.}(2002){Bromm}, {Coppi}, \& {Larson}}]{BCL02}
{Bromm}, V., {Coppi}, P.~S., \& {Larson}, R.~B. 2002, \apj, 564, 23

\bibitem[{{Bromm} \& {Loeb}(2003)}]{Bromm2003}
{Bromm}, V., \& {Loeb}, A. 2003, \nat, 425, 812

\bibitem[{{Bryan}(1999)}]{Bryan99}
{Bryan}, G.~L. 1999, {Comput. Sci. Eng.}, 46, 1

\bibitem[{{Cen}(2003)}]{Cen03_postWMAP}
{Cen}, R. 2003, \apjl, 591, L5

\bibitem[{{Ciardi} {et~al.}(2000){Ciardi}, {Ferrara}, \& {Abel}}]{CFA00}
{Ciardi}, B., {Ferrara}, A., \& {Abel}, T. 2000, \apj, 533, 594

\bibitem[{{Efstathiou}(1992)}]{Efstathiou92}
{Efstathiou}, G. 1992, \mnras, 256, 43P

\bibitem[{{Eisenstein} \& {Hu}(1999)}]{EH99}
{Eisenstein}, D.~J., \& {Hu}, W. 1999, \apj, 511, 5

\bibitem[{{Eisenstein} \& {Hut}(1998)}]{EH98}
{Eisenstein}, D.~J., \& {Hut}, P. 1998, \apj, 498, 137

\bibitem[{{Fan} {et~al.}(2006){Fan}, {Carilli}, \& {Keating}}]{FCK06}
{Fan}, X., {Carilli}, C.~L., \& {Keating}, B. 2006, ARA\&A, in press, preprint
  astro-ph/0602375

\bibitem[{{Galli} \& {Palla}(1998)}]{GP98}
{Galli}, D., \& {Palla}, F. 1998, \aap, 335, 403

\bibitem[{{Gnedin} \& {Abel}(2001)}]{GA01}
{Gnedin}, N.~Y., \& {Abel}, T. 2001, New Astronomy, 6, 437

\bibitem[{{Haiman} {et~al.}(2001){Haiman}, {Abel}, \& {Madau}}]{HAM01}
{Haiman}, Z., {Abel}, T., \& {Madau}, P. 2001, \apj, 551, 599

\bibitem[{{Haiman} {et~al.}(2000){Haiman}, {Abel}, \& {Rees}}]{HAR00}
{Haiman}, Z., {Abel}, T., \& {Rees}, M.~J. 2000, \apj, 534, 11

\bibitem[{Haiman \& Bryan(2006)}]{HB06}
Haiman, Z., \& Bryan, G.~L. 2006, ApJL, submitted, preprint astro-ph/0603541

\bibitem[{{Haiman} \& {Holder}(2003)}]{HH03}
{Haiman}, Z., \& {Holder}, G.~P. 2003, \apj, 595, 1

\bibitem[{{Haiman} {et~al.}(1996){Haiman}, {Rees}, \& {Loeb}}]{HRL96}
{Haiman}, Z., {Rees}, M.~J., \& {Loeb}, A. 1996, \apj, 467, 522

\bibitem[{{Haiman} {et~al.}(1997){Haiman}, {Rees}, \& {Loeb}}]{HRL97}
---. 1997, \apj, 484, 985

\bibitem[{{Iliev} {et~al.}(2006){Iliev}, {Ciardi}, {Alvarez}, {Maselli},
  {Ferrara}, {Gnedin}, {Mellema}, {Nakamoto}, {Norman}, {Razoumov},
  {Rijkhorst}, {Ritzerveld}, {Shapiro}, {Susa}, {Umemura}, \&
  {Whalen}}]{Iliev06}
{Iliev}, I.~T. {et~al.} 2006, \mnras, submitted, preprint astro-ph/0603199

\bibitem[{{Iliev} {et~al.}(2005){Iliev}, {Shapiro}, \& {Raga}}]{ISR05}
{Iliev}, I.~T., {Shapiro}, P.~R., \& {Raga}, A.~C. 2005, \mnras, 361, 405

\bibitem[{{Jenkins} {et~al.}(2001){Jenkins}, {Frenk}, {White}, {Colberg},
  {Cole}, {Evrard}, {Couchman}, \& {Yoshida}}]{Jenkins01}
{Jenkins}, A., {Frenk}, C.~S., {White}, S.~D.~M., {Colberg}, J.~M., {Cole}, S.,
  {Evrard}, A.~E., {Couchman}, H.~M.~P., \& {Yoshida}, N. 2001, \mnras, 321,
  372

\bibitem[{{Kitayama} {et~al.}(2004){Kitayama}, {Yoshida}, {Susa}, \&
  {Umemura}}]{KYSU04}
{Kitayama}, T., {Yoshida}, N., {Susa}, H., \& {Umemura}, M. 2004, \apj, 613,
  631

\bibitem[{{Kuhlen} \& {Madau}(2005)}]{KM05}
{Kuhlen}, M., \& {Madau}, P. 2005, \mnras, 363, 1069

\bibitem[{{Lacey} \& {Cole}(1993)}]{LC93}
{Lacey}, C., \& {Cole}, S. 1993, \mnras, 262, 627

\bibitem[{{Liddle} {et~al.}(1996){Liddle}, {Lyth}, {Viana}, \&
  {White}}]{Liddle96}
{Liddle}, A.~R., {Lyth}, D.~H., {Viana}, P.~T.~P., \& {White}, M. 1996, \mnras,
  282, 281

\bibitem[{{Machacek} {et~al.}(2001){Machacek}, {Bryan}, \& {Abel}}]{MBA01}
{Machacek}, M.~E., {Bryan}, G.~L., \& {Abel}, T. 2001, \apj, 548, 509

\bibitem[{{Machacek} {et~al.}(2003){Machacek}, {Bryan}, \& {Abel}}]{MBA03}
---. 2003, \mnras, 338, 273

\bibitem[{MacIntyre {et~al.}(2005)MacIntyre, Santoro, \& Thomas}]{MST05}
MacIntyre, M.~A., Santoro, F., \& Thomas, P.~A. 2005, astro-ph/0510074

\bibitem[{{Mesinger} \& {Haiman}(2004)}]{MH04}
{Mesinger}, A., \& {Haiman}, Z. 2004, \apjl, 611, L69

\bibitem[{{Norman} \& {Bryan}(1999)}]{NB99}
{Norman}, M.~L., \& {Bryan}, G.~L. 1999, in ASSL Vol. 240: Numerical
  Astrophysics, 19--+

\bibitem[{{Oh} \& {Haiman}(2002)}]{OH02}
{Oh}, S.~P., \& {Haiman}, Z. 2002, \apj, 569, 558

\bibitem[{{Oh} \& {Haiman}(2003)}]{OH03}
---. 2003, \mnras, 346, 456

\bibitem[{{Omukai}(2000)}]{Omukai2000}
{Omukai}, K. 2000, \apj, 534, 809

\bibitem[{{O'Shea} {et~al.}(2005){O'Shea}, {Abel}, {Whalen}, \&
  {Norman}}]{OShea05}
{O'Shea}, B.~W., {Abel}, T., {Whalen}, D., \& {Norman}, M.~L. 2005, \apjl, 628,
  L5

\bibitem[{{Page} {et~al.}(2006){Page}, {Hinshaw}, {Komatsu}, {Nolta},
  {Spergel}, {Bennett}, {Barnes}, {Bean}, {Dore'}, {Halpern}, {Hill},
  {Jarosik}, {Kogut}, {Limon}, {Meyer}, {Odegard}, {Peiris}, {Tucker}, {Verde},
  {Weiland}, {Wollack}, \& {Wright}}]{Page06}
{Page}, L. {et~al.} 2006, \apj, submitted

\bibitem[{{Ricotti} {et~al.}(2002{\natexlab{a}}){Ricotti}, {Gnedin}, \&
  {Shull}}]{RGS02a}
{Ricotti}, M., {Gnedin}, N.~Y., \& {Shull}, J.~M. 2002{\natexlab{a}}, \apj,
  575, 33

\bibitem[{{Ricotti} {et~al.}(2002{\natexlab{b}}){Ricotti}, {Gnedin}, \&
  {Shull}}]{RGS02b}
---. 2002{\natexlab{b}}, \apj, 575, 49

\bibitem[{{Scannapieco} {et~al.}(2002){Scannapieco}, {Ferrara}, \&
  {Madau}}]{SFM02}
{Scannapieco}, E., {Ferrara}, A., \& {Madau}, P. 2002, \apj, 574, 590

\bibitem[{{Scannapieco} {et~al.}(2003){Scannapieco}, {Schneider}, \&
  {Ferrara}}]{SSF03}
{Scannapieco}, E., {Schneider}, R., \& {Ferrara}, A. 2003, \apj, 589, 35

\bibitem[{{Schaerer}(2002)}]{Schaerer02}
{Schaerer}, D. 2002, \aap, 382, 28

\bibitem[{{Shapiro} {et~al.}(2004){Shapiro}, {Iliev}, \& {Raga}}]{SIR04}
{Shapiro}, P.~R., {Iliev}, I.~T., \& {Raga}, A.~C. 2004, \mnras, 348, 753

\bibitem[{{Shapiro} \& {Kang}(1987)}]{SK87}
{Shapiro}, P.~R., \& {Kang}, H. 1987, \apj, 318, 32

\bibitem[{{Sheth} \& {Tormen}(1999)}]{ST99}
{Sheth}, R.~K., \& {Tormen}, G. 1999, \mnras, 308, 119

\bibitem[{{Spergel} {et~al.}(2006{\natexlab{a}}){Spergel}, {Bean}, {Dore'},
  {Nolta}, {Bennett}, {Hinshaw}, {Jarosik}, {Komatsu}, {Page}, {Peiris},
  {Verde}, {Barnes}, {Halpern}, {Hill}, {Kogut}, {Limon}, {Meyer}, {Odegard},
  {Tucker}, {Weiland}, {Wollack}, \& {Wright}}]{Spergel05}
{Spergel}, D.~N. {et~al.} 2006{\natexlab{a}}, ApJ, submitted, preprint
  astro-ph/0603449

\bibitem[{{Spergel} {et~al.}(2006{\natexlab{b}}){Spergel}, {Verde}, {Peiris},
  {Komatsu}, {Nolta}, {Bennett}, {Halpern}, {Hinshaw}, {Jarosik}, {Kogut},
  {Limon}, {Meyer}, {Page}, {Tucker}, {Weiland}, \& {Wollack}}]{Spergel06}
---. 2006{\natexlab{b}}, {\apj, submitted}

\bibitem[{{Spergel} {et~al.}(2003){Spergel}, {Verde}, {Peiris}, {Komatsu},
  {Nolta}, {Bennett}, {Halpern}, {Hinshaw}, {Jarosik}, {Kogut}, {Limon},
  {Meyer}, {Page}, {Tucker}, {Weiland}, {Wollack}, \& {Wright}}]{Spergel03}
---. 2003, \apjs, 148, 175

\bibitem[{{Susa} {et~al.}(1998){Susa}, {Uehara}, {Nishi}, \& {Yamada}}]{Susa98}
{Susa}, H., {Uehara}, H., {Nishi}, R., \& {Yamada}, M. 1998, Progress of
  Theoretical Physics, 100, 63

\bibitem[{{Whalen} {et~al.}(2004){Whalen}, {Abel}, \& {Norman}}]{WAN04}
{Whalen}, D., {Abel}, T., \& {Norman}, M.~L. 2004, \apj, 610, 14

\bibitem[{{Wyithe} \& {Loeb}(2003)}]{WL03_postWMAP}
{Wyithe}, J.~S.~B., \& {Loeb}, A. 2003, \apjl, 588, L69

\bibitem[{{Yoshida} {et~al.}(2003){Yoshida}, {Sokasian}, {Hernquist}, \&
  {Springel}}]{Yoshida03}
{Yoshida}, N., {Sokasian}, A., {Hernquist}, L., \& {Springel}, V. 2003, \apjl,
  591, L1

\end{thebibliography}

\begin{appendix}

\section{Comparing Semi-Analytic and Simulation Mass Functions}
\label{sec:test_fcol}

Although it is not directly relevant to the radiative feedback
processes analyzed in the body of this paper, it is interesting to
examine the mass function of dark matter halos we find in our
simulations, and compare these to semi-analytic models.  Such a
comparison is especially interesting, since our simulation corresponds
to a biased region that is overdense on the scale of the simulation
box (rather than fixed to the cosmic mean density).  We are not aware
of a previous study of the halo mass function derived in such a biased
region. Here we present a preliminary comparison, and postpone more
detailed work for a future paper.

According to extended Press-Schechter formalism (EPS), the
contribution of halos with masses greater than $M_{\rm min}$ to the
mass fraction inside regions of mass scale
$M$$\equiv$(4/3)$\pi$$R^3$$\rhoO$ and extrapolated\footnote{We adopt
the standard convention to work with the density field linearly
extrapolated to $z=0$, i.e. $\delta(z=0) = \delta(z)/D(z)$, where D(z)
is the linear growth factor normalized so that $D(z=0)$ = 1 (e.g.,
\citealt{Liddle96}).} mean overdensity $\delbias$, can be expressed as
(e.g., \citealt{Bond91, LC93}):
\begin{equation}
\label{eq:f_col_PS}
\fcol = {\rm erfc} \left\{ \frac{\delc - \delbias}{\sqrt{2 [S(\Mmin) -
S(M_0)]}} \right\},
\end{equation}
where $\delc$ is the critical linear overdensity at halo
virialization, and $S(M)$ is the variance of the present-day linear
overdensity field filtered on scale $M$.  For comparison purposes
below, we chose $\Mmin=10^5\Msun$.

To compare our numerical results with equation (\ref{eq:f_col_PS}), we
chose our biasing scale, $M_0$, such that the enclosed spherical
volume is equal to the volume of the refined region in the simulation,
(4/3)$\pi$$R_0^3$ = $l_{\rm refined}^3$ = 1/64 $(h^{-1} ~ \rm Mpc)^3$,
or equivalently, a mass scale of $M_0$ = $6.4 \times 10^8 ~ \Msun$.
Keeping in the spirit of the linear, Lagrangian nature of EPS, we
obtain our overdensity bias, $\delbias$, from the initial redshift of
our simulation, $\zinit = 99$, before the refined region in our
Eulerian code has a chance to become too contaminated by inflow from
outside particles.  We obtain $\delta(\zinit)=0.1637$ in our refined
region, making our linear overdensity bias $\delbias = \delta(\zinit)
/ D(\zinit) = 12.8$.

\begin{figure}
\vspace{+0\baselineskip}
\myputfigure{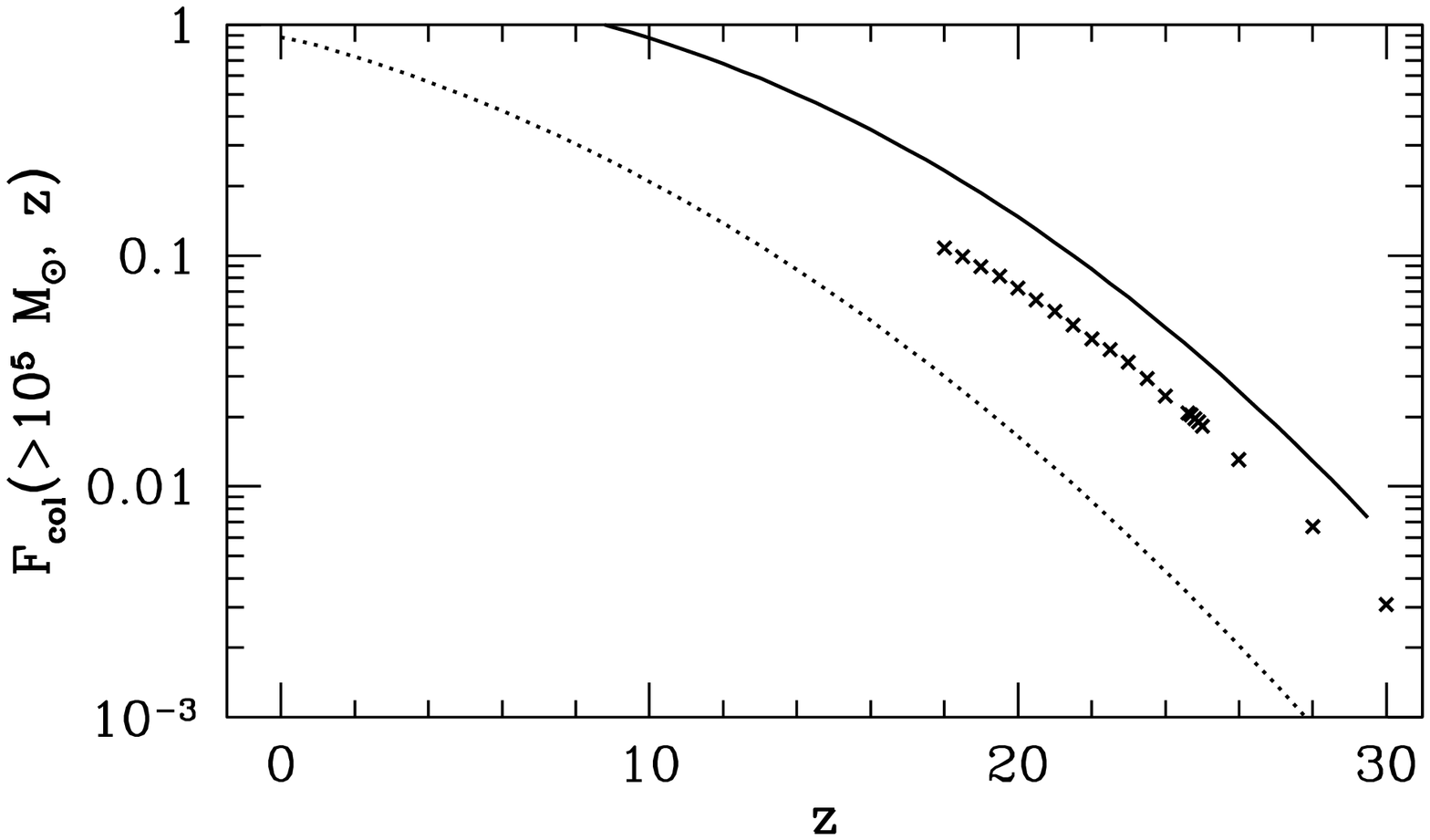}{7}{0.5}{.}{0.}
\vspace{-1\baselineskip} \figcaption{Fraction of mass of the biased
refined volume contained in halos with masses greater than
$10^5\Msun$.  The solid line is determined from the extended
Press--Schechter formalism, using equation (\ref{eq:f_col_PS}).  The
points denote the values obtained from the cosmological simulation by
summing over halos found with the HOP algorithm, in the case of no
background radiation field.  The dashed line is obtained by equation
(\ref{eq:f_col_PS}), but for a mean sample of the universe
(i.e. $\delbias=0$, $S(M_0)=0$).
\label{fig:f_col}}
\vspace{-1\baselineskip}
\end{figure}

We show the comparison of this biased mass fraction with our
simulation in Figure \ref{fig:f_col}.  The solid line is determined by
equation (\ref{eq:f_col_PS}), using the values quoted above.  The
points denote the values obtained from the cosmological simulation by
summing over halos found with the HOP algorithm, in the case of no
background radiation field.  The dashed line is obtained by equation
(\ref{eq:f_col_PS}), but for a mean sample of the universe
(i.e. $\delbias=0$, $S(M_0)=0$).

\begin{figure}
\vspace{+0\baselineskip} \myputfigure{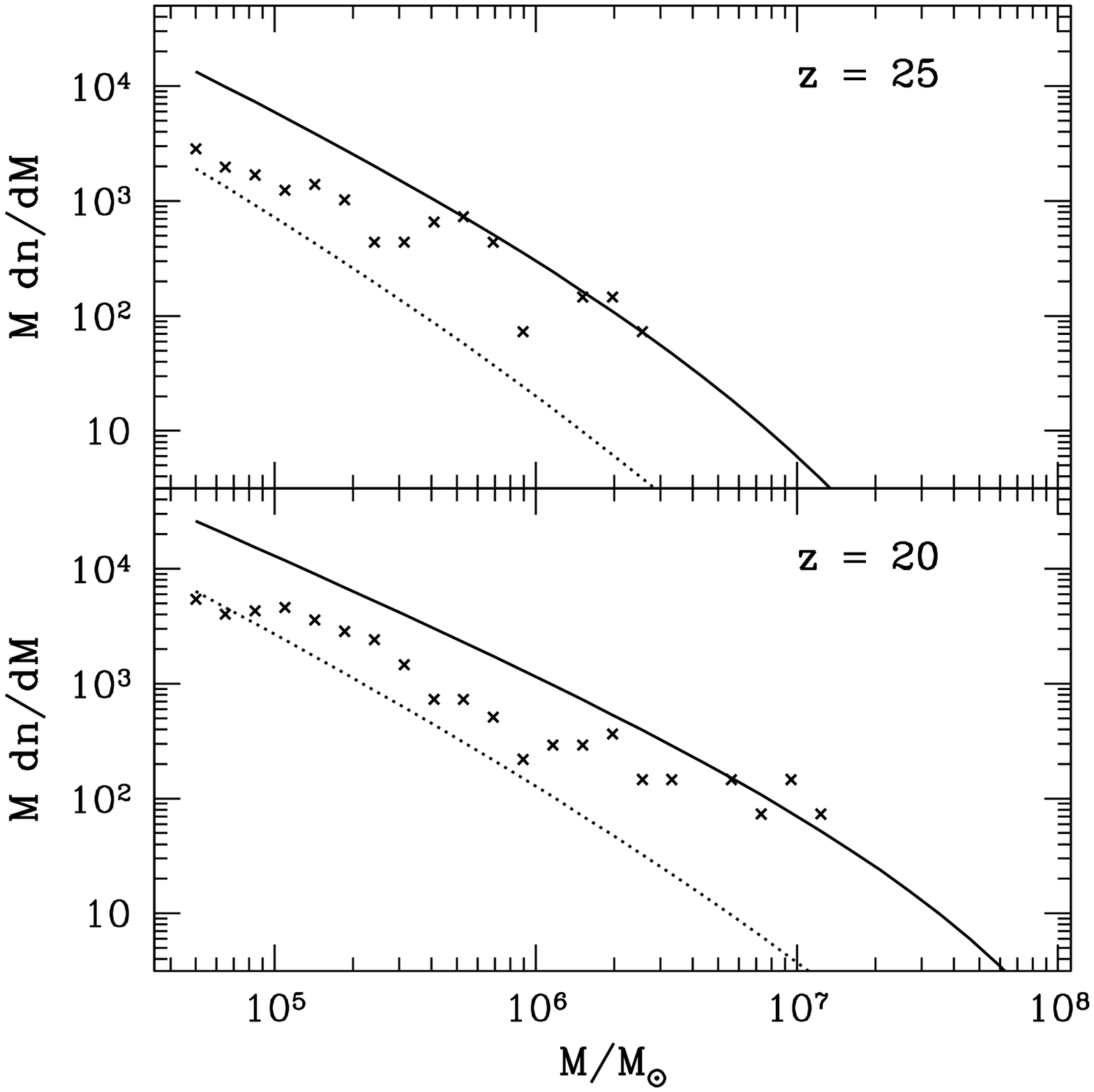}{7}{0.5}{.}{0.}
\vspace{-1\baselineskip} \figcaption{ Mass functions for the models
analogous to Figure \ref{fig:f_col} at $z=25$ ({\it top panel}), and
$z=20$ ({\it bottom panel}).
\label{fig:mf}}
\vspace{-1\baselineskip}
\end{figure}

We note that mass fractions obtained from the simulation are a factor
of $\sim2$ lower than those obtained by EPS.  To further probe this
discrepancy, in Figure \ref{fig:mf} we plot the analogous mass
functions at $z=25$ ({\it top panel}) and $z=20$ ({\it bottom panel}).
It is evident from the figure that abundance of low--mass halos is
over-predicted by EPS with respect to the simulation, while the
abundance of the largest halos in the simulation fits the EPS
prediction fairly well.

This discrepancy might be due to several reasons.  Firstly, it is
already known that EPS mass-functions in the low--redshift regime
suffer from similar over-predictions of low-mass halos, as well as an
under-prediction of high-mass halos, where ``low'' and ``high'' are
defined with respect to the characteristic collapse scale at $z$
\citep{Jenkins01, ST99}, with the mass functions differing by up to a
factor of $\sim 1.6$.  Another contributing factor to the discrepancy
could be the fact that the refined region of our simulation is a {\it
cubical} perturbation, while parameters in the standard EPS are
derived assuming {\it spherical} perturbations of the real-space
density field.
Finally, it has been shown that high--resolution cosmological
simulations are too small to provide accurate mass functions at high
redshifts, since they artificially cut-off the density modes larger
than their box sizes \citep{BL04}.  In particular, \citet{BL04} show
that at $z=20$, the true cosmic mean mass function can be a factor of
several higher than would be derived from a (1 Mpc)$^3$ simulation
box, with periodic boundary conditions normalized to the mean density.
Assuming that our biased simulation box suffers from a similar
underestimate, the mass function could be consistent with the EPS
prediction (after applying the correction proposed by \citealt{ST99}).

\citet{Yoshida03} have obtained a good fit at high redshifts
($z\sim20$) between EPS mass functions and those obtained from
simulations.  \citet{BL04} show that after correcting for the above
missing large--scale power, that their results are consistent with the
EPS mass function with the Sheth--Tormen correction.  However, these
results describe an {\it unbiased} simulation at mean density.  We
present the first comparison of the {\it biased} EPS and numerical
mass functions at high redshifts.  A detailed study on such a
comparison is beyond the scope of this paper, nor does it have an
impact on our main results.

\end{appendix}

\end{document}